\documentclass[10pt,twocolumn,letterpaper]{article}

\usepackage[pagenumbers]{cvpr}   

\usepackage[most]{tcolorbox}
\usepackage{xparse}
\usepackage{soul}

\definecolor{grey}{RGB}{245,245,245}

\newtcolorbox{promptbox}[1]{
    enhanced,
    breakable,
    colback=grey,
    colframe=black,
    fonttitle=\bfseries,
    fontupper=\small\ttfamily,
    before upper={\raggedright\emergencystretch=1em},
    arc=0mm,
    boxrule=0.8pt,
    bottomrule at break=0.4pt,
    toprule at break=0.4pt,
    left=6pt, right=6pt, top=4pt, bottom=6pt,
    width=\linewidth,
    title=#1,
    sharp corners
}

\definecolor{cvprblue}{rgb}{0.21,0.49,0.74}
\usepackage[breaklinks,colorlinks,allcolors=cvprblue]{hyperref}

\def\confName{CVPR}
\def\confYear{2027}

\title{Show Me the Answer: Visual Grounding on Dashboards}

\author{Yuefei Zhang\textsuperscript{1} \quad Songyin Tang\textsuperscript{1} \quad Yile Cai\textsuperscript{1} \quad Yuqi Bu\textsuperscript{2}\\
Liuwu Li\textsuperscript{3} \quad Yanna Lin\textsuperscript{4} \quad Zikun Deng\textsuperscript{1} \quad Yi Cai\textsuperscript{1}\\[0.5em]
{\small\textsuperscript{1}South China University of Technology \quad \textsuperscript{2}Shenzhen Polytechnic University}\\
{\small\textsuperscript{3}Guangxi Normal University \quad \textsuperscript{4}University of Waterloo}
}

\hypersetup{
  pdftitle={Show Me the Answer: Visual Grounding on Dashboards},
  pdfauthor={Yuefei Zhang, Songyin Tang, Yile Cai, Yuqi Bu, Liuwu Li, Yanna Lin, Zikun Deng, Yi Cai}
}

\begin{document}
\makeatletter
\let\supp@orig@addcontentsline\addcontentsline
\renewcommand{\addcontentsline}[3]{}
\makeatother
\maketitle

\begin{abstract}
Natural language interaction has become an increasingly important paradigm for accessing complex visual information systems. In visual data analytics environments, such as dashboards, a key capability is to ground user queries to specific visual elements, enabling systems to identify the visual evidence relevant to the query.
However, existing visual grounding research has primarily focused on natural images, while grounding in dashboards remains underexplored due to their complex multi-chart layouts and fine-grained visual elements.
In this paper, we study the task of \textbf{V}isual \textbf{G}rounding on \textbf{D}ashboards (\textbf{VGD}):
Given a dashboard screenshot and a natural language query, the goal is to localize the visual element corresponding to the query using a bounding box.
We propose VGD-Net, a two-stage grounding framework that reflects how users typically locate information in dashboards, by first performing query-aware chart retrieval and then fine-grained in-chart grounding.
To systematically evaluate VGD-Net's performance on this task, we construct VGD-Bench, a benchmark specifically designed for VGD, covering three complementary settings: chart retrieval, in-chart grounding, and overall VGD.
Experimental results show that our two-stage VGD-Net outperforms strong multimodal baselines on the VGD task, achieving a mean IoU of 57.06 and a success rate of 66.82, while also delivering superior performance over existing methods in both stages of chart retrieval and in-chart grounding.
These results highlight the value of stage decomposition for reliable and interpretable VGD, and its potential to support query-driven visual analysis.
\end{abstract}

\section{Introduction}

\begin{figure}[t]
    \centering
    \includegraphics[width=0.95\columnwidth]{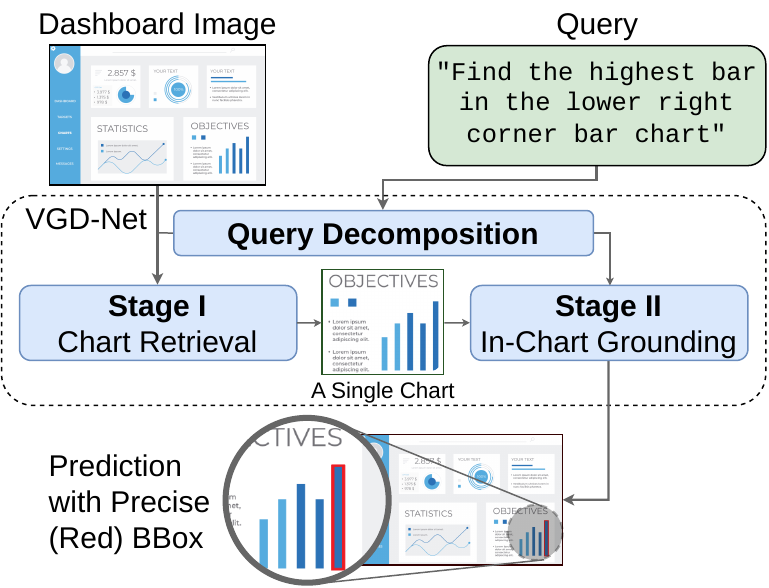}
    \caption{Overview of visual grounding on a dashboard.}
    \label{fig:teaser}
\end{figure}

Dashboards are widely used for presenting complex analytical results through collections of charts and visual components~\cite{sarikaya2019visual}. A fundamental task in such environments is to locate elements relevant to analysis (e.g., locating the highest bar in a chart).
While existing multimodal systems, such as visual question answering~\cite{masry2025chartgemma}, can provide textual answers, they primarily focus on question answering and often overlook explicit grounding of the visual elements~\cite{reich-schultz-2024-uncovering}. Hence, the spatial correspondence between answers and their supporting visual evidence remains unclear.

In this paper, we formulate a task called \textbf{VGD} (\textbf{V}isual \textbf{G}rounding on \textbf{D}ashboards), which localizes elements referred to by natural language queries in dashboard images. Given a dashboard and a query, the system predicts the target element using a bounding box. As shown in Fig.~\ref{fig:teaser}, when presented with a query such as ``Find the highest bar in the lower right corner bar chart,'' the system must comprehend the dashboard layout and output a precise bounding box around the target, as shown by the red box in Fig.~\ref{fig:teaser}.

By achieving VGD, we aim to improve the overall visual data analysis from three aspects.
First, VGD improves \textbf{interpretability} by providing explicit visual references for predicted answers.
Grounding relevant visual elements establishes clear correspondence between textual queries and structured visual information, making model outputs more transparent and trustworthy~\cite{yu2026sciegqadatasetscientificevidencegrounded}.
Second, it can further facilitate \textbf{semantic alignment}, as prior work~\cite{INFODET2025, RefChartQA2025} suggests that grounding relevant elements could improve the performance of reasoning tasks by enabling models to focus on the correct visual evidence.
Finally, VGD also enhances \textbf{accessibility} by enabling more flexible interaction processes in interactive visual analytics environments.
By supporting flexible natural language inputs, users can directly express their analytical intent for efficient information localization, which in turn facilitates downstream interactions such as selection and filtering~\cite{Qian2024VisualUI}, and further supports diverse multi-user collaborative scenarios with mixed input modalities.

While visual grounding (VG) has been extensively studied in natural images~\cite{xiao2026vg_survey}, dashboards differ fundamentally in their visual and structural characteristics, making VGD non-trivial.
Dashboards are structured collections of charts and visual elements with explicit spatial relationships. Multiple charts of the same type and style often co-exist and are arranged in a regular layout across the dashboard, while visual elements within each chart (e.g., bars) can appear highly similar.
Existing VG methods fail to exploit this structured layout while handling visual similarity across charts and elements.
In addition, VGD is driven by analysis-oriented queries that refer to chart content rather than isolated objects, further increasing the difficulty of VGD~\cite{ChartAB}.
When human users locate visual elements in a visual information system, they naturally exploit the underlying hierarchical structure~\cite{DBLP:conf/vl/Shneiderman96,Hornof01092004}: first identifying the relevant chart and then focusing on visual elements within it.
Inspired by this strategy, we propose \textbf{VGD-Net}, a \textbf{two-stage} framework that decomposes VGD into chart retrieval and in-chart grounding, enabling structured and query-driven reasoning over dashboards.
To align linguistic semantics with this decomposition, VGD-Net reformulates each user query into stage-specific sub-queries~\cite{khot2023decomposed}, namely a chart-retrieval query and an in-chart grounding query, which capture different levels of intent and inform the grounding process.
In Stage I, VGD-Net performs query-aware chart retrieval by detecting candidate charts and ranking them based on their relevance to the chart-level query, effectively narrowing down the search space.
In Stage II, conditioned on the selected chart, VGD-Net employs multi-level cross-modal features to localize the fine-grained query element.



To enable the development and evaluation of VGD-Net for VGD, we construct the first VGD-specific dataset, \textbf{VGD-Bench}.
It contains paired dashboards, natural language queries, and bounding box annotations for the target visual elements, reflecting realistic visual analysis tasks.
The dataset is also further organized to match the two-stage decomposition of our method, with annotations suitable for both chart retrieval (Stage I) and in-chart grounding (Stage II).
In addition, we design evaluation metrics tailored to the VGD task, complementing standard grounding measures (IoU and accuracy) to account for small targets and complex layouts.
As a benchmark, VGD-Bench enables comparison of methods, including multimodal large language models as well as specialized visual grounding approaches, providing a testbed for both stage-wise and overall VGD evaluation.

Extensive experimental results demonstrate that VGD-Net consistently outperforms strong existing baselines, which further underscores the value of stage decomposition for achieving more trustworthy visual grounding, and its potential for query-driven visual analysis and data exploration.

In summary, our contributions are:
\begin{itemize}
  \item We formulate \textbf{VGD} (Visual Grounding on Dashboards) as a task of localizing associated visual elements in multi-chart dashboards based on given natural language queries.
  \item We propose \textbf{VGD-Net}, a two-stage VGD framework inspired by how humans search dashboards, which decomposes grounding into chart retrieval and in-chart grounding to leverage the hierarchical structure of dashboards.
  \item We introduce \textbf{VGD-Bench}, the first human-curated real-world VGD benchmark with 777 dashboards and 2167 natural language queries, and chart retrieval and in-chart grounding annotations for systematic evaluation of VGD.

\end{itemize}

\section{Related Work}

\textbf{Chart and Dashboard Understanding}.
Early chart understanding focused on parsing and data recovery.
ChartOCR~\cite{DBLP:conf/wacv/LuoLWL21}
extracts chart primitives and reconstruct underlying numerical data. 
Recent approaches such as ChartParser~\cite{ChartParser2022}, StructChart~\cite{xia2026structchart}, and ChartGemma~\cite{masry2025chartgemma} combine visual perception with structural reasoning for chart visual question answering but do not localize visual evidence.
INFODET~\cite{INFODET2025} provides bounding boxes for infographics, but it focuses on layout detection rather than grounding fine-grained visual elements like bars. 
Datasets such as RefChartQA~\cite{RefChartQA2025} and START~\cite{START2025} provide bounding box annotations but target single-chart tasks rather than VGD with multiple charts.
As a result, existing chart and dashboard understanding methods are insufficient for VGD.

\begin{figure*}[tb!]
    \centering
    \includegraphics[width=\textwidth]{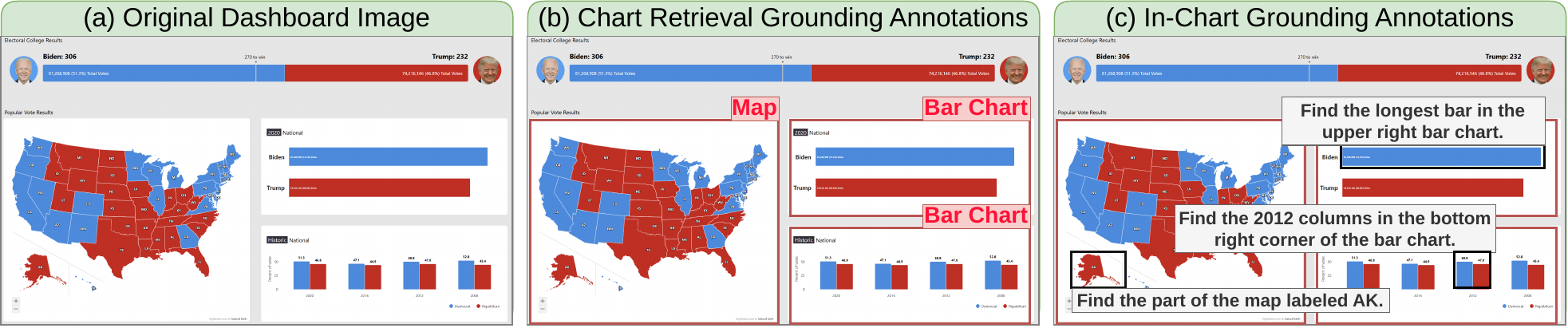}
    \caption{\textbf{VGD-Bench annotation workflow.} 
    (a) A sample dashboard adapted from a Highcharts demo~\cite{HighchartsElection}; 
    (b) Chart Retrieval Grounding Annotations and (c) In-Chart Grounding Annotations, showing VGD-Bench's hierarchical annotation scheme for localizing chart elements.}

    \label{fig:dataset_annotation}
\end{figure*}

\textbf{Object Detection and Visual Grounding}.
In multi-chart dashboards, VGD is challenging due to the co-existence of multiple, visually similar charts and fine-grained elements within each chart. 
Identifying the target chart can be formulated as an object detection problem.
CNN-based detectors (e.g., YOLOv8) and transformer-based architectures (e.g., RT-DETR~\cite{RTDETR2024}) have demonstrated strong capabilities for localizing objects in complex scenes.
Locating specific visual elements described by natural language requires fine-grained grounding.
Visual grounding methods, like TransVG~\cite{TransVG2021}, SeqTR~\cite{SeqTR2022}, SimVG~\cite{SimVG2024}, OneRef~\cite{xiao2024oneref} and BARE~\cite{li2026bare} have shown robustness on object-centric natural images. 
However, VGD application remains challenging: visual elements are often abstract, and separated by subtle geometric differences; deep features may lose the fine-grained cues necessary to components.
Open-world approaches like Grounding DINO~\cite{liu2024groundingdino} further achieve strong performance on natural images but are similarly limited in VGD. GUI grounding methods~\cite{cheng2024seeclick} can locate abstract screen components based on semantics and spatial cues.
Yet, VGD requires multi-chart grounding with quantitative reasoning.
In this work, we address VGD with a two-stage framework that leverages model strengths for multi-chart and fine-grained grounding.

\section{Dataset: VGD-Bench}
\label{sec:dataset}

We construct \textbf{VGD-Bench}, a benchmark comprising 777 real-world dashboards and 2,167 human-written queries. Unlike prior chart-related datasets, VGD-Bench is designed to evaluate the precise spatial localization of visual evidence. It provides (i) chart retrieval grounding annotations and (ii) query-aligned in-chart grounding annotations, enabling the evaluation and fine-tuning of both isolated stages and overall VGD, while supporting diverse approaches. Further details are provided in the supplementary material.

\begin{table}[tb!]
\centering
\caption{Detailed distribution of chart types across 6,259 total charts and 2,167 total queries derived from the 777 dashboards in VGD-Bench. 
The Charts and Queries columns represent chart retrieval annotations and the query-associated charts, respectively.}
\label{tab:data_distribution}
\fontsize{9}{10.8}\selectfont
\setlength{\tabcolsep}{2.6pt}
\resizebox{\columnwidth}{!}{
\begin{tabular}{l c c | l c c}
\toprule
\textbf{Category} & \textbf{Charts} & \textbf{Queries} & \textbf{Category} & \textbf{Charts} & \textbf{Queries} \\
\midrule
Card              & 2,174 (34.7\%) & 363 (16.8\%)   & Progress Bar    & 76 (1.2\%)     & 9 (0.4\%)     \\
Bar Chart         & 1,260 (20.1\%) & 712 (32.9\%)   & Dial Chart      & 66 (1.1\%)     & 21 (1.0\%)    \\
UI                & 549 (8.8\%)    & 32 (1.5\%)     & Image           & 50 (0.8\%)     & 8 (0.4\%)     \\
Line Chart        & 520 (8.3\%)    & 311 (14.4\%)   & Scatter Plot    & 46 (0.7\%)     & 34 (1.6\%)    \\
Data Table        & 337 (5.4\%)    & 151 (7.0\%)    & Heatmap         & 46 (0.7\%)     & 25 (1.2\%)    \\
Share             & 331 (5.3\%)    & 32 (1.5\%)     & Bubble Chart    & 41 (0.7\%)     & 37 (1.7\%)    \\
Donut Chart       & 244 (3.9\%)    & 190 (8.8\%)    & Radar Chart     & 28 (0.4\%)     & 24 (1.1\%)    \\
Area Chart        & 220 (3.5\%)    & 73 (3.4\%)     & Calendar        & 12 (0.2\%)     & 5 (0.2\%)     \\
Map               & 171 (2.7\%)    & 88 (4.1\%)     & Gantt Chart     & 6 (0.1\%)      & 1 (0.0\%)     \\
Pie Chart         & 77 (1.2\%)     & 48 (2.2\%)     & Time Line       & 5 (0.1\%)      & 3 (0.1\%)     \\
\bottomrule
\end{tabular}
}
\end{table}

\subsection{Data Collection and Statistical Insights}

To ensure the benchmark reflects real-world complex scenarios, dashboards are collected from diverse public platforms (e.g., Dribbble and Tableau). 
Several representative samples with their annotations are provided at the end of the supplementary material.
As shown in Table~\ref{tab:data_distribution}, the dataset comprises \textbf{6,259} bounding boxes for chart retrieval and \textbf{2,167} query-aligned for in-chart grounding. 

A key characteristic of VGD-Bench is its dense distribution of visual elements, typical of real-world dashboards. In the chart retrieval stage (see Table~\ref{tab:data_distribution}, Charts column), components such as Numeric Cards (34.7\%) and UI elements (8.8\%) dominate the visual space. However, in the in-chart grounding stage (see Table~\ref{tab:data_distribution}, Queries column), annotated queries primarily target data-driven visualizations, such as Bar Charts (32.9\%) and Line Charts (14.4\%). This change highlights a notable challenge: models must first navigate through visual distractions to isolate the appropriate charts.

\subsection{Hierarchical Annotation and Queries}

We implement visual grounding using a bounding box paradigm, which is effective for identifying structurally well-defined chart elements. To construct high-quality annotations, we developed a systematic, two-stage hierarchical annotation pipeline conducted by trained annotators familiar with chart reasoning tasks (illustrated in Fig.~\ref{fig:dataset_annotation}):

\textbf{Stage I: Chart Retrieval Grounding Annotation}. In the first stage, annotators carefully draw bounding boxes and assign type labels to all independent components within the dashboard, including various chart types, numeric cards, maps, textual components, and UI elements. This step establishes the structural foundation of the overall dashboard.

\textbf{Stage II: In-Chart Grounding Annotations}. In the second stage, annotators formulate 2--3 natural language queries on average for each dashboard while simultaneously identifying and precisely localizing their corresponding answers. To simulate realistic user intents and standardize the linguistic structure, the formulated queries are primarily localization-oriented (e.g., expressions such as \emph{Where}, \emph{Find}, or \emph{Locate}) and comprise two essential components:

\begin{itemize}
    \item \textbf{Target Chart Description:} Information to disambiguate the chart from others (e.g., chart type, position, or title).
    \item \textbf{In-chart Target Description:} Fine-grained visual or data attributes of the target element (e.g., ``the highest bar'').
\end{itemize}

For each structured query (e.g., \emph{``Where is the highest bar in the left bar chart?''}), annotators meticulously draw a single bounding box that localizes the \textit{minimal visual region} required to answer it. This joint annotation process ensures that the queries inherently demand both macro-level chart selection and fine-grained micro-level detail discrimination.

\subsection{Quality Assurance and Expert Review}

Given the complex nature of dashboard layouts and the potential for subjective ambiguity in grounding minimal regions, we avoided crowdsourcing platforms that often yield noisy bounding boxes. Instead, the annotation was entirely conducted by an expert group of the authors. 
To ensure consistency and high quality, each annotated dashboard (including bounding box tightness and query clarity) was independently reviewed by at least one other annotator (see supplementary material).
Only 124 of 2,167 annotations (5.7\%) were revised after review.
Disagreements were discussed among annotators and resolved through consensus.

\section{VGD-Net}
\label{sec:method}

\begin{figure*}[tb!]
    \centering
    \includegraphics[width=\textwidth]{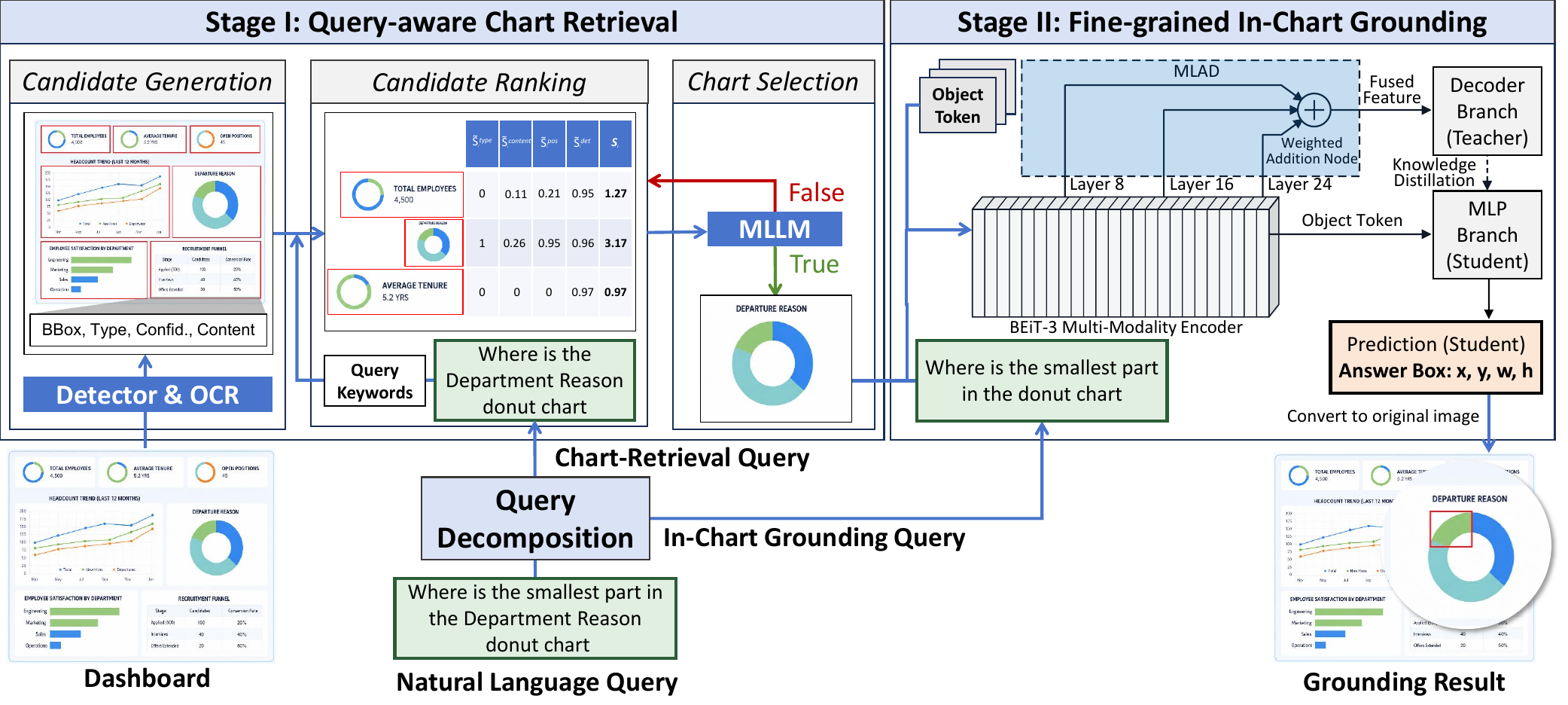}
    \caption{Architecture of the \textbf{VGD-Net} framework.}
    \label{fig:pipeline}
\end{figure*}

\subsection{Task Definition}
We define Visual Grounding on Dashboards (\textbf{VGD}) as the task of predicting a bounding box $b$ that localizes the visual evidence corresponding to a natural language query $Q$ within a dashboard image $I_d$: $(I_d, Q) \mapsto b$,
where $b$ denotes the bounding box enclosing the target region in the image.

Inspired by the hierarchical visual information-seeking behaviors of human users in visual information systems~\cite{DBLP:conf/vl/Shneiderman96,Hornof01092004}, we decompose the VGD task into two sequential grounding stages:
\textbf{(Stage I) Query-aware Chart Retrieval}, which selects the chart relevant to the query, and 
\textbf{(Stage II) Fine-grained In-Chart Grounding}, which grounds the target visual element within the selected chart.


\subsection{VGD-Net Architecture}

Due to the unique challenges of VGD, we propose \textbf{VGD-Net} (Fig.~\ref{fig:pipeline}), a framework incorporating the two-stage decomposition described above. VGD-Net operates on complex multi-chart dashboards, requiring the most relevant chart to be retrieved before subsequent fine-grained grounding. The overall framework is composed of three major functional modules, namely, query decomposition, query-aware chart retrieval, and fine-grained in-chart grounding.

\textbf{Query Decomposition}.
A valid natural language query that can locate a visual element in a dashboard typically specifies \emph{which chart} to attend to and \emph{which target element} to locate within it, either explicitly or implicitly.
To align with our two-stage grounding framework, we introduce a Query Decomposition module (Fig.~\ref{fig:pipeline}) that splits a query into stage-specific sub-queries:
$Q \rightarrow \left( Q^{\text{chart}},\; Q^{\text{in-chart}} \right)$, 
where $Q^{\text{chart}}$ captures chart retrieval intent for Stage I, and $Q^{\text{in-chart}}$ captures in-chart grounding intent for Stage II.
This task can be performed by a lightweight model (e.g., T5~\cite{raffel2020t5}) or an LLM (e.g., DeepSeek~\cite{DeepSeek}, as adopted in our implementation; see the supplementary material for the prompts).

This decomposition is necessary because a VGD query often inherently entangles chart-retrieval and in-chart grounding, introducing ambiguity for stage-specific reasoning. 
For example, given the query \emph{``Find the highest bar in the top-right bar chart,''} the two stages receive potentially conflicting signals: 
Stage I should identify the chart rather than a specific bar, while Stage II operates on a cropped chart where the reference to ``top-right'' is no longer valid. After decomposition, the query is reformulated as 
$Q^{\text{chart}}$: \emph{``Find the top-right bar chart''} for Stage I, and 
$Q^{\text{in-chart}}$: \emph{``Find the highest bar in the chart''} for Stage II. 
This separation provides each stage with a well-defined objective, reducing ambiguity and better decoupling the two stages~\cite{Yu_2018_CVPR}.

\textbf{Stage I: Query-aware Chart Retrieval}.
Given a dashboard image $I_d$, the goal is to select the chart corresponding to the query $Q^{\text{chart}}$. This stage performs \textit{query-aware chart retrieval}, corresponding to the \textbf{Stage I} module in Fig.~\ref{fig:pipeline}.

\textit{Candidate Generation.}
We first generate a set of candidate charts from the dashboard image.
This stage aims to achieve high recall by locating potential charts.
Formally, given a dashboard image $I_d$, we obtain a set of candidates: $\mathcal{D} = \{(b_i, c_i, s_i^{det})\}_{i=1}^{N}$, 
where $b_i$ denotes the bounding box, $c_i$ the predicted chart type, and $s_i^{det}$ the confidence score.
In our implementation, we instantiate this procedure using the RT-DETR-x detector~\cite{RTDETR2024}, a transformer-based object detector, due to its consistently notable performance in detecting chart-like structures.
A comparatively low confidence threshold (e.g., $\tau = 0.1$) is applied to preserve high recall.




\textit{Candidate Ranking.}
Given multiple candidate charts, we need to rank them according to their relevance to the query.
This requires matching chart properties with the query regarding the chart type, spatial position, and textual content.

To this end, we first parse the query $Q^{\text{chart}}$ to extract query keywords related to chart type, spatial position, and textual content.
Meanwhile, we apply OCR to extract textual information from the dashboard, which is associated with candidate charts via spatial heuristics.
Subsequently, each candidate chart is scored using four components.
First, we compute a type score $s_i^{type}$ by comparing the predicted chart type with query keywords.
Second, we compute a content score $s_i^{content}$ by measuring the textual similarity between OCR-extracted text and query keywords using the longest matched substring.
Third, we compute a position score $s_i^{pos}$ using a Gaussian distance-to-anchor score on normalized box-center coordinates ~\cite{Zhou2019Objects}, where candidates closer to the query-implied spatial position receive higher scores.
Finally, we incorporate the detector confidence $s_i^{det}$ produced by the detector.
The final ranking score is:
\begin{equation}
S_i = \tilde{s}_i^{det} + \tilde{s}_i^{type} + \tilde{s}_i^{pos} + \tilde{s}_i^{content},
\end{equation}
where each $\tilde{s}_i$ corresponds to a normalized score in $[0,1]$.



\textit{Chart Selection.}
The top-ranked candidate is visualized as a red bounding box on the dashboard and verified by the model Qwen3-VL-Plus~\cite{Qwen3VL2025}, using $Q^{\text{chart}}$ (MLLM check module in Fig.~\ref{fig:pipeline}).
The detailed prompt is provided in the supplementary material.
If the result is \textbf{True}, the candidate is accepted; otherwise, the next candidate is checked until a valid box is found or all candidates are exhausted~\cite{li-etal-2024-llatrieval}. After verification, the final chart is determined as follows:

\begin{equation}
\hat{i} = \arg\max_{i \in \mathcal{D}} S_i, \qquad b^{\text{chart}} = b_{\hat{i}}.
\end{equation}

The image is then cropped according to $b^{\text{chart}}$ to obtain the chart, which is passed to Stage II for fine-grained in-chart grounding. Hence, Stage I first grounds the correct chart, largely addressing multi-chart challenges in VGD.

\textbf{Stage II: Fine-grained In-Chart Grounding}.
Based on the chart selected by Stage I, Stage II performs fine-grained in-chart grounding to precisely locate the specific target element corresponding to $Q^{\text{in-chart}}$, as illustrated in the \textbf{Stage II} module of Fig.~\ref{fig:pipeline}.
This task requires capturing both multi-level visual elements (e.g., point, bar, donut slice) and language semantics (e.g., ``the highest value,'' ``the red bar,'' or ``the point at 2010'') in semantic alignment with the query.

To this end, we select the SimVG framework~\cite{SimVG2024}, a VLP-based visual grounding method, based on preliminary comparisons with several alternative grounding models (see the middle columns of Table~\ref{tab:all_results}).
SimVG leverages a one-tower pre-trained backbone (BEiT-3~\cite{wang2023beit3}) to encode visual and textual inputs in a shared feature space, and we preserve its native $16\times16$ patch projection.
Different layers capture information at varying abstraction levels, from fine-grained details to high-level semantics.
To better utilize these multi-level representations, we introduce a lightweight Multi-Layer Aggregation Module (MLAD) (the blue MLAD module in Fig.~\ref{fig:pipeline}), which aggregates intermediate features from BEiT-3 layers 8, 16, and 24.
By integrating features across layers, MLAD enhances the model's ability to incorporate both visual details and contextual cues, specifically addressing the multi-level grounding challenge highlighted above.


Formally, given the cropped chart image $I^{*}$ obtained from $b^{\text{chart}}$ and the in-chart grounding query $Q^{\text{in-chart}}$, the model outputs the bounding box of the target element as: 
\begin{equation}
b = f_{\theta}(I^{*}, Q^{\text{in-chart}}),
\end{equation}
where $f_{\theta}$ is the MLAD-enhanced SimVG model, and $b = (x, y, w, h)$ is the bounding box. 
MLAD assigns a scalar weight to each selected intermediate feature.
Let $\mathcal{L}=\{8,16,24\}$. The normalized weights are $\alpha_l=\exp(w_l)/\sum_{j\in\mathcal{L}}\exp(w_j)$, 
and the fused feature is
$F_{\text{fused}}=\sum_{l\in\mathcal{L}}\alpha_lF_l$, 
where $w_l$ is jointly optimized during training.


\section{Experiments}
\label{sec:experiments}

\subsection{Dataset and Implementation}
Following VGD-Bench, we split it into train/val/test sets (80/10/10) by dashboards to avoid content overlap and use decomposed queries.
The dataset was split using random partitioning, with the selected split maximizing chart-type coverage across splits.
Stage~I uses AdamW (learning rate $3\times10^{-4}$, batch size 4), while Stage~II uses Adam (learning rate $5\times10^{-5}$, batch size 1), on a single RTX~4090 (24\,GB) GPU.
During inference, each stage applies its own preprocessing. Details are provided in the supplementary material.

\textbf{Stage I:} Full dashboard images with VGD-Bench chart retrieval grounding annotations are used at their original image resolution, with the chart-retrieval query $Q^{\text{chart}}$ as input.

\textbf{Stage II:} Chart crops are obtained using the ground-truth chart bounding boxes from VGD-Bench, with In-Chart Grounding boxes as supervision. Each crop is resized to $640 \times 640$ without preserving the original aspect ratio, and the input text is the in-chart grounding query $Q^{\text{in-chart}}$.

Data splitting and training information are provided in the supplementary material.

\subsection{Evaluation Metrics}

To fully evaluate VGD-Net, we utilize standard localization metrics alongside specialized metrics to address the bias of IoU against small or fine-grained visual elements~\cite{LeJeune2023SIoU}. 
\begin{itemize}
    \item \textbf{IoU \& Acc@0.5 (Acc)}: Mean IoU of predicted vs. ground-truth boxes; Acc@0.5 = fraction with IoU $> 0.5$.
    \item \textbf{Dot Distance (DotD)}~\cite{Xu2021DotDistance}: measures the center alignment between boxes by computing
    $D=\|\mu_p-\mu_g\|_2$,
    where $\mu_p$ and $\mu_g$ are the predicted and ground-truth box centers. The resulting score is
    $\mathrm{DotD}=\exp(-D/S)$,
    where $S$ is the average bounding box size in VGD-Bench.

    \item \textbf{Normalized Wasserstein Distance (NWD)}~\cite{Wang2021NWD}: NWD models each bounding box $b=(c_x,c_y,w,h)$ as a 2D Gaussian distribution $\mathcal{N}(\mu,\Sigma)$, where $\mu=(c_x,c_y)$ is the center and $\Sigma$ is a covariance matrix constructed from the box dimensions. The similarity between the predicted distribution $\mathcal{N}_p$ and the ground truth $\mathcal{N}_g$ is measured by: 
    \begin{equation}
        W(\mathcal{N}_p,\mathcal{N}_g) = \sqrt{\|\mu_p-\mu_g\|_2^2+\left\|\Sigma_p^{1/2}-\Sigma_g^{1/2}\right\|_F^2}.
    \end{equation}
    The distance is normalized to $\mathrm{NWD}=\exp(-W/C)$, where $C$ is a dataset-specific constant in practice overall.
    
    \item \textbf{Scale-aware Average Precision (AP)}: Following the COCO protocol~\cite{Lin2017FPN}, $\text{AP}_s$, $\text{AP}_m$, and $\text{AP}_l$ are reported in our experiments for small, medium, and large targets across IoU thresholds from 0.50 to 0.95 in steps of 0.05.
\end{itemize}

While IoU and Acc are used as standard metrics across all settings, DotD, NWD, and scale-aware AP complement IoU/Acc for fine-grained and small-object localization. Together, these metrics provide a comprehensive evaluation of grounding quality across various scales. For clarity and consistency, all results are reported on a scale of \textbf{0 to 100}.

\subsection{Baselines / Model Categories}
We group all evaluated methods into three categories:

    \textbf{General MLLMs}: We select Mistral Medium 3\cite{MistralMedium3_2025}, InternVL2.5~\cite{chen2026internvl25}, GPT-4o~\cite{GPT4o2024}, GPT-5~\cite{OpenAI_GPT5_2026}, Gemini-2.5 Pro~\cite{Gemini2_5_2025}, Claude 4 Sonnet~\cite{Claude4Sonnet2025}, Llama-4 Maverick~\cite{Llama4_2025}, GLM-4.5V~\cite{GLM4_5V2025}, Qwen2.5-VL~\cite{bai2025qwen25vl}, and Qwen3-VL-Plus~\cite{Qwen3VL2025}.
    They are accessed via APIs or used as released. Prompts are given in the supplementary material.

    \textbf{Chart-specific / Literature Models (No fine-tuning)}: 
    We select some key vision-language understanding and visual grounding models, including CLIP~\cite{Radford2021CLIP}, OWLv2~\cite{minderer2023scaling}, KOSMOS-2~\cite{peng2023kosmos2}, and Florence-2~\cite{xiao2024florence2}.
    These models are all evaluated zero-shot, as adapting them to box supervision requires architectural changes (e.g., CLIP has no box head).

    \textbf{Fine-tuned on VGD-Bench}: 
    We select visual grounding models and other models that can be fine-tuned on our dataset, including TransVG~\cite{TransVG2021}, MDETR~\cite{MDETR2021}, SeqTR~\cite{SeqTR2022}, Grounding DINO~\cite{liu2024groundingdino}, LLaVA~\cite{liu2023visual}, LLaVA-NeXT~\cite{liu2024llavanext}, PaliGemma~\cite{paligemma2024}, OneRef~\cite{xiao2024oneref}, InstanceVG~\cite{dai2026instancevg}, SimVG~\cite{SimVG2024}, and BARE~\cite{li2026bare}.
    We also fine-tune InternVL2.5~\cite{chen2026internvl25} and Qwen2.5-VL~\cite{bai2025qwen25vl}.
    To ensure fairness, model-specific settings follow their original recommendations, with minor adaptations for VGD-Bench. Details of the adaptations are provided in the supplementary material.

We evaluate all methods on VGD-Bench in three settings: 
    (1) \textbf{Stage I} (Query-aware Chart Retrieval), input is the full dashboard and the output is the chart bounding box, evaluated using IoU and Acc; 
    (2) \textbf{Stage II} (Fine-grained In-Chart Grounding), input is the cropped chart (oracle) and output is the target element's bounding box, evaluated by IoU, Acc, DotD, NWD, and AP$_s$/AP$_m$/AP$_l$; 
    (3) \textbf{Overall VGD}, performing full-task grounding given the entire dashboard and original query with the same metrics as Stage II. 

Table~\ref{tab:all_results} reports the results of all methods in each setting. 
For all methods other than VGD-Net, we also evaluate their performance on overall VGD using a two-stage pipeline, with the results reported in the supplementary material.

\subsection{Main Results and Baseline Comparison}
\begin{table*}[t]
\centering
\caption{
Performance comparison on VGD-Bench.
Stage I performs query-aware chart retrieval.
Stage II performs fine-grained in-chart grounding.
Overall VGD directly predicts the answer region from the dashboard.
The full VGD-Net comprises ``RT-DETR-x + Ranking + MLLM check'' for Stage I and ``SimVG + MLAD + $16\times16$ patch'' for Stage II.
Methods marked with $^{\star}$ use resized inputs rather than the original image resolution; detailed resolutions and all implementation details for this table are provided in the supplementary material.
}
\label{tab:all_results}
\fontsize{9}{10.8}\selectfont
\setlength{\tabcolsep}{1.8pt}
\begin{tabular}{lcc|ccccccc|ccccccc}
\toprule

&
\multicolumn{2}{c|}{\textbf{Stage I}}
&
\multicolumn{7}{c|}{\textbf{Stage II}}
&
\multicolumn{7}{c}{\textbf{Overall VGD}}
\\

\cmidrule(lr){2-3}
\cmidrule(lr){4-10}
\cmidrule(l){11-17}

\textbf{Method}
&
\textbf{IoU}
&
\textbf{Acc}
&
\textbf{IoU}
&
\textbf{Acc}
&
\textbf{DotD}
&
\textbf{NWD}
&
\textbf{AP$_s$}
&
\textbf{AP$_m$}
&
\textbf{AP$_l$}
&
\textbf{IoU}
&
\textbf{Acc}
&
\textbf{DotD}
&
\textbf{NWD}
&
\textbf{AP$_s$}
&
\textbf{AP$_m$}
&
\textbf{AP$_l$}
\\

\midrule

\multicolumn{17}{l}{\emph{General MLLMs}}\\

Mistral Medium 3 &
35.62 & 25.39 & 
10.29 & 4.67 & 50.59 & 65.87 & 0.00 & 0.19 & 5.07 & 
5.73 & 2.35 & 32.12 & 49.08 & 0.00 & 0.00 & 0.00\\

InternVL2.5-8B &
12.12 & 3.23 &
6.53 & 4.61 & 30.75 & 37.71 & 0.00 & 0.28 & 6.84 &
2.50 & 1.38 & 8.34 & 15.73 & 0.00 & 0.19 & 0.88\\

GPT-4o & 38.50 & 35.98 & 
17.80 & 8.41 & 58.72 & 70.23 & 0.00 & 0.78 & 6.93 & 
7.70 & 3.29 & 31.61 & 45.34 & 0.00 & 0.00 & 2.21 \\

GPT-5 & 54.80 & 61.66 & 
27.13 & 22.90 & 68.14 & 78.04 & 0.00 & 7.86 & 11.87 & 
6.28 & 3.76 & 26.88 &  40.67 & 0.00 & 0.19 & 2.34\\

Gemini-2.5 Pro & 71.42 & 82.14 & 
40.67 & 44.24 & 78.28 & 83.63 & 1.72 & 17.52 & 28.35 & 23.63 & 20.10 & 58.42 & 69.61 & 0.00 & 4.48 & 12.86\\

Claude 4 Sonnet & 26.81 & 20.21 & 
11.12 & 7.94 & 46.45 & 64.13 & 0.00 & 1.65 & 6.67 &
3.92 & 0.47 & 24.95 & 42.47 & 0.00 & 0.00 & 0.26 \\

Llama-4 Maverick &
46.37 & 46.84 &
14.43 & 7.48 & 55.29 & 68.08 & 0.00 & 0.68 & 6.53 &
7.84 & 2.82 & 33.60 & 48.66 & 0.00 & 0.19 & 2.47\\

GLM-4.5V &
53.51 & 57.48 &
15.97 & 12.15 & 52.59 & 64.71 & 0.00 & 2.23 & 12.00 &
6.83 & 3.76 & 25.19 & 36.82 & 0.00 & 0.00 & 5.45\\

Qwen2.5-VL-7B &
18.73 & 12.24 &
5.74 & 2.30 & 34.31 & 49.66 & 0.00 & 0.28 & 3.16 &
1.88 & 0.00 & 14.60 & 25.49 & 0.00 & 0.00 & 0.00\\

Qwen3-VL-Plus &
23.95 & 19.16 &
18.60 & 13.55 & 54.49 & 65.90 & 0.00 & 2.14 & 10.67 &
10.99 & 7.98 & 34.27 & 44.55 & 0.00 & 1.43 & 5.97\\

\midrule

\multicolumn{17}{l}{\emph{Chart-specific / Literature Models (no fine-tuning)}}\\

CLIP (ViT-B/32)$^\star$ &
27.13 & 25.39 &
13.61 & 7.48 & 46.42 & 50.65 & 0.00 & 1.36 & 7.60 &
7.89 & 4.69 & 29.33 & 32.69 & 0.00 & 1.24 & 3.38\\

OWLv2 (ViT-B/16)$^\star$ &
18.00 & 12.50 &
18.68 & 14.95 & 51.69 & 54.99 & 0.00 & 7.71 & 10.00 &
7.73 & 5.61 & 22.13 & 28.17 & 0.00 & 1.62 & 5.90\\

KOSMOS-2$^\star$ &
11.40 & 4.15 &
18.44 & 15.89 & 49.53 & 50.79 & 0.32 & 4.67 & 12.31 &
2.00 & 0.93 & 10.34 & 9.16 & 0.00 & 0.00 & 1.79\\

Florence-2$^\star$ &
14.28 & 4.66 &
23.41 & 22.90 & 55.28 & 58.10 & 0.00 & 11.71 & 17.56 &
4.70 & 3.27 & 13.92 & 14.87 & 0.00 & 1.90 & 4.10\\

\midrule

\multicolumn{17}{l}{\emph{Fine-tuned on VGD-Bench}}\\

TransVG (ResNet50)$^\star$&
35.56 & 41.45 &
29.64 & 29.91 & 63.98 & 73.46 & 0.00 & 14.56 & 24.40 &
5.95 & 0.93 & 26.05 & 32.96 & 0.00 & 0.00 & 0.77\\

MDETR (ResNet101)$^\star$ &
20.20 & 22.28 &
29.64 & 34.11 & 57.24 & 65.03 & 1.29 & 20.10 & 25.64 &
3.79 & 2.34 & 16.94 & 26.70 & 0.00 & 0.48 & 1.92\\

SeqTR (DarkNet53)$^\star$ &
47.48 & 52.55 &
44.38 & 48.60 & 71.33 & 78.47 & 1.39 & 30.87 & 46.53 &
10.05 & 7.48 & 34.63 & 46.64 & 0.00 & 0.48 & 6.79\\

Grounding DINO (Swin-T)$^\star$ &
52.41 & 55.44 &
44.59 & 49.53 & 70.23 & 76.75 & 5.16 & 37.62 & 42.31 &
27.20 & 28.50 & 44.73 & 52.61 & 0.97 & 18.38 & 28.72\\

LLaVA (ViT-L/14)$^\star$ &
50.18 & 56.99 &
26.69 & 21.50 & 61.90 & 71.60 & 0.00 & 6.19 & 17.69 &
10.53 & 6.54 & 37.70 & 52.95 & 0.00 & 1.52 & 4.36\\

LLaVA-NeXT (ViT-L/14)$^\star$ &
58.99 & 70.98 &
40.16 & 42.99 & 75.50 & 82.25 & 0.65 & 17.33 & 34.23 &
14.38 & 9.81 & 48.62 & 64.04 & 0.00 & 0.67 & 8.46\\

PaliGemma (SigLIP)$^\star$ &
50.05 & 54.40 &
20.94 & 17.76 & 55.87 & 68.50 & 0.00 & 3.62 & 16.41 &
9.97 & 5.14 & 36.10 & 51.82 & 0.32 & 0.67 & 2.44\\

OneRef (BEiT-3)$^\star$ &
51.63 & 61.75 &
47.60 & 53.46 & 74.63 & 79.85 & 0.00 & 9.68 & 38.55 &
14.14 & 9.68 & 36.41 & 46.88 & 0.00 & 3.20 & 10.00\\

InstanceVG (BEiT-3)$^\star$ &
47.02 & 52.33 &
41.79 & 46.73 & 71.36 & 77.61 & 1.29 & 31.52 & 32.82 &
9.22 & 6.07 & 31.85 & 45.08 & 0.00 & 0.57 & 4.87\\

InternVL2.5-8B$^\star$ &
64.89 & 74.19 &
40.50 & 43.32 & 75.55 & 78.49 & 0.00 & 14.31 & 34.94 &
15.58 & 8.29 & 50.66 & 63.47 & 0.00 & 1.78 & 5.37\\

Qwen2.5-VL-7B$^\star$ &
63.77 & 78.80 &
37.94 & 43.32 & 72.35 & 81.18 & 0.00 & 13.85 & 29.87 &
18.04 & 14.75 & 50.65 & 64.85 & 0.00 & 1.47 & 9.11\\

BARE (BEiT-3)$^\star$ &
44.42 & 54.38 &
18.84 & 14.75 & 50.77 & 50.36 & 0.00 & 0.00 & 9.71 &
7.58 & 2.30 & 24.38 & 32.98 & 0.00 & 0.56 & 2.00\\

SimVG (BEiT-3)$^\star$ &
56.59 & 62.24 &
50.55 & 59.35 & 78.67 & 84.42 & 2.90 & 33.81 & 48.46 &
17.00 & 15.21 & 45.67 & 57.74 & 0.00 & 3.06 & 14.87\\


VGD-Net w/o Stage I$^\star$ &
--&--&
--&--&--&--&--&--&--&
26.63 & 26.17 & 58.10 & 67.41 & 0.00 & 8.10 & 20.90\\

VGD-Net (Stage I + Stage II$^\star$) &
\textbf{86.78} & \textbf{92.06} &
\textbf{58.23} & \textbf{67.76} & \textbf{82.97} & \textbf{87.40} & \textbf{7.10} & \textbf{43.71} & \textbf{57.69} &
\textbf{57.06} & \textbf{66.82} & \textbf{82.40} & \textbf{86.79} & \textbf{6.13} & \textbf{43.14} & \textbf{53.33}\\

\bottomrule
\end{tabular}
\end{table*}

\textbf{Main Results}.
VGD-Net achieves strong performance across both stages: Stage I reaches 86.78 IoU and 92.06 Acc (left columns of Table~\ref{tab:all_results}), while Stage II with MLAD and the native $16\times16$ patch projection attains 67.76 Acc, 82.97 DotD, and 87.40 NWD (middle columns of Table~\ref{tab:all_results}).
All query decompositions were manually verified by three human annotators without errors.
For overall VGD, VGD-Net achieves \textbf{57.06} IoU and \textbf{66.82} Acc on test, and maintains a high NWD of \textbf{86.79} (the right columns of Table~\ref{tab:all_results}).
Notably, without the two-stage design, many strong Stage II models perform poorly on VGD, whereas VGD-Net maintains high performance.
In the overall VGD setting, VGD-Net also achieves the best AP$_s$, AP$_m$ and AP$_l$ among all methods (6.13, 43.14 and 53.33), showing robust localization ability across target scales.
We measured inference latency per component, totaling 1113.3 ms/query: Candidate Generation (44.51 ms), Candidate Ranking (987.61 ms), Chart Selection (58.97 ms), and In-chart Grounding (22.21 ms). Candidate Ranking latency comes from the OCR process.

\textbf{Baseline Comparison}.
We compare three model categories across stages, highlighting VGD-Net’s advantages.

\textit{Stage I.}
As shown in Table~\ref{tab:all_results}, MLLMs achieve strong performance (e.g., Gemini-2.5 Pro), but remain limited for accurate chart retrieval. Fine-tuned grounding models such as SimVG and SeqTR also lag behind detector-based pipelines~\cite{xiao2026vg_survey}. In contrast, Stage I of VGD-Net improves performance by integrating multiple cues (i.e., chart type, spatial position, and textual content), showing effectiveness.

\begin{figure}[tb!]
    \centering
    \includegraphics[width=1\linewidth]{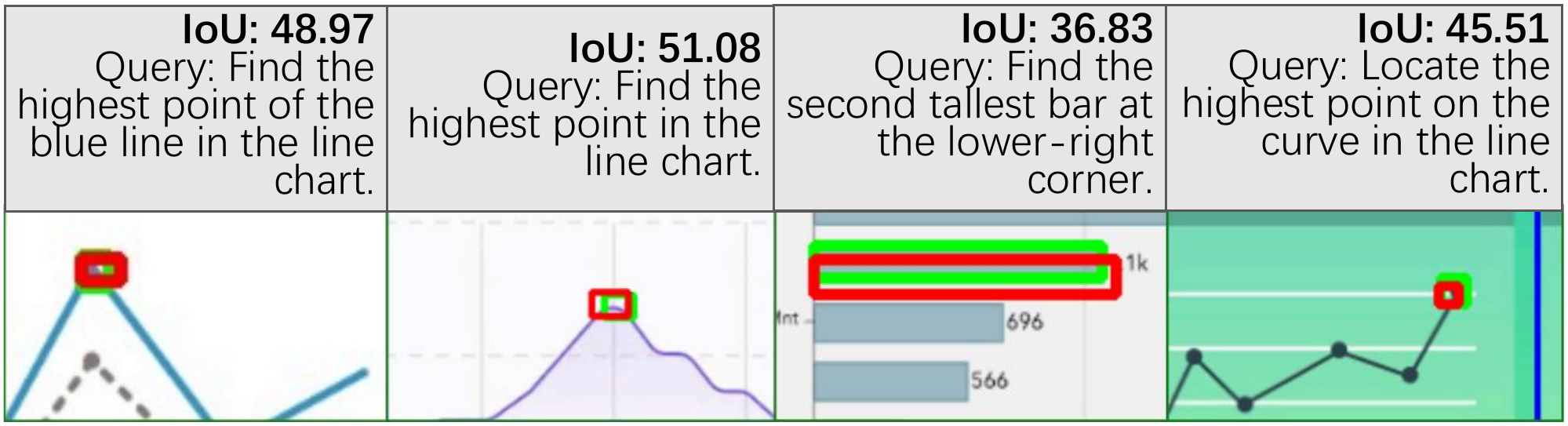}
    \caption{IoU limitation: aligned predictions (red) may score low vs. ground truth (green) for small or elongated objects.}
    \label{fig:iou_limit}
\end{figure}

\textit{Stage II.}
A different trend emerges in Stage II (the middle columns of Table~\ref{tab:all_results}): general MLLMs degrade noticeably~\cite{yang-etal-2025-visual}, while fine-tuned models improve greatly~\cite{xu2025mcbench}, with SimVG~\cite{SimVG2024}, OneRef~\cite{xiao2024oneref} and Grounding DINO~\cite{liu2024groundingdino} providing solid baselines, demonstrating our VGD-Bench’s utility. 
Our Stage II (MLAD + native patch 16) further improves performance across all metrics, not only highlighting the importance of multi-layer feature aggregation for localizing small, dense elements, but also revealing a limitation of IoU, which may underestimate the prediction quality for these elements (see Fig.~\ref{fig:iou_limit}), motivating the use of DotD, NWD, and the three AP metrics (AP$_s$, AP$_m$, and AP$_l$).

\textit{VGD performance.}
As the right columns of Table~\ref{tab:all_results} show, all baselines exhibit large performance drops in the overall VGD setting (i.e., full dashboards and original queries).
The drops stem from the difficulty of grounding targets from full dashboards.
In contrast, VGD-Net maintains strong performance, validating its two-stage design.
For a fairer comparison, we also evaluate all baselines under the two-stage setting and report results in the supplementary material, where VGD-Net still achieves better performance.

\subsection{Ablation Analysis}
We conduct ablation studies to evaluate the contribution of VGD-Net components, analyzing the impact of each setting. All results of the ablation studies conducted on the validation set are provided in the supplementary material.

\textbf{Stage I}. As described in VGD-Net, Stage I scores candidates using detector confidence, chart type, spatial position, and textual content ($s_i^{\text{det}}, s_i^{\text{type}}, s_i^{\text{pos}}, s_i^{\text{content}}$). 
In Table~\ref{tab:ablation_stage1_resize}, all rows except last three correspond to the highest-score selections. 
The results show that each component contributes to performance improvements. 
Adding an MLLM check further improves results, while comparisons in Table~\ref{tab:all_results} indicate that the gains mainly come from the Stage I design rather than relying on MLLMs, as some MLLMs achieve limited grounding performance when used alone (GLM-4.5V IoU 53.51, Acc 57.48; Qwen3-VL-Plus IoU 23.95, Acc 19.16).
The last row of Table~\ref{tab:ablation_stage1_resize} also shows that direct MLLM checking (Qwen3-VL-Plus) without candidate scoring yields lower performance (IoU 79.85, Acc 84.11).

\begin{table}[tb!]
\centering
\caption{Ablation study for Stage I. ``M'': MLLM check.}
\fontsize{9}{10.8}\selectfont
\setlength{\tabcolsep}{2.15pt}
\label{tab:ablation_stage1_resize}
\begin{tabular}{lcc}
\toprule
Method & IoU & Acc \\
\midrule
RT-DETR-x ($s_i^{\text{det}}$) & 23.36 & 24.77 \\
+ $s_i^{\text{type}}$ & 52.13 & 55.14 \\
+ $s_i^{\text{type}}$ + $s_i^{\text{pos}}$ & 58.93 & 62.62 \\
+ $s_i^{\text{type}}$ + $s_i^{\text{pos}}$ + $s_i^{\text{content}}$ & 76.30 & 81.31 \\
+ $s_i^{\text{type}}$ + $s_i^{\text{pos}}$ + $s_i^{\text{content}}$ + GLM-4.5V(M) & 86.43 & 91.59 \\
+ $s_i^{\text{type}}$ + $s_i^{\text{pos}}$ + $s_i^{\text{content}}$ + Qwen3-VL-Plus(M)(Ours) & \textbf{86.78} & \textbf{92.06} \\
+ Qwen3-VL-Plus(M) & 79.85 & 84.11 \\
\bottomrule
\end{tabular}
\end{table}

\textbf{Stage II}.
Table~\ref{tab:ablation_stage2} independently isolates the Stage II components on the test set. Starting from the original SimVG, adding MLAD raises IoU/Acc from 50.55/59.35 to 55.05/64.49. Keeping the native $16\times16$ patch projection and giving MLAD mixing weights their own learning rate further raises them to 58.23/67.76. The largest AP$_s$ gain comes from the native patch projection (3.87 to 7.10), while AP$_m$ and AP$_l$ improve steadily across the two components.
\begin{table}[tb!]
\centering
\caption{Ablation study for Stage II. ``P16'': native $16\times16$ patch. }
\fontsize{9}{10.8}\selectfont
\setlength{\tabcolsep}{3pt}
\label{tab:ablation_stage2}
\resizebox{\columnwidth}{!}{
\begin{tabular}{lccccccc}
\toprule
Method & IoU & Acc & DotD & NWD & AP$_s$ & AP$_m$ & AP$_l$ \\
\midrule
SimVG & 50.55 & 59.35 & 78.67 & 84.42 & 2.90 & 33.81 & 48.46 \\
+ MLAD & 55.05 & 64.49 & 80.82 & 85.75 & 3.87 & 41.14 & 50.64 \\
+ MLAD + P16 (Ours)& \textbf{58.23} & \textbf{67.76} & \textbf{82.97} & \textbf{87.40} & \textbf{7.10} & \textbf{43.71} & \textbf{57.69} \\
\bottomrule
\end{tabular}}
\end{table}

\textbf{VGD Performance}.  
In Table~\ref{tab:all_results}, the second-to-last row shows that VGD-Net without Stage I achieves an IoU of 26.63 and Acc of 26.17, with relatively low performance across all metrics. Adding Stage I, VGD-Net improves performance with IoU of 57.06, Acc of 66.82, and an increase in all other metrics, showing the effectiveness of the two-stage design. We also report the results for applying the two-stage design to all other baselines in the supplementary material, providing further evidence of its effectiveness.

\subsection{Error Propagation Analysis}
We analyze the performance of Stage II conditioned on the success of Stage I.
Across the 414 validation and test samples, Stage I successfully retrieves the target chart (IoU $>$ 0.5) for 376 samples, on which the pipeline achieves a success rate of 60.11\% (226/376). When Stage I fails to retrieve the target chart (38 samples, IoU $\le$ 0.5), the success rate drops to 36.84\% (14/38). In short, accurate chart retrieval is crucial for subsequent fine-grained grounding.
A detailed analysis is provided in the supplementary material.




\section{Conclusion, Limitations and Future Work}
\label{sec:conclusion}

In this paper, we study \textbf{Visual Grounding on Dashboards (VGD)}, which localizes visual elements based on natural language queries in multi-chart dashboards. To tackle the challenges of structured, repetitive layouts, we propose \textbf{VGD-Net}, a two-stage framework that first performs query-aware chart retrieval and then fine-grained in-chart grounding.
To support this task, we construct \textbf{VGD-Bench}, a dataset of \textbf{777} dashboard images and \textbf{2167} human-written queries, with chart retrieval and in-chart grounding labels fully annotated manually by human annotators.
VGD-Bench enables evaluation in three settings: chart retrieval, in-chart grounding, and overall VGD.
Experimental results show that VGD-Net consistently outperforms strong baselines, including MLLMs and existing grounding approaches, across all evaluation settings. In the overall VGD, VGD-Net achieves a grounding accuracy of \textbf{66.82} with a mean IoU of \textbf{57.06}, while also improving \textbf{DotD} (\textbf{82.40}), \textbf{NWD} (\textbf{86.79}), and \textbf{AP$_s$/AP$_m$/AP$_l$}, demonstrating better grounding quality across various scales. These findings confirm the effectiveness of VGD-Net’s design in handling VGD, indicating that clearly separating the task into two stages and considering the structure of dashboard are important for reliable VGD performance. We believe this work will provide a foundation for future research on grounding-centric dashboard understanding and enable effective natural language-driven interaction in visual analytics systems.

Although VGD-Net performs well, errors made in Stage I may affect the results of Stage II. Moreover, the method relies on annotations, which limits scalability in data-sparse settings~\cite{Zhang2022WSOLWSOD_Survey}. Future work may focus more on applying few-shot and weakly supervised learning to visual grounding.

 \section{LLM Usage Statement}
DeepSeek and Qwen3-VL-Plus are used in VGD-Net for query decomposition and candidate verification, respectively. Other LLMs and vision-language models are also used as baselines.
We use OpenAI models only to assist with editing experimental code.
No generative AI is used to generate research data or develop the scientific claims and proofs in this work. All AI-assisted code is reviewed by the authors. The authors take responsibility for the paper and the supplementary materials.

\bibliographystyle{ieeenat_fullname}
\bibliography{sections_CVPR/References}

\clearpage

\renewcommand{\underline}[1]{\ul{#1}}

\twocolumn[{%
  \vspace*{25pt}
  \begin{center}
    {\Large\bfseries \thetitle}\\[0.6em]
    {\large Supplementary Material}
  \end{center}
  \vspace{1.0em}
}]

\renewcommand{\thesection}{\Alph{section}}
\renewcommand{\thesubsection}{\thesection.\arabic{subsection}}
\renewcommand{\theHsection}{supp.\Alph{section}}
\renewcommand{\contentsname}{Table of Contents}
\setcounter{tocdepth}{2}
\setcounter{section}{0}

\makeatletter
\let\addcontentsline\supp@orig@addcontentsline
\makeatother

{\footnotesize
\tableofcontents
}
\vspace{0.6em}

\section{Details of VGD-Bench}
\label{sec:annotations}

This supplement provides detailed information about VGD-Bench, complementing the overview presented in the main text. We describe the annotation categories, format, protocol, and quality assurance procedures in full.

\subsection{Annotation Categories and Examples}

We categorize dashboard charts in VGD-Bench into multiple visual types, as shown in Table~\ref{tab:dashboard_categories}. The exact counts and distribution of each category are provided in the main text. Real examples of each category are shown in Figure~\ref{fig:examples}.

\begin{table}[htbp]
\centering
\caption{Dashboard visual element categories and examples.}
\label{tab:dashboard_categories}
\resizebox{\columnwidth}{!}{
\begin{tabular}{ll}
\toprule
\textbf{Type} & \textbf{Description} \\
\midrule
Card & Shows a key metric or summary. \\
Bar Chart & Shows categories with bars proportional to values. \\
Line Chart & Shows trends with connected lines. \\
UI & UI components like buttons, filters, or dropdowns. \\
Data Table & Data in a structured grid. \\
Donut Chart & Circular chart with hollow center for proportions. \\
Share & Shows percentage values compactly. \\
Area Chart & Line chart with filled area. \\
Map & Shows spatial/geographic data. \\
Progress Bar & Shows task or metric progress. \\
Dial Chart & Gauge showing a value within a range. \\
Pie Chart & Circular chart with slices showing proportions. \\
Heatmap & Values encoded by color intensity. \\
Image & Static images or icons. \\
Scatter Plot & Values plotted as points. \\
Bubble Chart & Points sized by an additional variable. \\
Radar Chart & Multivariate data in a radial layout. \\
Calendar & Shows temporal data in calendar layout. \\
Timeline & Events shown linearly over time. \\
Gantt Chart & Bars showing task schedules over time. \\
\bottomrule
\end{tabular}
}
\end{table}

\subsection{Annotation Format}
VGD-Bench annotations are organized into three JSON files corresponding to the dataset splits: train, val, and test.
Each JSON file contains multiple entries, where each entry represents a single dashboard image.
Each image entry includes basic metadata (image\_id, file\_name, width, height) and a list of \textbf{regions}.
A \textbf{region} corresponds to a visual element on the dashboard, such as a chart, card, or UI component, and contains a bounding box and optional query information.
The \textbf{bbox} field follows the COCO format \([x, y, w, h]\), where \((x, y)\) is the top-left corner, and \(w, h\) are the width and height of the box. This standard format can be easily converted to other bounding box representations if needed.
A region may include one or more \textbf{targets}, which store queries or reasoning steps associated with that region.
Each target contains a smaller \textbf{bbox}, indicating the precise location of the \textbf{original\_query} ($Q$), and optionally \textbf{model\_splits}, which provide a stepwise decomposition of the query for sequential model grounding. \textbf{model\_splits} has three fields: \textbf{s1\_text}, the Chart-Retrieval Query ($Q^{\text{chart}}$); \textbf{s2\_text}, the In-Chart Grounding Query ($Q^{\text{in-chart}}$); and \textbf{method}, which indicates the model used for query decomposition.
A region may have multiple targets (e.g., multiple questions about the same chart), or no targets if no queries were assigned.

This structure ensures that VGD-Bench annotations can be used directly for training, evaluation, or transformed to suit different model input formats. The general structure of a single image entry is illustrated below:

\begin{promptbox}{VGD-Bench JSON Format}
\begin{lstlisting}[breaklines=true, columns=fullflexible, frame=none]
{
    "image_id": "image_0001",
    "file_name": "image_0001.png",
    "width": 1149,
    "height": 651,
    "regions": [
        {
            "bbox": [222, 186, 216, 460],
            "type": "Bar Chart",
            "targets": [
                {
                    "bbox": [389, 238, 35, 365],
                    "original_query": "Find the tallest bar in the bar chart at the lower-left corner.",
                    "model_splits": {
                        "s1_text": "Find the bar chart at the lower-left corner",
                        "s2_text": "Find the tallest bar in the bar chart",
                        "method": "deepseek-api"
                    }
                }
            ]
        }
    ]
}
\end{lstlisting}
\end{promptbox}

For Stage I (Query-aware Chart Retrieval), the original dashboard image and the annotations of outer \textbf{regions} (i.e., \textbf{bbox} and \textbf{type}, if required by the model) are used along with \textbf{s1\_text} as the query.

For Stage II (Fine-grained In-Chart Grounding), each region is cropped from the dashboard image using the \textbf{region's bbox} to obtain a single chart image, and the corresponding \textbf{targets' bbox} and \textbf{s2\_text} are used for grounding within the chart.

For overall VGD, the full dashboard image is used together with each \textbf{target's bbox} and \textbf{original\_query}.

In all cases, the \textbf{bbox} values serve as the expected outputs: during training, they provide supervision; during inference, the predicted \textbf{bbox} is compared against the ground truth to compute metrics and measure performance.

\subsection{Annotation Protocol}

We designed a rigorous two-stage expert annotation protocol specifically tailored to the hierarchical nature of VGD. The protocol emphasizes annotation quality, consistency, and reproducibility through detailed guidelines, iterative calibration, and multi-layer quality assurance.

\begin{figure*}[t]
  \centering
  \includegraphics[width=\textwidth]{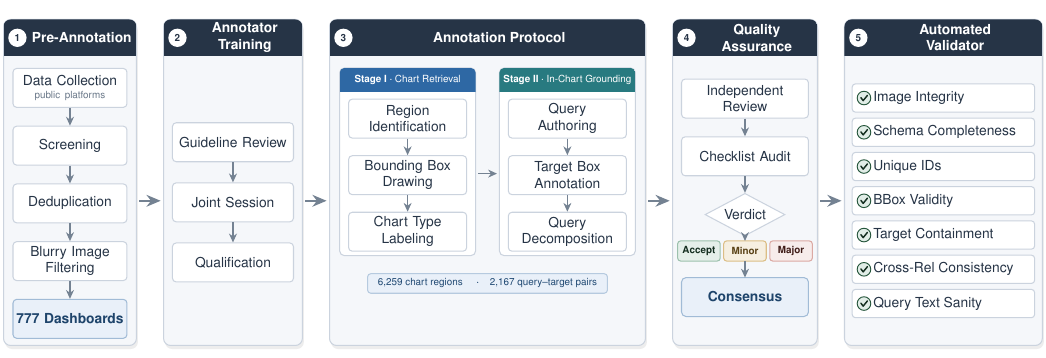}
  \caption{Overview of the VGD-Bench annotation pipeline.}
  \label{fig:annotation_pipeline}
\end{figure*}

\noindent\textbf{Pre-Annotation Preparation.}

\noindent\textbf{Data Collection.} Dashboards were sourced from public design platforms (e.g., Dribbble) and commercial visualization galleries (e.g., Tableau Public), spanning domains including business intelligence, finance, marketing, healthcare, and operations. We prioritized dashboards with heterogeneous chart types and dense layouts to maximize inter-chart ambiguity and visual complexity. All collected dashboards were exported or screenshotted at the highest available resolution while preserving original aspect ratios. Copyright-compliant images were retained; any dashboard with personally identifiable information was excluded.

\noindent\textbf{Screening Criteria.} Each candidate dashboard underwent a manual screening process against the following criteria: (i) minimum resolution of $600 \times 400$ pixels to ensure legibility of chart elements; (ii) presence of at least two visually distinct chart-type components to enable non-trivial chart retrieval; (iii) absence of severe visual artifacts (e.g., watermarks obscuring data, heavy compression noise); (iv) complete, non-truncated layout with all chart titles, axis labels, and legends visible; and (v) no substantial duplicate content with already-admitted dashboards (measured by perceptual hash similarity). Dashboards failing any criterion were discarded. After screening, 777 dashboards were retained from an initial pool of over 1{,}200 candidates.

\noindent\textbf{Annotator Training and Calibration.}

\noindent\textbf{Annotator Qualifications.} All annotation was conducted by an expert group of the authors, all of whom hold graduate degrees in computer science or related fields and have prior experience with visual annotation tasks and chart reasoning. No external crowdsourcing was used, as pilot experiments revealed that crowd workers unfamiliar with chart semantics produced inconsistent boxes and ambiguous queries.

\noindent\textbf{Training Phase.} Before formal annotation began, all annotators completed a training phase consisting of: (i) a detailed review of the annotation guidelines document, which specified category definitions, bounding box conventions, and query formulation rules; (ii) a joint annotation session on 10 representative dashboards, where annotators discussed edge cases and calibrated their understanding; and (iii) an individual qualification round on 15 dashboards, after which annotations were collectively reviewed and discrepancies discussed. This iterative calibration ensured that all annotators converged on a shared interpretation of the annotation standards before proceeding to the full dataset.

\noindent\textbf{Stage I: Chart Retrieval Grounding Annotation.}

\noindent\textbf{Region Identification.} For each dashboard, annotators exhaustively identified all independent visual components. An independent visual component was defined as a perceptually distinct region that conveys a coherent unit of information; this includes data charts (bar, line, pie, donut, area, scatter, bubble, radar, funnel, heatmap, and gauge charts), numeric or textual cards (KPI indicators, summary statistics), tables, maps, and salient UI elements (filters, dropdowns, legends, titles, logos). Each component was assigned exactly one bounding box and one type label.

\noindent\textbf{Bounding Box Guidelines.} All bounding boxes were drawn as tight, axis-aligned rectangles using a custom annotation interface. The following rules were enforced: (i) boxes must tightly enclose the visual extent of the component, including its title, axis labels, and legend if these are integral to the component; (ii) for charts with non-rectangular visual encodings (e.g., pie, donut, gauge), the box must cover the full circular or arc-shaped rendering area; (iii) overlapping components (e.g., a legend positioned over a chart area) were annotated as separate regions with potentially overlapping boxes; (iv) boxes must not extend beyond the dashboard image boundary. Annotators were instructed to zoom in to pixel level when adjusting box edges to ensure tightness.

\noindent\textbf{Type Taxonomy.} Each region was labeled with one of 20 fine-grained component types. The full taxonomy is detailed in Table~\ref{tab:dashboard_categories}. Annotators were provided with a visual reference sheet containing prototypical examples of each type to reduce labeling ambiguity. In cases where a component did not clearly belong to any predefined type, annotators could propose a new type, which was then discussed and either adopted into the taxonomy or mapped to the closest existing type.

\noindent\textbf{Stage I Output.} Stage I produced 6{,}259 bounding boxes across the 777 dashboards (mean: 8.1 regions per dashboard; range: 2--27). Each region record includes a unique region identifier, bounding box coordinates, and a component type label.

\noindent\textbf{Stage II: In-Chart Grounding Annotation.}

\noindent\textbf{Query Authoring Guidelines.} After Stage I, annotators authored 2--3 queries per dashboard. Queries were required to satisfy the following constraints:
\begin{itemize}
  \item \textbf{Localization Orientation:} Every query must ask for a spatial location. Queries were required to begin with a localization-oriented interrogative such as \emph{Where}, \emph{Find}, \emph{Locate}, \emph{Point to}, or \emph{Identify the position of}. Purely descriptive or analytical queries (e.g., ``What is the trend?'' or ``Summarize the dashboard'') were disallowed.
  \item \textbf{Chart Disambiguation:} Each query must contain sufficient information to uniquely identify the target chart among all charts on the dashboard. Disambiguation could be achieved through chart type (e.g., ``the donut chart''), spatial position (e.g., ``the chart in the top-right corner''), title or label content (e.g., ``the chart titled `Monthly Revenue'\,''), or a combination thereof. When multiple charts of the same type appeared on the same dashboard, spatial position or title was required.
  \item \textbf{In-Chart Target Specification:} Each query must describe the target element within the chart with sufficient precision to localize a minimal bounding box. Acceptable descriptions include: relative rankings (``the highest bar'', ``the second-lowest value''), specific data values (``the bar representing 2021'', ``the value at 45\%''), spatial descriptors (``the leftmost cluster'', ``the segment at the top of the pie''), or legend-referenced elements (``the blue line'', ``the red segment'').
  \item \textbf{Answerability:} The target must be unambiguously identifiable from the visual content of the chart alone, without requiring external knowledge or numerical computation beyond reading the chart.
  \item \textbf{Diversity:} Queries for the same dashboard should target different chart types or different in-chart elements when targeting the same chart, to ensure diverse coverage.
\end{itemize}

\noindent\textbf{Target Box Annotation.} For each query, annotators drew a single bounding box that enclosed the minimal visual region sufficient to answer the query. The following rules applied: (i) for a bar in a bar chart, the box tightly enclosed that individual bar (including its top edge); (ii) for a line at a specific x-position, the box enclosed the line segment and marker in the vicinity of that x-value; (iii) for a pie/donut segment, the box tightly enclosed the arc segment; (iv) for a numeric card value, the box enclosed the numeric text. Annotators were explicitly instructed to minimize box area while ensuring completeness: the box should contain exactly the target element and nothing more.

\noindent\textbf{Query Decomposition.} For stage-specific supervision, each original query $Q$ is decomposed by DeepSeek-API into a chart-retrieval sub-query $Q^{\text{chart}}$ and an in-chart grounding sub-query $Q^{\text{in-chart}}$, using the prompt in Section~\ref{sec:prompts}. The procedure is: the model identifies the portion of $Q$ that describes the target chart (e.g., ``the bar chart at the lower-left corner'') and outputs it as $Q^{\text{chart}}$; the remaining portion describing the in-chart target (e.g., ``the tallest bar'') is formulated as a self-contained localization query $Q^{\text{in-chart}}$. Where the chart description is implicit or ambiguous, the prompt instructs the model to make it explicit in $Q^{\text{chart}}$. All 2{,}167 decompositions were audited by three human annotators to ensure correctness and self-containedness. {For every target in all splits, the stored \textbf{s1\_text} and \textbf{s2\_text} are identical to the raw DeepSeek completions after removing the list numbering (``1.''/``2.''), and the raw API response is retained in the annotation file. An earlier description mentioned a 500-query human/LLM comparison subset; it does not correspond to a version used in the results.}

\noindent\textbf{Quality Assurance and Expert Review.}

\noindent\textbf{Independent Review Protocol.} Every annotated dashboard (both Stage I and Stage II annotations) was independently reviewed by at least one additional annotator who was not the original annotator. The reviewer examined each dashboard against a standardized checklist covering:
\begin{itemize}
  \item \textbf{Bounding Box Tightness:} Are all boxes tightly drawn without excessive padding or clipping of the target element? For Stage I, does each box cover the full component including titles and labels? For Stage II, does each target box enclose the minimal answer region?
  \item \textbf{Component Type Accuracy:} Is each region assigned the correct type label according to the taxonomy? Are there any components that were missed or double-counted?
  \item \textbf{Query Linguistic Quality:} Is each query grammatically correct, unambiguous, and natural? Does it follow the localization-oriented format? Are the chart disambiguation and in-chart target descriptions both sufficient and necessary?
  \item \textbf{Query--Box Correspondence:} Does the target box correctly correspond to the query? Is the box the minimal region that answers the query? Could the query be answered by a different element on the dashboard?
  \item \textbf{Decomposition Correctness:} Does $Q^{\text{chart}}$ correctly identify the target chart? Does $Q^{\text{in-chart}}$ correctly describe the in-chart target? Are both sub-queries self-contained and answerable?
\end{itemize}

\noindent\textbf{Review Outcomes and Revision.} The reviewer assigned one of three outcomes to each annotation: (i) \emph{Accepted}: no changes needed; (ii) \emph{Minor Revision}: small adjustments required (e.g., tightening a box edge, clarifying query wording); (iii) \emph{Major Revision}: significant rework needed (e.g., wrong target box, ambiguous query, incorrect type label). The original annotator addressed all flagged issues. The revision rate was low: only 124 of 2{,}167 query--box pairs (5.7\%) required revision, of which 89 (4.1\%) were minor and 35 (1.6\%) were major. The most common were: box not tight enough (42 cases), query ambiguity (31 cases), incorrect chart disambiguation (24 cases), type label correction (18 cases), and missed region (9 cases).

\noindent\textbf{Consensus Resolution.} Cases where the reviewer and original annotator disagreed on the appropriate annotation were escalated to a group discussion involving at least three annotators. The group examined the dashboard and the disputed annotation jointly, discussed the applicable guideline, and reached a consensus decision. All consensus decisions were documented to inform future annotation and to refine the guidelines. Only 18 cases (0.8\%) required escalation to full consensus resolution.

\noindent\textbf{Automated Validation.} Before finalizing the dataset, we implemented a suite of automated validation scripts to detect structural and semantic errors that may have been missed during manual review. The validation pipeline included the following checks:
\begin{itemize}
  \item \textbf{Image Integrity:} All referenced image files must exist on disk, be decodable by standard image libraries (PIL/OpenCV), and have dimensions matching the metadata (width, height) recorded in the JSON annotation.
  \item \textbf{Schema Completeness:} Every JSON entry must contain all required fields (image\_id, file\_name, width, height, regions). Every region must contain bbox and type. Every target must contain bbox and original\_query. Missing fields trigger an immediate rejection.
  \item \textbf{Unique Identifiers:} All image\_id values must be unique across the dataset. All region identifiers within a dashboard must be unique. All target identifiers within a region must be unique.
  \item \textbf{Bounding Box Validity:} Every bbox must satisfy $x \geq 0$, $y \geq 0$, $w > 0$, $h > 0$, and $x+w \leq$ width, $y+h \leq$ height. Zero-area and out-of-bounds boxes are rejected.
  \item \textbf{Region--Target Containment:} Every target bbox must be strictly contained within its parent region bbox. This is the annotation guideline; violations indicate an annotation error, are flagged by the validator, and were corrected in the post-hoc audit (re-associated to the correct chart, or excluded when the query cannot be grounded in the chart).
  \item \textbf{Cross-Reference Consistency:} Any cross-references between annotations (e.g., the same chart appearing in multiple queries) must be consistent in their type labels and bounding box coordinates.
  \item \textbf{Query Text Sanity:} Every query string must be non-empty, contain at least three words, and not contain any disallowed patterns (e.g., placeholder text, non-ASCII characters).
\end{itemize}
The automated validation ran after each revision round. Any dashboard failing any check was returned to the annotation team for correction. A subsequent manual re-audit of the annotations identified a small number of region--target association errors, in which a target was attached to a neighbouring region; these were corrected before the final evaluation reported in this paper.

\noindent\textbf{Data Splits.} Train/validation/test partitions were created at the dashboard level to prevent content leakage: all annotations (regions and queries) belonging to the same dashboard were assigned to the same split. This ensures that chart images and query patterns from the same dashboard do not appear across splits, providing a rigorous test of generalization to unseen dashboards. The split ratio was 80\%/10\%/10\% (train/val/test). The dataset was split using random partitioning, with the selected split maximizing chart-type coverage across splits.

\begin{figure*}[t]
  \centering
  \includegraphics[width=\textwidth]{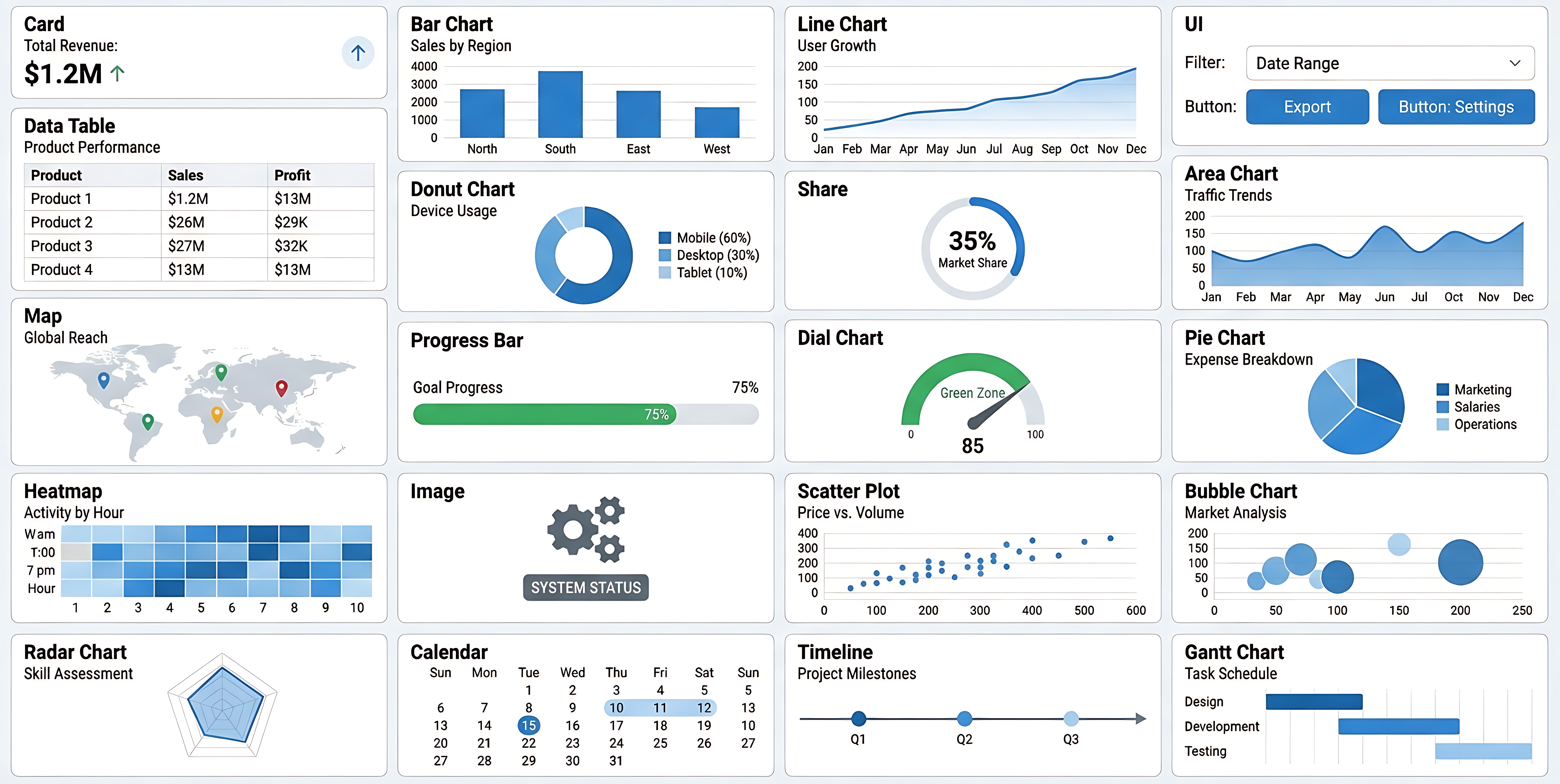}
  \caption{Example visualizations of all chart categories in VGD-Bench.}
  \label{fig:examples}
\end{figure*}

\section{Prompt Design for MLLMs}
\label{sec:prompts}

All prompts in this section are designed for use with the official APIs of multimodal large language models (MLLMs). 
They are directly provided as input to the models without additional tool invocation or intermediate processing.

Underlined parts are instance-dependent placeholders (image, query, candidate actions), instantiated at inference.

\subsection{Query Decomposition}
In VGD-Net, we adopt DeepSeek as the large language model for this task. The designed prompts are not model-specific and generalize to other large language models.

\begin{promptbox}{Prompt for DeepSeek}

\textbf{System Prompt}

You are a data annotation assistant.

Your task:\\
Given a single instruction prompt for chart/map/dashboard understanding,
split it into TWO ordered steps.

Rules:\\
- Step 1: locate or identify the relevant chart/map/visual context\\
- Step 2: perform the specific analytical task within that context\\
- Keep the language simple and imperative\\
- Do NOT add extra steps\\
- Output strictly in JSON with keys: completion1, completion2\\
- If the original instruction only mentions locating a chart/map/visual context with no further task,
  then completion2 should be identical to completion1\\

Examples:\\
Input: Find the blue lake on the map\\
Output:\\
\{\\
  "completion1": "1. Find the map",\\
  "completion2": "2. Find the blue lake on the map"\\
\}\\

Input: Find the highest point of the green line in the line graph on the left\\
Output:\\
\{\\
  "completion1": "1. Find the line graph on the left",\\
  "completion2": "2. Find the highest point of the green line in the line graph"\\
\}\\

Input: Find the smallest bubble in the right scatter plot\\
Output:\\
\{\\
  "completion1": "1. Find the right scatter plot",\\
  "completion2": "2. Find the smallest bubble in the scatter plot"\\
\}\\

Input: Find the smallest bubble of the scatter plot on the right\\
Output:\\
\{\\
  "completion1": "1. Find the right scatter plot",\\
  "completion2": "2. Find the smallest bubble in the scatter plot"\\
\}\\

Input: Find the maximum value area in the left area chart\\
Output:\\
\{\\
  "completion1": "1. Find the left area chart",\\
  "completion2": "2. Find the maximum value area in the area chart"\\
\}\\

Input: Find the maximum value in the area chart on the left\\
Output:\\
\{\\
  "completion1": "1. Find the left area chart",\\
  "completion2": "2. Find the maximum value in the area chart"\\
\}\\

Input: Find the darkest region in the middle heatmap\\
Output:\\
\{\\
  "completion1": "1. Find the middle heatmap",\\
  "completion2": "2. Find the darkest colored region in the heatmap"\\
\}\\

Input: Locate the day with the highest sales revenue.\\
Output:\\
\{\\
  "completion1": "1. Locate the day with the highest sales revenue.",\\
  "completion2": "2. Locate the day with the highest sales revenue."\\
\}\\

Input: Locate the line chart at the bottom.\\
Output:\\
\{\\
  "completion1": "1. Locate the line chart at the bottom.",\\
  "completion2": "2. Locate the line chart at the bottom."\\
\}\\

Input: Positioned to ``Avg Satisfaction''.\\
Output:\\
\{\\
  "completion1": "1. Positioned to ``Avg Satisfaction'',\\
  "completion2": "2. Positioned to ``Avg Satisfaction''\\
\}\\

Input: Where is the value related to Total Treatments?.\\
Output:\\
\{\\
  "completion1": "1. Where is the value related to Total Treatments?",\\
  "completion2": "2. Where is the value related to Total Treatments?"\\
\}\\

Input: In the bar chart titled \"heart rate,\" where is the tallest bar located?.\\
Output:\\
\{\\
  "completion1": "1. Where is the bar chart titled \"heart rate,\"",\\
  "completion2": "2. Where is the tallest bar located?"\\
\}\\

Input: Where is the donut chart?\\
Output:\\
\{\\
  "completion1": "1. Where is the donut chart?",\\
  "completion2": "2. Where is the donut chart?"\\
\}\\

Input: Where is the highest point of the orange solid line in the line chart?\\
Output:\\
\{\\
  "completion1": "1. Where is the line chart?",\\
  "completion2": "2. Where is the highest point of the orange solid line?"\\
\}\\

Input: Where is the number for Total Balance?\\
Output:\\
\{\\
  "completion1": "1. Where is the number for Total Balance?",\\
  "completion2": "2. Where is the number for Total Balance?"\\
\}\\

\textbf{User Prompt}

\underline{Where is the largest part in the right donut chart?}

\end{promptbox}

\subsection{Query Keywords Extraction}
To obtain structured query keywords used in Stage I, we employ a prompt-based parsing strategy.
Specifically, given a natural language query $Q^{\text{chart}}$, we use a large language model (DeepSeek) to extract structured fields, including chart type, spatial position and textual content.
We adopt its text-only variant, as the task requires purely linguistic understanding without visual input.

These keywords supply the type, position and content scores in the candidate ranking module, matched against candidate metadata and OCR-extracted text (Section~\ref{sec:implementation}).

The detailed prompt used for query keywords extraction is shown below.

\begin{promptbox}{Prompt for Query Keywords Extraction}

\textbf{System Prompt}

You are a dashboard query parsing expert. Your task is to extract structured information from natural language queries about dashboards.\\

Given a query, extract the following information:\\

1. **chart\_type**: The type of chart/visual element mentioned in the query\\
   Possible values: Bar Chart, Line Chart, Pie Chart, Donut Chart, Area Chart, Scatter Plot, Bubble Chart, Heatmap, Data Table, Card, Map, Radar Chart, Dial Chart, Progress Bar, Calendar, Gantt Chart, Time Line, Waterfall Chart, Mind Map, UI, Image, or null (if not specified)

2. **position**: The spatial position of the chart on the dashboard\\
   Possible values: top-left, top-right, bottom-left, bottom-right, top, bottom, left, right, center, or null (if not specified)

3. **ordinal\_position**: If the query specifies an ordinal position (e.g., ``first'', ``second'')\\
   Format: \texttt{\{"index": 1-5, "axis": "left|right|top|bottom"\}} or null

4. **content\_keywords**: A SINGLE content keyword string from the query (for OCR matching against chart titles/labels)\\
   CRITICAL RULES:\\
   - Return EXACTLY ONE string in the list, or an empty list \texttt{[]} if no content is mentioned\\
   - The keyword must be the COMPLETE, CONTINUOUS content phrase as it would appear in the chart title or label or something that reflects the content of the chart\\
   - Keep connectors like ``vs'', ``and'', ``\&'', ``/'', ``by'' that are part of the chart content\\
     (e.g. ``customer age vs income'' is ONE keyword, not two)\\
   - Exclude functional words: find, locate, show, where, is, the, a, in, on, at, of, with, for\\
   - Exclude chart type words: chart, graph, plot, card, table, dashboard, visualization\\
   - Exclude position words: left, right, top, bottom, corner, center, first, second\\
   - Preserve special characters such as \$, \%, ``vs'', and ``\&'' when they appear in the query content\\
   - If the query only mentions chart type or position but no specific content, return \texttt{[]}\\

Return format (JSON only, no other text):\\
\{\\
\quad "chart\_type": "Bar Chart" or null,\\
\quad "position": "bottom-right" or null,\\
\quad "ordinal\_position": \{"index": 1, "axis": "left"\} or null,\\
\quad "content\_keywords": ["one complete phrase"] or []\\
\}\\

Examples:\\

1. ``Find the bar chart at the bottom right showing Total Sales''\\
   $\rightarrow$ \texttt{\{"chart\_type": "Bar Chart", "position": "bottom-right", "ordinal\_position": null, "content\_keywords": ["Total Sales"]\}}

2. ``Locate the first line chart from the left with Q1 Revenue data''\\
   $\rightarrow$ \texttt{\{"chart\_type": "Line Chart", "position": null, "ordinal\_position": \{"index": 1, "axis": "left"\}, "content\_keywords": ["Q1 Revenue"]\}}

3. ``Where is the pie chart titled Market Share?''\\
   $\rightarrow$ \texttt{\{"chart\_type": "Pie Chart", "position": null, "ordinal\_position": null, "content\_keywords": ["Market Share"]\}}

4. ``Find the card showing \$100M revenue''\\
   $\rightarrow$ \texttt{\{"chart\_type": "Card", "position": null, "ordinal\_position": null, "content\_keywords": ["\$100M revenue"]\}}

5. ``Show me the highest value in the area chart''\\
   $\rightarrow$ \texttt{\{"chart\_type": "Area Chart", "position": null, "ordinal\_position": null, "content\_keywords": []\}}

6. ``scatter plot bottom left showing customer age vs income''\\
   $\rightarrow$ \texttt{\{"chart\_type": "Scatter Plot", "position": "bottom-left", "ordinal\_position": null, "content\_keywords": ["customer age vs income"]\}}

7. ``Find the line chart with Monthly Revenue vs Target data''\\
   $\rightarrow$ \texttt{\{"chart\_type": "Line Chart", "position": null, "ordinal\_position": null, "content\_keywords": ["Monthly Revenue vs Target"]\}}

8. ``heatmap showing user retention by region''\\
   $\rightarrow$ \texttt{\{"chart\_type": "Heatmap", "position": null, "ordinal\_position": null, "content\_keywords": ["user retention by region"]\}}\\

\textbf{User Prompt}

\underline{Where is the right donut chart?}

\end{promptbox}

\subsection{MLLM Check}
As the final selection step in Stage I, we introduce a multimodal large language model (MLLM) to determine the correct chart from the top-ranked candidates.
Given the query and candidate regions, the MLLM performs holistic reasoning to select the chart that matches the query intent.

In VGD-Net, we adopt Qwen3-VL-Plus as the default MLLM due to its strong multimodal understanding capability.
To ensure a fair and comprehensive evaluation, we also explore alternative MLLMs (e.g., GLM-4.5V) in the ablation study, following the input and prompting protocol.

The detailed prompt used for the MLLM check module is shown below.

\begin{promptbox}{Prompt for MLLM Check}

\textbf{System Prompt}

I have marked a region with a red bounding box on this dashboard image.\\
Check the following:\\
1. Does the red box tightly enclose the chart/element the query refers to?\\
2. Is the chart TYPE inside the box correct (e.g. bar chart, line chart, card)?\\
3. Is the POSITION on the dashboard correct (e.g. bottom-right, left side)?\\
4. Does the TITLE or CONTENT visible inside the box match the query?\\
If the red box correctly identifies the queried region, answer True.\\
If the box is on the wrong chart or wrong position, answer False.\\
Answer only True or False.

\textbf{User Prompt}

\underline{Where is the right donut chart?}\\

\begin{center}
\includegraphics[width=0.8\linewidth]{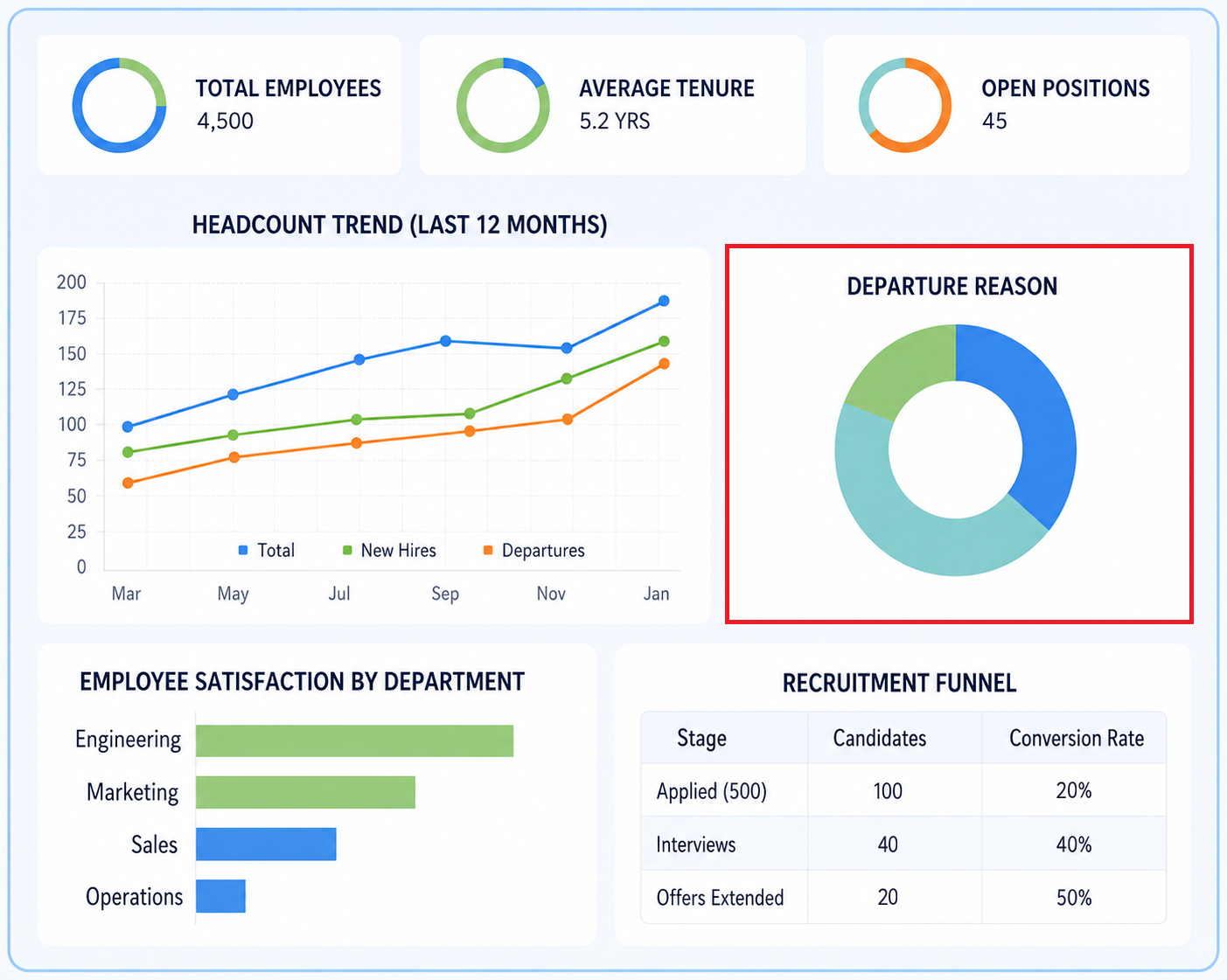}
\end{center}

\end{promptbox}

\subsection{Baseline Evaluation}
For Chart Retrieval, In-Chart Grounding, and overall VGD, we conduct baseline evaluations with MLLMs. To this end, we design three distinct sets of prompts, each corresponding to one of these tasks.

\begin{promptbox}{Prompt for Chart Retrieval}

\textbf{System Prompt}

You are an expert AI assistant specialized in Dashboard Understanding and Chart Localization.\\
You are given a dashboard image composed of multiple card-style chart containers.\\

Your goal is to locate the chart container described by the user query and output its bounding box.\\

TASK\\
- The user query describes a specific chart in the dashboard (e.g., "Find the bar chart at the bottom right").\\
- You must identify the correct chart card/container that matches the description.\\
- Output the bounding box of the entire chart container (the full card, including its axes and title, but NOT including adjacent cards).\\

COORDINATE SYSTEM (CRITICAL)\\
- The coordinate system is a 0-1000 normalized scale.\\
- Top-left corner is (0, 0); bottom-right corner is (1000, 1000).\\
- Bounding box format MUST be: [x, y, w, h]\\
- (x, y) is the top-left corner; w and h are the box width and height.
- Do NOT output absolute pixel values.
- All values MUST be integers in [0, 1000].\\

DASHBOARD LAYOUT REASONING (MANDATORY)\\
- First, identify how many charts are in the dashboard and their grid positions (e.g., 2x2, 1x3).\\
- Then match the user's description (e.g., "top-right", "bottom line chart") to the correct card.\\
- The bounding box must cover the full chart container — do NOT cut off axes, titles, or legends.\\
- Do NOT include parts of adjacent cards.\\

OUTPUT FORMAT (STRICT)\\
Return STRICT JSON only, exactly in the following format:\\
\{\\
  "reasoning": "brief justification of which chart was selected and why",\\
  "box\_2d": [x, y, w, h]\\
\}\\
- Coordinates must be integers in [0, 1000].\\
- Do NOT include markdown code blocks.\\
- Do NOT include any extra text outside the JSON object.\\

\textbf{User Prompt}

\underline{Where is the right donut chart?}\\

\begin{center}
\includegraphics[width=0.8\linewidth]{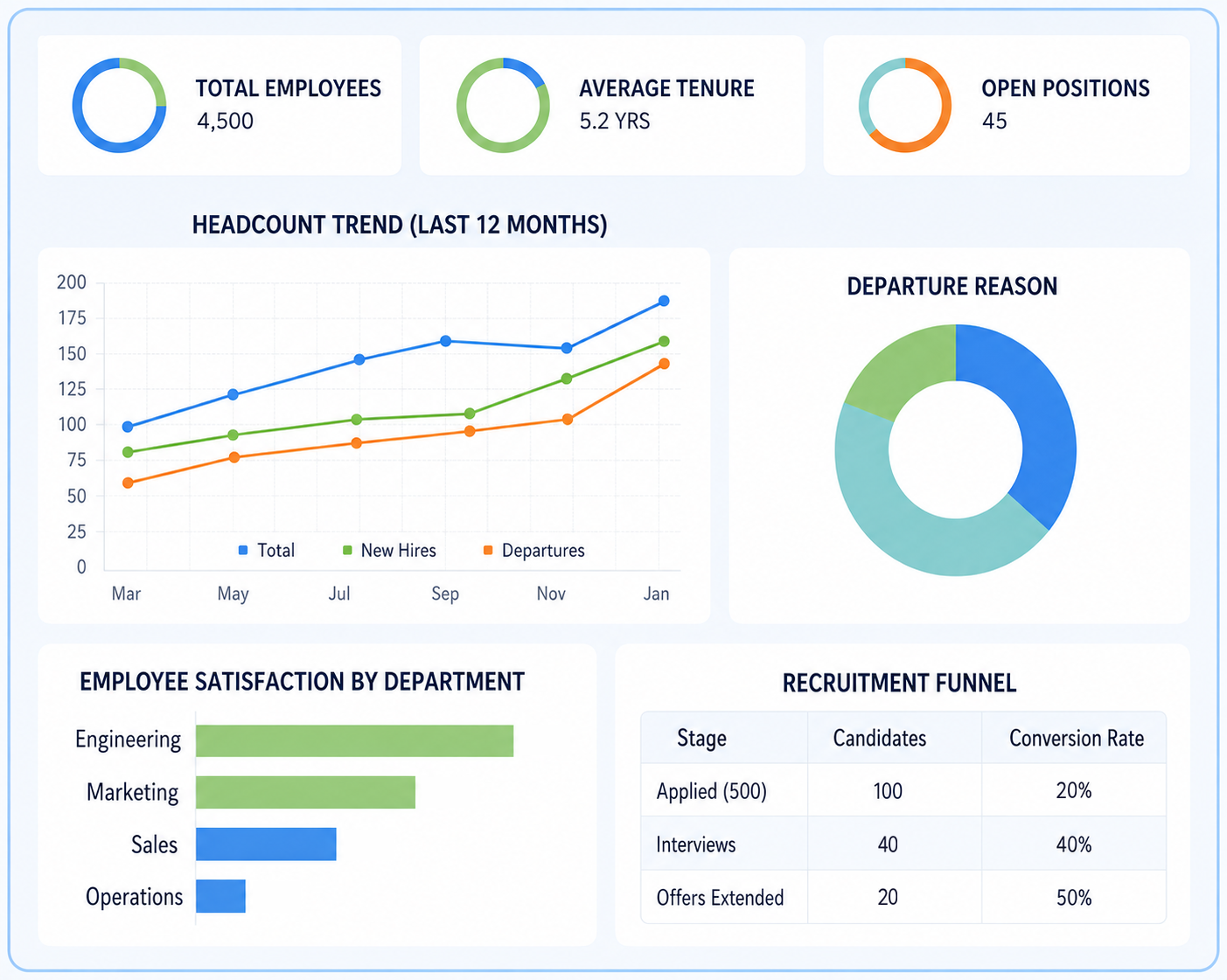}
\end{center}

\end{promptbox}

\begin{promptbox}{Prompt for In-Chart Grounding}

\textbf{System Prompt}

You are an expert AI assistant specialized in Visual Grounding and Chart Understanding.
You are given a single cropped chart image (already isolated from a dashboard).\\

Your goal is to locate the specific visual element described by the user query within this chart and output its bounding box.\\

TASK\\
- The input image is a single chart (bar chart, line chart, pie chart, scatter plot, etc.).\\
- The user query describes a specific element within this chart (e.g., "Find the highest point on the line", "Find the second highest bar").\\
- Locate the target element using visual evidence only.\\

COORDINATE SYSTEM (CRITICAL)\\
- The coordinate system is a 0-1000 normalized scale relative to THIS cropped image.\\
- Top-left corner of the cropped image is (0, 0); bottom-right is (1000, 1000).\\
- Bounding box format MUST be: [x, y, w, h]\\
- (x, y) is the top-left corner; w and h are the box width and height.
- Do NOT output absolute pixel values.\\
- All values MUST be integers in [0, 1000].\\

CHART UNDERSTANDING RULES\\
- Identify the chart type.\\
- Reason strictly within the plotting area of the chart.\\

For quantitative charts (e.g., bar, line, area, pie/donut):\\
- Identify all visual elements representing values (e.g., bars, points, segments).\\
- Compare their magnitudes visually (e.g., height, position, or area/angle).\\
- Rank elements from largest to smallest when needed.\\
- Do NOT rely on color, legend order, or labels unless explicitly requested.\\

BOUNDING BOX QUALITY\\
- Use exactly ONE axis-aligned rectangular bounding box.\\
- The box must tightly enclose ONLY the target element.\\
- Do NOT include neighboring elements, grid lines, axis labels, or titles.\\
- Do NOT let the box spill outside the chart area.\\

OUTPUT FORMAT (STRICT)\\
Return STRICT JSON only, exactly in the following format:\\
\{\\
  "reasoning": "brief justification",\\
  "box\_2d": [x, y, w, h]\\
\}\\
- Coordinates must be integers in [0, 1000].\\
- Do NOT include markdown code blocks.\\
- Do NOT include any extra text outside the JSON object.\\

\textbf{User Prompt}

\underline{Where is the largest part in the donut chart?}\\

\begin{center}
\includegraphics[width=0.6\linewidth]{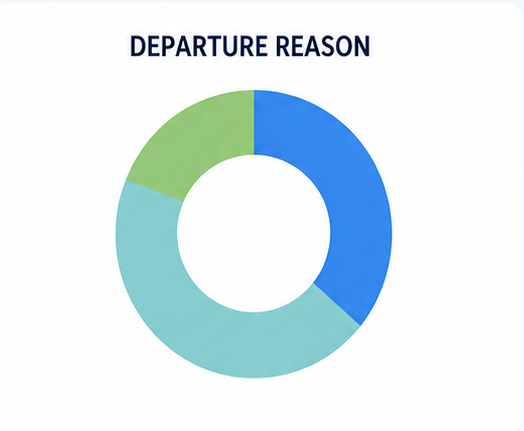}
\end{center}

\end{promptbox}

\begin{promptbox}{Prompt for Overall VGD}

\textbf{System Prompt}

You are an expert AI assistant specialized in Visual Grounding, UI Element Localization, and Chart Understanding.
You are particularly strong at dashboard-style layouts composed of multiple card-based containers.\\

Your goal is to precisely locate visual elements in complex dashboards and output accurate bounding boxes that strictly respect UI boundaries.\\

TASK\\
1. Analyze the user query and identify:
   - The target chart container (e.g., top-right card, bottom-left panel).
   - The target visual element inside that container (e.g., a specific bar, line point, axis label).\\
2. Locate the target element in the image using visual evidence only.\\
3. Output the bounding box of the target element.\\

COORDINATE SYSTEM (CRITICAL)\\
- The coordinate system is a 0-1000 normalized scale.\\
- Top-left corner is (0, 0); bottom-right corner is (1000, 1000).\\
- Bounding box format MUST be: [x, y, w, h]\\
- (x, y) is the top-left corner; w and h are the box width and height.
- Do NOT output absolute pixel values.\\
- All values MUST be integers in [0, 1000].\\

DASHBOARD \& CONTAINER AWARENESS (MANDATORY)\\
- The image is a dashboard composed of multiple card-style containers.\\
- Each card typically has a rounded rectangle boundary, subtle drop shadow, and uniform padding.\\
- You MUST reason at the container level before reasoning about chart content.\\

Container rules:\\
- Do NOT include header bars, navigation panels, side menus, or adjacent cards.\\
- Do NOT let the bounding box spill outside the card containing the target chart.\\
- Do NOT cut off any part of the target element.\\
- Do NOT include unnecessary background whitespace.\\

CHART UNDERSTANDING RULES\\
- Identify the chart type (bar, line, scatter, etc.).\\
- Reason strictly within the plotting area of the chart.\\

For quantitative charts (e.g., bar, donut/pie, line chart/area chart):\\
- Identify all visual elements representing values (e.g., bars or segments).\\
- Compare their magnitudes visually (e.g., height, position, or area/angle).\\
- Rank elements from largest to smallest.\\
- If the query asks for an ordinal (e.g., "second largest"), select based on this ranking.\\
- Do NOT rely on color, legend order, labels, or center text unless explicitly requested.\\

BOUNDING BOX QUALITY\\
- Use exactly ONE axis-aligned rectangular bounding box.\\
- The box must tightly enclose ONLY the target element.\\
- Do NOT include neighboring bars, grid lines, axis labels, or titles.\\
- Do NOT let the box spill outside the chart area.\\

OUTPUT FORMAT (STRICT)\\
Return STRICT JSON only, exactly in the following format:\\
\{\\
  "reasoning": "brief justification",\\
  "box\_2d": [x, y, w, h]\\
\}\\
- Coordinates must be integers in [0, 1000].\\
- Do NOT include markdown code blocks.\\
- Do NOT include any extra text outside the JSON object.\\

\textbf{User Prompt}

\underline{Where is the largest part in the right donut chart?}\\

\begin{center}
\includegraphics[width=0.8\linewidth]{images/example.png}
\end{center}

\end{promptbox}

\subsection{Directly Outputting the Drawn Image}
To demonstrate that requiring the model to output bounding box coordinates does not adversely affect its performance, we also design prompts that instruct the model to directly generate annotated images and output the resulting visualizations. This comparison serves to justify the use of bounding box outputs as a reasonable evaluation interface.

\begin{promptbox}{Prompt for Chart Retrieval (Visualization)}

\textbf{System Prompt}

You are an expert AI assistant specialized in Dashboard Understanding and Chart Localization.\\
You are given a dashboard image composed of multiple card-style chart containers.\\

Your goal is to locate the chart container described by the user query and output an image with the bounding box drawn.\\

TASK\\
- The user query describes a specific chart in the dashboard (e.g., "Find the bar chart at the bottom right").\\
- You must identify the correct chart card/container that matches the description.\\
- Output the bounding box of the entire chart container (the full card, including its axes and title, but NOT including adjacent cards) by drawing it on the image.\\

DASHBOARD LAYOUT REASONING (MANDATORY)\\
- First, identify how many charts are in the dashboard and their grid positions (e.g., 2x2, 1x3).\\
- Then match the user's description (e.g., "top-right", "bottom line chart") to the correct card.\\
- The bounding box must cover the full chart container — do NOT cut off axes, titles, or legends.\\
- Do NOT include parts of adjacent cards.\\

OUTPUT FORMAT (STRICT)\\
Return ONLY the original image with the predicted bounding box drawn on it.\\
- The bounding box must be drawn clearly on the image.\\
- Do NOT output JSON, coordinates, or any text.\\
- Do NOT include markdown code blocks.\\
- Do NOT include any extra explanation.\\

\textbf{User Prompt}

\underline{Where is the right donut chart?}\\

\begin{center}
\includegraphics[width=0.8\linewidth]{images/example.png}
\end{center}

\end{promptbox}

\begin{promptbox}{Prompt for In-Chart Grounding (Visualization)}

\textbf{System Prompt}

You are an expert AI assistant specialized in Visual Grounding and Chart Understanding.
You are given a single cropped chart image (already isolated from a dashboard).\\

Your goal is to locate the specific visual element described by the user query within this chart and output an image with the bounding box drawn.\\

TASK\\
- The input image is a single chart (bar chart, line chart, pie chart, scatter plot, etc.).\\
- The user query describes a specific element within this chart (e.g., "Find the highest point on the line", "Find the second highest bar").\\
- Locate the target element using visual evidence only.\\

CHART UNDERSTANDING RULES\\
- Identify the chart type.\\
- Reason strictly within the plotting area of the chart.\\

For quantitative charts (e.g., bar, line, area, pie/donut):\\
- Identify all visual elements representing values (e.g., bars, points, segments).\\
- Compare their magnitudes visually (e.g., height, position, or area/angle).\\
- Rank elements from largest to smallest when needed.\\
- Do NOT rely on color, legend order, or labels unless explicitly requested.\\

BOUNDING BOX QUALITY\\
- Use exactly ONE axis-aligned rectangular bounding box.\\
- The box must tightly enclose ONLY the target element.\\
- Do NOT include neighboring elements, grid lines, axis labels, or titles.\\
- Do NOT let the box spill outside the chart area.\\

OUTPUT FORMAT (STRICT)\\
Return ONLY the original cropped chart image with the predicted bounding box drawn on it.\\
- The bounding box must be drawn clearly on the image.\\
- Do NOT output JSON, coordinates, or any text.\\
- Do NOT include markdown code blocks.\\
- Do NOT include any extra explanation.\\

\textbf{User Prompt}

\underline{Where is the largest part in the donut chart?}\\

\begin{center}
\includegraphics[width=0.6\linewidth]{images/example_stage2.png}
\end{center}

\end{promptbox}

\begin{promptbox}{Prompt for Overall VGD (Visualization)}

\textbf{System Prompt}

You are an expert AI assistant specialized in Visual Grounding, UI Element Localization, and Chart Understanding.
You are particularly strong at dashboard-style layouts composed of multiple card-based containers.\\

Your goal is to precisely locate visual elements in complex dashboards and output an image with the bounding box drawn that strictly respect UI boundaries.\\

TASK\\
1. Analyze the user query and identify:
   - The target chart container (e.g., top-right card, bottom-left panel).
   - The target visual element inside that container (e.g., a specific bar, line point, axis label).\\
2. Locate the target element in the image using visual evidence only.\\
3. Output the bounding box of the target element by drawing it on the image.\\

DASHBOARD \& CONTAINER AWARENESS (MANDATORY)\\
- The image is a dashboard composed of multiple card-style containers.\\
- Each card typically has a rounded rectangle boundary, subtle drop shadow, and uniform padding.\\
- You MUST reason at the container level before reasoning about chart content.\\

Container rules:\\
- Do NOT include header bars, navigation panels, side menus, or adjacent cards.\\
- Do NOT let the bounding box spill outside the card containing the target chart.\\
- Do NOT cut off any part of the target element.\\
- Do NOT include unnecessary background whitespace.\\

CHART UNDERSTANDING RULES\\
- Identify the chart type (bar, line, scatter, etc.).\\
- Reason strictly within the plotting area of the chart.\\

For quantitative charts (e.g., bar, donut/pie, line chart/area chart):\\
- Identify all visual elements representing values (e.g., bars or segments).\\
- Compare their magnitudes visually (e.g., height, position, or area/angle).\\
- Rank elements from largest to smallest.\\
- If the query asks for an ordinal (e.g., "second largest"), select based on this ranking.\\
- Do NOT rely on color, legend order, labels, or center text unless explicitly requested.\\

BOUNDING BOX QUALITY\\
- Use exactly ONE axis-aligned rectangular bounding box.\\
- The box must tightly enclose ONLY the target element.\\
- Do NOT include neighboring bars, grid lines, axis labels, or titles.\\
- Do NOT let the box spill outside the chart area.\\

OUTPUT FORMAT (STRICT)\\
Return ONLY the original dashboard image with the predicted bounding box drawn on it.\\
- The bounding box must be drawn clearly on the image.\\
- Do NOT output JSON, coordinates, or any text.\\
- Do NOT include markdown code blocks.\\
- Do NOT include any extra explanation.\\

\textbf{User Prompt}

\underline{Where is the largest part in the right donut chart?}\\

\begin{center}
\includegraphics[width=0.8\linewidth]{images/example.png}
\end{center}

\end{promptbox}
\section{Implementation Details}
\label{sec:implementation}

\subsection{Hardware and Software Environment}
All experiments ran on a single NVIDIA RTX 4090 (24\,GB VRAM) with a 16-core Intel Xeon Platinum 8481C CPU and 80\,GB of RAM, under Ubuntu 22.04 LTS. The code is written in Python 3.10 with PyTorch and CUDA 11.8, and all dependencies are pinned in an isolated Conda environment for reproducibility.

\subsection{Training Configurations}
The proposed VGD-Net is composed of two distinct stages. To optimize task-specific feature representations, the two stages are trained independently.

\subsubsection{Stage I: Chart-Level Grounding}
For Stage I, we fine-tune RT-DETR-x on the VGD-Bench training set for query-aware chart retrieval. The model is initialized from the official RT-DETR-x checkpoint pretrained on Objects365. The input image resolution is fixed at $1024 \times 1024$, and we set the nominal batch size (nbs) to 64 with an effective batch size of 4 on a single NVIDIA RTX 4090 GPU (24\,GB VRAM). Training runs for 100 epochs with the AdamW optimizer, using an initial learning rate of $3 \times 10^{-4}$ and a cosine learning rate schedule with a final learning rate factor of 0.05. A linear warm-up phase spans the first 5 epochs, during which the bias learning rate starts from 0.0. The weight decay is set to $1 \times 10^{-4}$ and momentum to 0.9. Automatic mixed precision (AMP) is enabled, and multi-scale training is disabled. Early stopping triggers after 25 epochs without improvement on the validation set.

For data augmentation, we apply moderate HSV jittering ($\pm 0.02$ hue, $\pm 0.3$ saturation, $\pm 0.2$ value), random horizontal flipping with probability 0.5, translation up to 0.1 of the image size, and scaling by a factor up to 0.4. Mosaic augmentation is applied with probability 0.5, and MixUp with probability 0.05. Copy-Paste and random erasing are not used. Mosaic augmentation is disabled during the last 20 epochs (close\_mosaic=20) to stabilize late-stage fine-tuning. All random operations are controlled by a fixed seed of 42 with deterministic mode enabled for reproducibility. During validation, a confidence threshold of 0.001 and an IoU threshold of 0.6 are used for NMS.

\textbf{Stage-I inference and selection.} At inference we run the fine-tuned detector at $1024 \times 1024$ with a confidence threshold of 0.10 and NMS IoU 0.6, which keeps a high-recall candidate set. For every candidate we compute four scores: the detector confidence $s^{\text{det}}$; a binary type score $s^{\text{type}}$; a Gaussian position score $s^{\text{pos}} = \exp(-d^2 / 2\sigma^2)$ on normalized box centres, where candidates closer to the query-implied spatial position score higher; and a content score $s^{\text{content}}$ given by the longest-common-substring ratio between the query content keyword and the OCR text inside the candidate box (the maximum over texts whose centres fall inside the box). The four scores are combined, and the highest-scoring candidate is verified first. The verifier draws that box in red on the original-resolution dashboard and asks the multimodal model whether the box tightly encloses the requested chart; the candidate is accepted when the verifier answers True, otherwise the next candidate in rank order is checked. Qwen3-VL-Plus is the default verifier; the ablation uses GLM-4.5V with its thinking mode disabled. OCR is performed with PaddleOCR (PP-OCRv5 mobile detector and recognizer, images rescaled to 1600\,px on the long side).

We additionally audited the Stage-I annotations. For a small number of queries the annotated target box lies outside the chart region it was attached to; those queries were re-associated with the region that contains the target, and all Stage-I numbers reported in this paper are computed on this audited annotation set.

\subsubsection{Stage II: Fine-grained In-Chart Grounding}
For Stage II, we fine-tune SimVG with our proposed MLAD module and the native $16 \times 16$ patch projection for fine-grained in-chart grounding.

Training instances are constructed from ground-truth chart regions, with the corresponding in-chart query and target box as supervision. During training, chart crops undergo large-scale jitter with a scale range of 0.3--1.4 to simulate the localization offset introduced by Stage I predictions at inference time, and are then resized to $640 \times 640$ without preserving aspect ratio; validation and test crops are resized in the same manner. RGB inputs are normalized using mean (123.675, 116.28, 103.53) and standard deviation (58.395, 57.12, 57.375). Queries are tokenized with the BEiT-3 SentencePiece tokenizer and truncated or padded to 40 tokens.

\textbf{Backbone and fusion.} The grounding model uses a BEiT-3 Large visual-language encoder with its native $16 \times 16$ patch projection, a drop-path rate of 0.1, and no frozen encoder layers. SimVG resamples the pretrained $16 \times 16$ projection to $32 \times 32$ to reduce the token count; we keep the native projection, so a $640 \times 640$ crop produces a $40 \times 40$ token grid instead of $20 \times 20$, and the encoder operates at the resolution at which it was pretrained. MLAD extracts the one-based 8th, 16th, and 24th BEiT-3 layers. Three learnable logits, initialized as (0, 0, 1), are normalized by softmax and used to compute a weighted sum of the layer features. The fused features are passed to the SimVG grounding head, which uses a 256-dimensional embedding, one object query, six encoder layers, and three decoder layers; only the decoder branch is trained.

\textbf{Optimization.} The model is initialized from the officially released SimVG ViT-Large decoder-branch checkpoint and optimized for 50 epochs with Adam, a batch size of 1, and two data-loader workers. The batch size is reduced from 2 to 1 because the $40 \times 40$ token grid at $640 \times 640$ fills the memory of a single 24\,GB card. The base learning rate is $5 \times 10^{-5}$ and the BEiT-3 visual encoder uses $5 \times 10^{-6}$. The three MLAD logits receive a dedicated learning rate of $5 \times 10^{-4}$: under the original parameter grouping they inherit the backbone rate and remain close to their (0, 0, 1) initialization, which leaves the fusion close to the last-layer features; the dedicated group allows them to adapt. We use $\beta = (0.9, 0.98)$, $\epsilon = 10^{-9}$, zero weight decay, and AMSGrad. MultiStepLR-WarmUp uses two warm-up epochs, a decay factor of 0.1, and milestones at epochs 30 and 45. The gradient norm is clipped to 0.15. Training uses seed 42 without mixed precision and without an exponential moving average, and we report the final-epoch checkpoint.

At inference, Stage~II performs a single forward pass on the $640 \times 640$ crop and maps the predicted box back to chart coordinates using the crop origin and size, so the pipeline requires no second pass and no additional parameters.

\begin{figure*}[t]
\centering
\includegraphics[width=\textwidth]{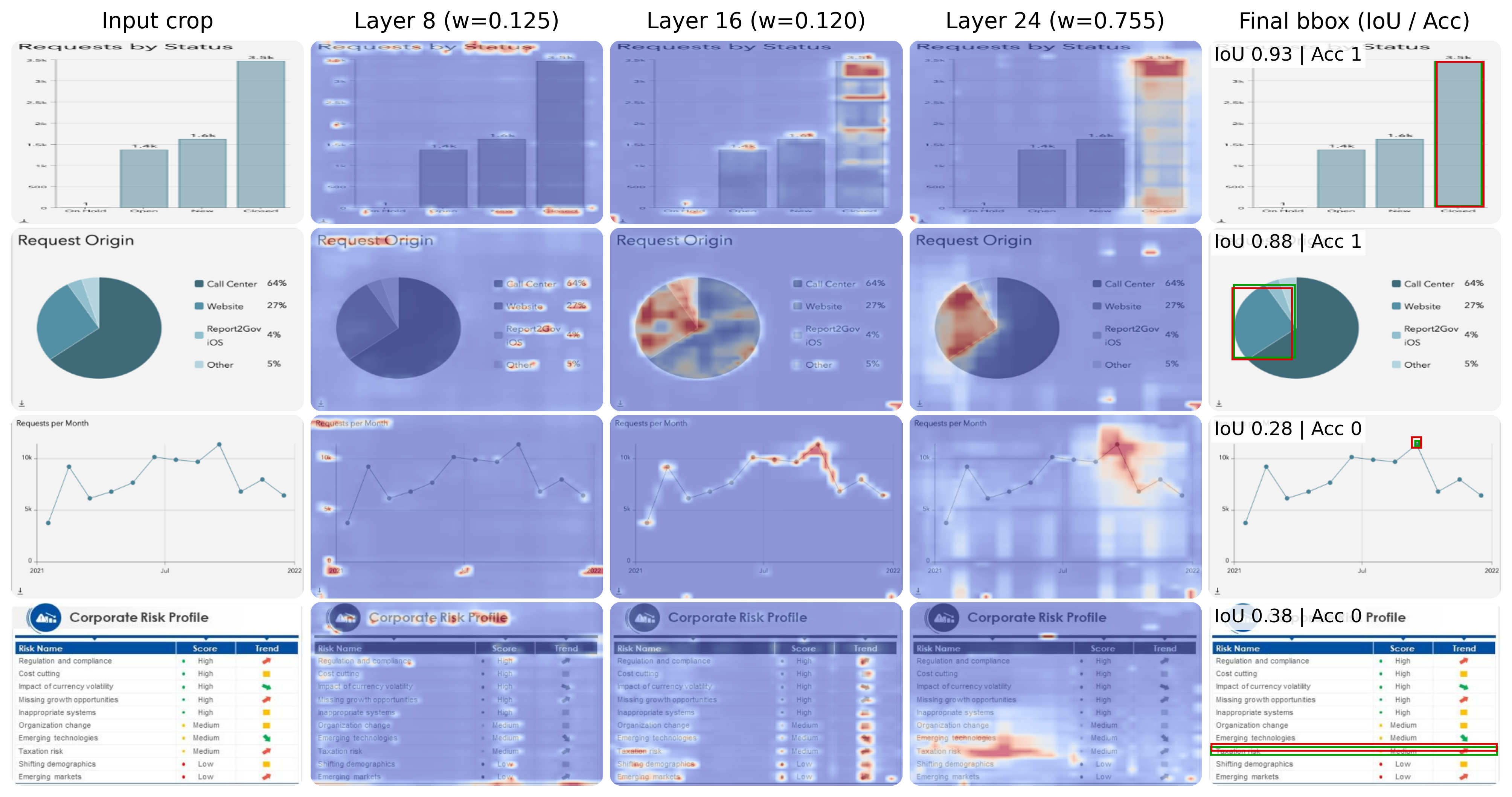}
\caption{MLAD layer attention on validation crops from the released Stage II model. Rows are single crops with their in-chart queries (bar, pie, line, and table charts from top to bottom); columns show the input, the query-token attention on the $40\times40$ patch grid at BEiT-3 layers 8, 16, and 24 (head-averaged, summed over query tokens, percentile-clipped per layer, blue = low, red = high), and the final prediction (red) against the ground truth (green) with its IoU and Acc@0.5 hit. The MLAD softmax weights are $0.125$, $0.120$, and $0.755$.}
\label{fig:mlad_attention}
\end{figure*}

\textbf{What MLAD fuses.} Figure~\ref{fig:mlad_attention} visualizes the attention of the three layers that MLAD combines, using crops and queries from the validation split and the released Stage II checkpoint. The progression is consistent across chart types: layer 8 spreads attention over the chart title, axes, and multiple elements; layer 16 concentrates on the region that contains the answer; and layer 24 aligns with the queried element itself (the tallest bar, the requested table row, the rightmost point of the line). MLAD assigns most of the fusion weight to layer 24 ($0.755$) while retaining the earlier layers as a small corrective component. This is a qualitative illustration of the fusion behaviour; the quantitative ablation in Table~\ref{tab:ablation_stage2_val} measures the effect of the module.

\subsubsection{Ablation Configurations}
We compare three cumulative configurations that differ only in the components under study; all of them use the same data, ground-truth-crop protocol and evaluation script.
\begin{itemize}
\item \textbf{SimVG}: BEiT-3 Large with the interpolated $32 \times 32$ projection, no MLAD, 50 epochs, batch size 2, base learning rate $5 \times 10^{-5}$, encoder learning rate $5 \times 10^{-6}$, milestones at epochs 30 and 45, deterministic seed 6666, no EMA.
\item \textbf{SimVG + MLAD}: adds the 8/16/24 fusion. As in the original parameter grouping, the three mixing logits are optimized with the backbone learning rate ($5 \times 10^{-6}$); all other settings are identical to SimVG.
\item \textbf{+ native patch and MLAD logits group}: keeps the native $16 \times 16$ projection, gives the MLAD logits their own $5 \times 10^{-4}$ group, trains for 50 epochs with batch size 1 (the $40 \times 40$ token grid does not fit batch size 2 on a 24\,GB card), and uses deterministic seed 42; all other settings are identical. This is our final Stage II configuration.
\end{itemize}
The ablation changes neither the architecture, the loss, nor the training data of any variant. The final configuration yields the largest AP$_s$ gain (test 3.87$\rightarrow$7.10; validation 7.80$\rightarrow$10.73), consistent with the finer native $16\times16$ token grid. Alongside the projection, this configuration also gives the MLAD logits their own $5\times10^{-4}$ learning-rate group and uses batch size 1 and seed 42; these implementation choices accompany the projection change and are kept identical for the final model.

The validation-set results of the three configurations are reported in Table~\ref{tab:ablation_stage2_val}; the corresponding test-set results are reported in the main paper.
\begin{table}[tb!]
\centering
\caption{Ablation study for Stage II on the validation set (n=200). ``P16'': native $16\times16$ patch projection. The corresponding test-set results are reported in the main paper.}
\label{tab:ablation_stage2_val}
\resizebox{\columnwidth}{!}{
\begin{tabular}{lccccccc}
\toprule
Method & IoU & Acc & DotD & NWD & AP$_s$ & AP$_m$ & AP$_l$ \\
\midrule
SimVG & 43.48 & 51.00 & 71.44 & 79.34 & 2.93 & 27.36 & 48.09 \\
+ MLAD & 45.98 & 52.00 & 71.91 & 79.02 & 7.80 & 31.98 & 49.56 \\
+ MLAD + P16 (Ours) & \textbf{49.25} & \textbf{53.50} & \textbf{74.07} & \textbf{80.70} & \textbf{10.73} & \textbf{36.26} & 48.68 \\
\bottomrule
\end{tabular}}
\end{table}

\section{Baseline Reproduction Details}
\label{sec:baseline_reproduction}

All baseline systems reported in the main paper were run by us with their official code, released checkpoints, or official APIs. We group the baselines into three categories: (i) \emph{General MLLMs}, evaluated zero-shot through their official APIs, except Qwen2.5-VL-7B and InternVL2.5-8B, which are evaluated locally without fine-tuning; (ii) \emph{Chart-specific / literature models} evaluated without fine-tuning; and (iii) \emph{models fine-tuned on VGD-Bench}. The shared protocol is documented first, followed by the configuration of each group and of the representative two-stage baselines.

\subsection{Shared Data and Evaluation Protocol}

\textbf{Data and annotation version.} Every fine-tuned method is trained on the same VGD-Bench training split. The test list (214 targets) and the image files are identical across methods; no method receives extra data, pseudo-labels, or test-time images. The validation split contains 200 targets and the test split 214. Every row in every table is scored on the same lists and from the same per-sample prediction files.

\textbf{Query decomposition.} For Stage I (chart-level retrieval) the query is \textbf{s1\_text} and the prediction target is the chart \textbf{region} bounding box. For Stage II (in-chart grounding) the query is \textbf{s2\_text}, the input is a single chart cropped from the dashboard, and the prediction target is the \textbf{target} bounding box expressed relative to that crop. For the overall setting the dashboard and the \textbf{original\_query} are used, and the answer region is predicted in the original image coordinate system. {The \textbf{s1\_text} and \textbf{s2\_text} fields are generated by DeepSeek-API (Section~\ref{sec:prompts}) and manually verified; all Stage I and Stage II entries in every table use these same fields. They are not human-rewritten: the stored sub-queries are the DeepSeek outputs with the list numbering removed and the raw API response retained (Section~\ref{sec:annotations}). The two-stage results, including the 66.82\% test Acc@0.5 in Table~\ref{tab:two_stage}, therefore use the unmodified decompositions, and no method has access to a human-corrected version; at inference the original query is decomposed automatically without human intervention.}

{\textbf{Input resolution.} All annotations and all metrics are defined in the original dashboard coordinate system. Each method receives the image in the format prescribed by its own architecture. API-based MLLMs receive the original-resolution dashboard, subject to the provider's internal preprocessing. VGD-Net resizes the dashboard to $1024\times1024$ for Stage I and each chart crop to $640\times640$ for Stage II (Section~\ref{sec:implementation}). Fixed-resolution baselines use their official input sizes, for example, $224\times224$ for BARE, $320\times320$ for InstanceVG (fine-tuned variant), $384\times384$ for OneRef, $640\times640$ for SeqTR and SimVG, and $800\times800$ for MDETR, while MLLM baselines use their native image preprocessing (e.g., Qwen2.5-VL's smart-resize). We deliberately do not force all methods to one common resolution: changing a baseline's official input size would alter its architecture-specific preprocessing and make the comparison less faithful to the published method. Regardless of input size, every predicted box is mapped back to the original dashboard coordinates before scoring, and every Stage II crop is cut from the original-resolution dashboard. Resolution therefore affects only each model's internal representation, not the evaluation protocol.}

\textbf{Composed two-stage evaluation.} When a two-stage method is evaluated end-to-end, Stage II receives the chart crop produced by its own Stage I rather than the GT crop. The predicted element box is mapped from crop coordinates back to the original dashboard coordinates before scoring, so that ground-truth crops and predicted crops are handled by exactly the same geometric mapping. All two-stage methods are evaluated under both settings (ground-truth crop and predicted crop) with the same code path.

\textbf{Target retention.} Annotation guidelines require every target box to be contained in its parent chart region; the post-hoc audit found a small number of violations, which were re-associated or, when the query cannot be grounded in the attached chart (two queries about line-chart elements), excluded from the validation set. Targets that are not fully contained are kept and scored normally rather than silently dropped or clipped, and a method that fails to return a box is assigned IoU $=0$. 

\textbf{Metrics.} All metrics are computed offline from saved per-sample predictions with one script, independently of the model code. IoU, Acc@0.5 and Acc@0.75 follow the standard definitions, DotD and NWD are the distance-based complements reported in the main paper, and AP@[.5:.95] is the hit rate averaged over the ten IoU thresholds $\{0.50,0.55,\dots,0.95\}$. The main paper follows the COCO protocol for the threshold schedule and scale bins; since every query yields a single box against a single ground-truth box, the precision--recall curve at a given threshold degenerates to a single point, so AP@$t$ coincides with the hit rate Acc@$t$ and AP@[.5:.95] is their average over the ten thresholds. AP$_s$, AP$_m$ and AP$_l$ stratify targets by ground-truth area ($32^2$ and $96^2$). No test-time augmentation, model ensembling, multi-crop voting, or per-method score threshold tuning is applied. Ground-truth boxes are loaded only to compute the metrics after prediction; they are never visible to the candidate ranking, the verifier, or any model input.

\subsection{General MLLMs}

\subsubsection{API-based MLLMs}

Mistral Medium 3, GPT-4o, GPT-5, Gemini-2.5 Pro, Claude 4 Sonnet, Llama-4 Maverick, GLM-4.5V and Qwen3-VL-Plus are evaluated zero-shot through the official API of their respective providers. No parameters are updated. All models are queried with a fixed grounding prompt that asks for a single bounding box, and the returned box is converted to the original dashboard coordinate system and scored with the same metric script as every other method. No method-specific prompt tuning, retry-based selection, or test-time ensemble is applied.

\subsubsection{Local non-fine-tuned models}

Qwen2.5-VL-7B and InternVL2.5-8B are evaluated locally from their released checkpoints without any fine-tuning or adapter; no parameters are updated for these two entries.

\textbf{Qwen2.5-VL-7B (zero-shot).} The image is processed by the model's native smart-resize with at most $512\times28^2$ pixels, and inference is run in bf16 with greedy decoding (no sampling or ensembling). A single fixed grounding prompt is used for all settings: the query is prefixed with ``Locate the target described by:'' and the model is asked to return only a JSON object \texttt{\{"bbox": [x1, y1, x2, y2]\}} in absolute pixel coordinates. Stage I receives the dashboard and s1\_text, Stage II receives the GT crop and s2\_text, the composed two-stage setting receives the Stage I predicted crop, and the overall setting receives the dashboard and original\_query. The predicted box is converted from the smart-resized space back to the original dashboard coordinates using the fine-tuned model's mapping.

\textbf{InternVL2.5-8B (zero-shot).} The released checkpoint is evaluated locally with its image preprocessing and the same zero-shot protocol: a fixed grounding prompt, greedy decoding, the same Stage I / Stage II / composed / overall inputs, and conversion of the predicted box to the original dashboard coordinates. No fine-tuning or adapter is used.

\subsection{Chart-specific / Literature Models (no fine-tuning)}

This group contains chart-specific and literature grounding models that are evaluated without fine-tuning, including CLIP, OWLv2, KOSMOS-2 and Florence-2. They use the same data splits, crop protocol and metric implementation as all other methods, and no parameters are updated. All four consume resized inputs rather than the original image resolution: CLIP scores each candidate after resizing it to $224\times224$, KOSMOS-2 resizes the whole image to its native $224\times224$ input, OWLv2 uses the $960\times960$ input of its official processor, and Florence-2 uses the $768\times768$ input of its official processor.

\subsubsection{CLIP}

CLIP is a contrastive image--text model that is trained to maximize the alignment between image and text embeddings; it has no bounding-box regression or detection head. Fine-tuning it into a grounding model would therefore require attaching a detection/grounding head and adding box supervision, which changes the model rather than adapting it. This is a property of the architecture, not of the annotation format: the VGD-Bench JSON is consumed directly, but CLIP cannot emit a box without external proposals. We consequently evaluate CLIP \textbf{zero-shot}.

\textbf{Common setup.} \texttt{openai/clip-vit-base-patch32}; regions from OpenCV Selective Search (fast mode, up to 1000 proposals, minimum side 32\,px) on CPU; each candidate is resized to CLIP's $224\times224$ input and scored individually against the text query (one forward pass per candidate), and the highest-scoring candidate is returned. fp32, no fine-tuning, post-processing or test-time augmentation. CLIP scoring runs on a single RTX 4090 GPU; this is numerically equivalent to CPU forward for the same fp32 model, so the device does not affect the numbers.

\textbf{Coordinate mapping.} For Stage I and Overall the prediction is already in dashboard coordinates. For Stage II the candidate region is predicted inside the chart crop and mapped back to the dashboard by adding the crop origin; the ground-truth target box is likewise expressed in crop coordinates by subtracting the crop origin. No threshold or non-maximum suppression is applied.

{\textbf{KOSMOS-2 (zero-shot).} We evaluate the grounding model \texttt{microsoft/kosmos-2-patch14-224} from its released checkpoint without fine-tuning or adapter. Images are resized to the model's native $224\times224$ input. Each task uses the prompt \texttt{<grounding> \{query\}} and greedy decoding with at most 128 new tokens; the emitted \texttt{<object><patch\_index\_...></object>} tokens are converted to a box with the official post-processing and mapped back to dashboard coordinates. The same protocol is used for Stage I (dashboard + s1\_text), Stage II (GT crop + s2\_text), and overall VGD (dashboard + original\_query); no prompt tuning, retry-based selection, or ensembling is applied.}

\subsection{Fine-tuned on VGD-Bench}

All methods in this group are initialized from their official checkpoints and fine-tuned on the VGD-Bench training split under the shared protocol above. We give the complete configurations of the representative baselines below: TransVG, MDETR, SeqTR, Grounding DINO, LLaVA-1.5, OneRef, InstanceVG, Qwen2.5-VL-7B (LoRA), BARE and SimVG. The remaining fine-tuned entries in the main table (LLaVA-NeXT, PaliGemma and InternVL2.5-8B) follow the same data split, training/evaluation protocol and metric implementation, using each model's official code and released checkpoint; their default training settings (input resolution, optimizer, schedule and augmentation) are kept from the original repositories, with only the batch size reduced where necessary to fit a single 24\,GB GPU.

{\subsubsection{TransVG (ResNet-50)}}

TransVG is fine-tuned from the RefCOCOg checkpoint released with the official implementation (ResNet-50 visual backbone, BERT-base text encoder, $640\times640$ padded input). The three settings are trained independently with AdamW (learning rate $1\times10^{-4}$ for the fusion head and $1\times10^{-5}$ for the backbone and BERT), batch size 4, 20 epochs, and the repository defaults otherwise; we report the best-validation checkpoint. All three settings follow the shared protocol; the two-stage Overall entry feeds the Stage I predicted crop to the Stage II model as is.

\subsubsection{MDETR}

MDETR is fine-tuned from the released ResNet-101 checkpoint pretrained on COCO+VG+Flickr. Inputs are resized to $800\times800$ without preserving the aspect ratio; queries are tokenized with a RoBERTa-base encoder (maximum length 256). The model uses a 256-dimensional hidden size, six encoder and six decoder layers, eight heads, a 2048-dimensional FFN, 100 object queries, sine positional embeddings and auxiliary decoding losses. All parameters are trainable; the checkpoint is loaded with \texttt{strict=False}.

\textbf{Common configuration.} AdamW, lr $1\times10^{-4}$ (backbone $5\times10^{-5}$, text encoder $1\times10^{-4}$), weight decay $1\times10^{-4}$, linear-with-warmup schedule, gradient clipping 0.1, batch size 2; loss CE\,1 + L1\,5 + GIoU\,2 + contrastive-align\,1 with EOS coefficient 0.1 (contrastive loss disabled); resize-only augmentation; fp32; final checkpoint; single RTX 4090 (24\,GB); seed 42.

\textbf{Settings.} Stage I uses EMA 0.9998 (peak 5.4\,GB) and Stage II trains without EMA (4.7\,GB); the direct Overall model follows the Stage I recipe.

\textbf{Coordinate mapping.} The dashboard or chart crop is resized directly to $800\times800$ and the target box is normalized in this frame. The predicted box is divided by the resize factors and, for Stage II, the crop origin is added. The text encoder receives the whole query; the positive map marks all query tokens as corresponding to the single target box.

{The MDETR entries in all tables use a corrected evaluation script that executes the MDETR decoder per query. An earlier legacy script reused the decoder's query prior without running the decoder, which produced much lower Stage I/II scores; those numbers are superseded. The direct Overall VGD model (full dashboard $+$ \texttt{original\_query}) is trained separately for 30 epochs with the same recipe and evaluated with the same decoder protocol.}

\subsubsection{SeqTR}

SeqTR with a DarkNet-53 backbone is fine-tuned in two independent stages from the official checkpoint released by the authors, which was pretrained on the merged referring-expression corpus. All parameters except the text embedding are initialized from this checkpoint (520 of 521 parameter tensors); the embedding is re-initialized because the VGD-Bench vocabulary differs from the original one.

\textbf{Common configuration.} Input $640\times640$, queries truncated to 15 tokens, coordinate predictor with 1{,}000 bins; AdamW, lr $5\times10^{-5}$ (cosine annealing), weight decay $1\times10^{-4}$, batch size 4, gradient clipping 1.0, no data augmentation; single RTX 4090 (24\,GB), Ubuntu 22.04, Python 3.10, PyTorch 2.0.1, CUDA 11.8, mmcv-full 1.7.2, mmdet 2.28.2; seed 42 with deterministic cuDNN.

\textbf{Settings.} The query vocabulary holds 742 tokens for Stage I and 896 for Stage II; Stage I keeps the checkpoint with the best validation chart Acc@0.5 and Stage II the one with the best validation loss. The Overall model uses the same schedule without augmentation.

\textbf{Coordinate mapping.} At inference the crop is resized to $640\times640$, the predicted box is divided by the scale factor, and the crop origin is added to give dashboard coordinates.

\textbf{Deviations from the original SeqTR recipe.} The original SeqTR paper trains from scratch with Adam, an initial learning rate of $5\times10^{-4}$, a batch size of 128, large-scale jittering (LSJ), and exponential moving averaging (EMA). Reproducing that schedule is neither necessary nor feasible here: the released checkpoint is already trained on a large referring-expression corpus, the 24\,GB GPU cannot hold a batch of 128 at $640\times640$, and all fine-tuned baselines share a short fine-tuning budget. We therefore fine-tune with AdamW at $5\times10^{-5}$ and batch size 4, and disable LSJ and EMA while keeping the official pretrained weights. These choices are applied uniformly to both stages and to the Overall setting.

{\subsubsection{Grounding DINO (Swin-T)}}

{Grounding DINO is fine-tuned with the Open-GroundingDino training code from the released \texttt{groundingdino\_swint\_ogc} Swin-T checkpoint. Instances are converted to the ODVG grounding format, in which every sample carries one phrase--box pair. The only deviation from the official recipe is the input resolution: both training and inference resize every input, dashboard or chart crop, to a fixed $640\times640$ grid without preserving the aspect ratio, matching the fixed-square input convention used by VGD-Net's Stage II, instead of the default $800$/$1333$ resize that preserves it; the batch size is $4$, the optimizer is AdamW with base learning rate $5\times10^{-5}$ and backbone rate $5\times10^{-6}$, the weight decay is $1\times10^{-4}$, the learning rate is multiplied by $0.1$ twice, the gradient norm is clipped to $0.1$, the BERT text encoder is frozen, and the seed is $42$. At inference we take the highest-scoring box, without a score threshold or NMS, and map it back through the crop origin for crop-based settings; we report the final checkpoint. The two-stage Overall entry crops the Stage I prediction and feeds it to the Stage II model, as in the shared protocol.}

{\subsubsection{LLaVA-1.5}}

{LLaVA-1.5 (\texttt{llava-hf/llava-1.5-7b-hf}; CLIP ViT-L/14-336 vision encoder and Vicuna-7B language model) is fine-tuned with LoRA (rank 8, $\alpha=32$, dropout 0.05) on the query/key/value/output and gate/up/down projections. We use AdamW with learning rate $1\times10^{-4}$ and cosine decay, batch size 1 with 16-step gradient accumulation, bf16 and gradient checkpointing. All three settings follow the shared protocol; each setting outputs a JSON object with a normalized \texttt{[x1,y1,x2,y2]} box (the native text-only interface of these MLLMs, which have no box-regression head), and the parsed box is scored. The two-stage Overall entry crops the dashboard with the predicted Stage I region and reuses the Stage II adapter.}

\subsubsection{OneRef}

OneRef is a single-tower BEiT-3 model that regresses the target box directly. We use the base variant with patch size 16 and a $384\times384$ input, and fine-tune the two stages independently from the official mixed REC pretraining checkpoint released by the authors (the \texttt{mixup} model in \texttt{rec\_mixup\_grounding\_pretraining\_base}). All parameters are trainable, and the optional segmentation and mask-constraint heads are disabled.

\textbf{Common configuration.} Queries tokenized with the released SentencePiece model (vocabulary 64010) and truncated to 64 tokens; AdamW, lr $3\times10^{-5}$ (cosine annealing), weight decay $1\times10^{-4}$, batch size 12, 20 epochs; L1 + GIoU box regression; resize/crop/translate augmentation; gradient clipping disabled; single RTX 4090 (24\,GB), Ubuntu 22.04, Python 3.10, PyTorch 2.0.1, CUDA 11.8; seed 42.

\textbf{Settings.} We keep the best chart Acc@0.5 for Stage I and the best element Acc@0.5 for Stage II; the Overall model uses the same schedule and augmentation.

\textbf{Coordinate mapping.} At inference the crop is resized to $384\times384$ with centered zero-padding, the predicted box is mapped through the same padding and scale factors, and the crop origin is added.

\textbf{Deviations from the original OneRef recipe.} The official recipe uses eight GPUs with batch size 32 (effective 256). Our single 24\,GB GPU holds a batch of 12 at $384\times384$ with all parameters trainable; we keep the same 20-epoch budget, optimizer, learning rate, input resolution, and query length. The official script additionally enables an auxiliary box-mask constraint on top of the box regression loss. We disable it because VGD-Bench provides bounding-box annotations only, whereas the constraint's mask targets are designed for pixel-level annotations of referring-segmentation datasets. The released code uses no warmup, and we follow it. These choices are applied identically to both stages and to the Overall setting.

\subsubsection{InstanceVG (fine-tuned variant)}

We fine-tune InstanceVG on VGD-Bench; this is the InstanceVG row reported in the main results table. The official training format is COCO/refcoco MixedSeg with instance masks; we convert the VGD-Bench annotations to this format and use a rectangle polygon derived from each ground-truth box as a pseudo-mask. The mask-loss weights are set to zero, so only the box branch is optimized; mask metrics are not used.

\textbf{Common configuration.} We start from \texttt{InstanceVG-B-refcoco}; BEiT-3 Base (patch 16) with \texttt{MIXRefUniModel}; \texttt{UniGRefDeformableHead} (hidden 256, one query, AQSM); BEiT-3 SentencePiece tokenizer, maximum query length 40; input $320\times320$ (aspect ratio not preserved); Adam, lr $1\times10^{-4}$ (backbone $1\times10^{-5}$), $\beta{=}(0.9,0.98)$, $\epsilon{=}10^{-9}$, zero weight decay; MultiStepLRWarmUp with 3 warm-up epochs and $\times0.1$ at epoch 15, 20 epochs; batch 4, fp32, no EMA; best val DetACC@0.5 retained; single RTX 4090 (24\,GB); seed 42.

\textbf{Coordinate mapping.} The dashboard or chart crop is resized directly to $320\times320$; the predicted box is divided by the resize factors and, for Stage II/two-stage, the crop origin is added. The conversion uses \texttt{model\_splits.s1\_text} / \texttt{s2\_text} (with \texttt{original\_query} fallback). The official training entry required a small patch to accept the extra validation return value; no other method-specific test-time tuning is applied.

\subsubsection{Qwen2.5-VL-7B (LoRA fine-tuned)}

Qwen2.5-VL-7B-Instruct is a general-purpose multimodal model without an official grounding fine-tuning recipe for VGD-Bench. We adapt it with LoRA on the VGD-Bench training split, independently for the two stages, starting from the released instruction-tuned checkpoint.

\textbf{Common configuration.} Vision tower frozen; LoRA adapters (rank 8, alpha 32, dropout 0.05) on all linear layers of the language model; AdamW (cosine annealing), weight decay $1\times10^{-4}$, batch 2 with gradient accumulation 8 (effective 16); images processed by the native smart-resize with at most $512\times28^2$ pixels; maximum sequence length 2048; bf16, gradient clipping 1.0; single RTX 4090 (24\,GB), peak $\approx$22\,GB; seed 42.

\textbf{Settings.} lr $1\times10^{-4}$ for Stage I (best validation chart Acc@0.5) and $7\times10^{-5}$ for Stage II (final checkpoint, best validation element Acc@0.5); the Overall model follows the Stage I recipe.

\textbf{Coordinate mapping.} At inference the crop is smart-resized with at most $512\times28^2$ pixels, the predicted $[x,y,w,h]$ box is converted to $[x_1,y_1,x_2,y_2]$, and the coordinates are mapped back through the resize and the crop origin.

\textbf{Deviations from full fine-tuning.} A full fine-tune of the 7B model requires more than 60\,GB for weights, gradients, and Adam states, which is not available on our single 24\,GB GPU. We therefore freeze the vision tower and attach LoRA adapters only to the language model. No test-time adaptation, ensembling, or multi-crop voting is used.

\subsubsection{BARE}

BARE is a one-tower visual grounding model built on BEiT-3. We fine-tune three independent BARE-B specialists from the public BEiT-3 Base checkpoint (patch 16, 224); the official BARE-B/L checkpoints are not released, so the BARE-specific modules (LSM, VBC, R$^2$E, the vision-language transformer and the box head) are implemented following the paper and trained from scratch.

\textbf{Common configuration.} Backbone BEiT-3 Base (12 layers, hidden 768) frozen except LoRA ($r{=}32$, $\alpha{=}16$, dropout 0.1 on the q/k/v/out projections); queries tokenized with the BEiT-3 SentencePiece model (vocabulary 64010), truncated to 77 tokens; AdamW, lr $1\times10^{-4}$ (cosine annealing), weight decay $1\times10^{-4}$; 60 epochs; loss L1 + GIoU + contrastive + RTCC + mask (box-derived masks); ColorJitter only; fp32; single RTX 4090 (24\,GB); Python 3.9.25, PyTorch 2.2.2+cu121, transformers 4.30.0, timm 0.4.12, peft 0.6.2; seed 3407.

\textbf{Settings.} All stages use $224\times224$ inputs; batch 8 with a 6.0\,GB peak for Stage I and Stage II, and batch 16 with a 10.7\,GB peak for the direct Overall model.

\textbf{Coordinate mapping.} BARE regresses a normalized $[x,y,w,h]$ box (with $(x,y)$ the box centre) in the padded $224\times224$ frame. At inference the padding is removed and the box is divided by the resize ratio; for Stage II the crop origin is added. No test-time augmentation, confidence thresholding or test-set tuning is used. Dependencies are fairscale 0.4.13, torchmetrics 0.11.4, einops 0.8.2 and numpy 1.21.2.

\subsubsection{SimVG}

SimVG is a one-tower grounding framework: a BEiT-3 backbone jointly encodes the image and the query, and a DETR-style head predicts the target box. We fine-tune the two stages independently from the officially released ViT-Large decoder-branch checkpoint (the \texttt{large-DB} RefCOCO model), obtained from the authors' ModelScope repository because the original links are no longer public.

\textbf{Common configuration.} Backbone BEiT-3 Large with $32\times32$ patches (obtained by interpolating the pretrained $16\times16$ projection), drop-path 0.1, no frozen layers; head with text-guided query generation, 256-dimensional embedding, one object query, six encoder and three decoder layers, only the decoder branch trained; queries tokenized with the BEiT-3 SentencePiece model (vocabulary 64010), truncated to 40 tokens; input $640\times640$ (aspect ratio not preserved); Adam, lr $5\times10^{-5}$ (backbone $5\times10^{-6}$), $\beta{=}(0.9,0.98)$, $\epsilon{=}10^{-9}$, AMSGrad, zero weight decay; MultiStepLR with 2-epoch warm-up and $\times0.1$ at epochs 30 and 45; batch size 2, 50 epochs; large-scale jitter (scale 0.3--1.4); gradient clipping 0.15; fp32, no EMA; single RTX 4090 (24\,GB); seed 6666.

\textbf{Settings.} Stage I keeps the best validation chart Acc@0.5 and Stage II the best validation element Acc@0.5; the Overall model uses the same recipe.

\textbf{Coordinate mapping.} At inference the crop is resized to $640\times640$ without preserving the aspect ratio, the predicted box is divided by the resize factor, and the crop origin is added back to obtain the box in dashboard coordinates.

\textbf{Deviations from the VGD-Net Stage II configuration.} The SimVG baseline is kept deliberately close to the original SimVG setting, so that it isolates the contribution of the Stage II components described in Section~\ref{sec:implementation}. Relative to the VGD-Net Stage II configuration, the baseline uses the interpolated $32\times32$ projection instead of the native $16\times16$ projection, omits MLAD, uses batch size 2 instead of 1, uses deterministic seed 6666 instead of 42, and keeps the validation-selected checkpoint instead of the final epoch; the data, the 50-epoch schedule, and the base and encoder learning rates are identical. Neither configuration uses an exponential moving average. Only the decoder branch is trained, matching the released checkpoint.

\subsection{Fairness Considerations}

Three properties make the comparison in the main paper reproducible and fair. First, the data pipeline is shared: identical images, identical annotations, identical query splits, and identical handling of out-of-region targets. The evaluation is model-agnostic: every method writes per-sample predictions to a common format, and a single script computes all metrics, including DotD, NWD, AP and its scale-stratified variants, from those files. Third, no method-specific tuning is performed at test time, and each method is trained once with a fixed seed; the reported numbers therefore reflect a single run rather than a selected best-of-$N$ result. We additionally report both the ground-truth-crop and the predicted-crop variants for every two-stage method, which isolates Stage II quality from Stage I errors.

\section{Additional Results}
\subsection{Validation Set Results}
\begin{table*}[t]
\centering
\caption{
Performance comparison on VGD-Bench (Validation set).
Stage I performs query-aware chart retrieval.
Stage II performs fine-grained in-chart grounding.
Overall VGD predicts the answer region directly from the dashboard for all methods except VGD-Net, whose two-stage Overall entry feeds the Stage I predicted crop to Stage II; the two-stage evaluation of every baseline is reported in Table~\ref{tab:two_stage}. All rows use the same 200-target validation list; the test list (214 targets) is identical for every method.
The full VGD-Net comprises ``RT-DETR-x + Ranking + MLLM check'' for Stage I and ``SimVG + MLAD + native patch 16'' for Stage II.
Methods marked with~$^{\star}$ use resized inputs rather than the original image resolution; detailed resolutions are given in Sections~C and~D.
}
\label{tab:val_results}
\fontsize{9}{10.8}\selectfont
\setlength{\tabcolsep}{1.8pt}
\begin{tabular}{lcc|ccccccc|ccccccc}
\toprule

&
\multicolumn{2}{c|}{\textbf{Stage I}}
&
\multicolumn{7}{c|}{\textbf{Stage II}}
&
\multicolumn{7}{c}{\textbf{Overall VGD}}
\\

\cmidrule(lr){2-3}
\cmidrule(lr){4-10}
\cmidrule(l){11-17}

\textbf{Method}
&
\textbf{IoU}
&
\textbf{Acc}
&
\textbf{IoU}
&
\textbf{Acc}
&
\textbf{DotD}
&
\textbf{NWD}
&
\textbf{AP$_s$}
&
\textbf{AP$_m$}
&
\textbf{AP$_l$}
&
\textbf{IoU}
&
\textbf{Acc}
&
\textbf{DotD}
&
\textbf{NWD}
&
\textbf{AP$_s$}
&
\textbf{AP$_m$}
&
\textbf{AP$_l$}
\\

\midrule

\multicolumn{17}{l}{\emph{General MLLMs}}\\

Mistral Medium 3 &
34.60 & 28.16 & 
8.59 & 2.97 & 42.20 & 59.72 & 0.00 & 0.24 & 4.18 & 
3.27 & 1.49 & 20.13 & 38.27 & 0.00 & 0.00 & 0.72\\

InternVL2.5-8B &
6.19 & 2.48 &
7.72 & 3.47 & 33.55 & 46.81 & 0.00 & 0.76 & 2.61 &
5.30 & 1.98 & 20.87 & 36.27 & 0.00 & 0.00 & 2.90\\

GPT-4o & 39.49 & 37.21 & 
15.24 & 9.41 & 48.72 & 62.76 & 0.80 & 1.76 & 7.91 & 
6.03 & 1.49 & 23.41 & 37.30 & 0.00 & 0.11 & 1.30\\

GPT-5 & 51.77 & 58.86 & 
22.53 & 18.81 & 58.44 & 70.34 & 0.80 & 5.18 & 10.60 & 
6.12 & 2.97 & 27.80 & 46.13 & 0.00 & 0.65 & 2.75\\

Gemini-2.5 Pro & 68.82 & 76.40 & 
35.71 & 36.14 & 69.18 & 76.72 & 2.60 & 16.12 & 30.15 & 
16.79 & 12.38 & 50.28 & 63.89 & 0.00 & 1.41 & 8.99\\

Claude 4 Sonnet & 29.75 & 26.29 & 
9.23 & 5.94 & 38.12 & 56.42 & 0.00 & 1.88 & 5.37 &
3.02 & 0.99 & 15.50 & 34.04 & 0.00 & 0.00 & 0.72\\

Llama-4 Maverick &
47.33 & 50.29 &
14.86 & 6.93 & 47.38 & 61.85 & 0.00 & 0.82 & 8.96 &
7.57 & 4.95 & 25.44 & 41.53 & 0.00 & 0.22 & 4.06\\

GLM-4.5V &
51.77 & 53.71 &
13.55 & 8.91 & 43.67 & 57.17 & 0.00 & 1.06 & 12.84 &
5.40 & 2.97 & 18.28 & 28.71 & 0.00 & 0.00 & 4.93\\

Qwen2.5-VL-7B &
29.97 & 28.98 &
18.22 & 13.86 & 50.07 & 59.60 & 0.49 & 4.57 & 8.99 &
6.31 & 2.48 & 23.45 & 37.01 & 0.00 & 0.22 & 2.32\\

Qwen3-VL-Plus &
21.94 & 16.48 &
15.22 & 10.40 & 46.90 & 58.83 & 2.00 & 1.65 & 6.87 &
9.93 & 5.94 & 28.35 & 39.19 & 0.00 & 1.09 & 6.67\\

\midrule

\multicolumn{17}{l}{\emph{Chart-specific / Literature Models (no fine-tuning)}}\\

CLIP (ViT-B/32)$^{\star}$ &
21.08 & 15.91 &
11.07 & 3.96 & 37.64 & 42.22 & 0.00 & 1.53 & 3.28 &
5.60 & 1.49 & 24.56 & 26.72 & 0.00 & 0.00 & 2.75\\

OWLv2 (ViT-B/16)$^{\star}$ &
18.90 & 10.80 &
16.97 & 11.88 & 41.84 & 45.90 & 0.00 & 10.22 & 8.84 &
5.95 & 3.47 & 17.69 & 21.40 & 0.00 & 2.93 & 1.45\\

KOSMOS-2 (1.6B)$^\star$ &
13.77 & 6.25 &
14.82 & 11.88 & 40.91 & 40.87 & 0.00 & 3.48 & 15.07 &
1.41 & 0.00 & 7.96 & 5.78 & 0.00 & 0.00 & 0.00\\

Florence-2 (0.77B)$^{\star}$ &
14.61 & 5.68 &
19.95 & 16.83 & 44.79 & 49.35 & 0.98 & 9.57 & 17.10 &
3.20 & 2.48 & 11.48 & 11.30 & 0.00 & 0.00 & 4.06\\

\midrule

\multicolumn{17}{l}{\emph{Fine-tuned on VGD-Bench}}\\

TransVG (ResNet50)$^\star$ &
42.21 & 50.00 &
24.17 & 25.25 & 56.26 & 67.96 & 1.20 & 9.06 & 20.15 &
5.20 & 2.48 & 19.56 & 26.91 & 0.00 & 0.00 & 1.88\\

MDETR (ResNet101)$^\star$ &
20.16 & 22.73 &
21.44 & 22.77 & 44.95 & 54.21 & 2.20 & 12.72 & 21.88 &
3.70 & 1.49 & 12.45 & 20.44 & 0.00 & 0.11 & 1.59\\

SeqTR (DarkNet53)$^\star$ &
52.27 & 57.87 &
36.95 & 40.10 & 64.13 & 72.07 & 6.00 & 23.53 & 36.27 &
9.49 & 5.94 & 30.09 & 41.73 & 0.00 & 2.50 & 2.32\\

Grounding DINO (Swin-T)$^{\star}$ &
55.92 & 59.66 &
42.93 & 46.53 & 66.19 & 72.82 & 15.37 & 33.26 & 44.93 &
25.00 & 26.24 & 39.78 & 46.10 & 6.34 & 22.72 & 17.83\\

LLaVA (ViT-L/14)$^{\star}$ &
49.46 & 57.39 &
22.62 & 18.81 & 55.82 & 66.94 & 0.49 & 6.41 & 15.36 &
6.69 & 1.49 & 30.30 & 46.46 & 0.00 & 0.00 & 2.61\\

LLaVA-NeXT (ViT-L/14)$^{\star}$ &
55.35 & 67.05 &
32.27 & 33.17 & 65.09 & 74.39 & 3.17 & 11.96 & 29.28 &
9.69 & 6.44 & 36.63 & 53.44 & 0.00 & 0.22 & 4.93\\

PaliGemma (SigLIP)$^{\star}$ &
48.81 & 53.98 &
16.22 & 11.88 & 44.22 & 58.64 & 0.00 & 3.59 & 12.46 &
8.12 & 4.95 & 29.17 & 45.63 & 0.00 & 1.63 & 3.04\\

OneRef (BEiT-3)$^\star$ &
53.31 & 61.27 &
36.42 & 40.69 & 66.95 & 75.41 & 0.00 & 3.94 & 30.73 &
11.16 & 9.31 & 30.14 & 41.13 & 0.00 & 2.52 & 12.59\\

InstanceVG (BEiT-3)$^\star$ &
53.24 & 59.66 &
33.54 & 36.63 & 61.33 & 70.16 & 4.63 & 18.37 & 34.06 &
11.05 & 6.44 & 34.29 & 45.66 & 0.00 & 0.54 & 3.77\\

InternVL2.5-8B$^{\star}$ &
59.72 & 71.28 &
32.86 & 34.16 & 64.23 & 72.93 & 2.20 & 12.24 & 29.25 &
12.57 & 8.91 & 42.47 & 58.44 & 1.67 & 0.66 & 6.52\\

Qwen2.5-VL-7B$^{\star}$ &
59.62 & 74.51 &
31.67 & 33.50 & 69.04 & 78.80 & 0.00 & 9.57 & 27.68 &
14.05 & 9.31 & 39.11 & 55.12 & 0.00 & 2.26 & 6.38\\

BARE (BEiT-3)$^\star$ &
46.18 & 58.82 &
11.97 & 6.86 & 40.00 & 40.38 & 0.00 & 0.00 & 5.73 &
6.07 & 1.47 & 19.98 & 28.94 & 0.00 & 0.18 & 5.00\\

SimVG (BEiT-3)$^\star$ &
56.49 & 61.80 &
41.72 & 46.04 & 70.07 & 78.49 & 5.63 & 29.07 & 42.79 &
13.37 & 10.78 & 38.80 & 52.50 & 0.73 & 2.47 & 7.14\\

VGD-Net w/o Stage I$^{\star}$ &
-- & -- &
-- & -- & -- & -- & -- & -- & -- &
21.39 & 21.00 & 48.67 & 59.86 & 0.73 & 6.37 & 18.24\\

VGD-Net (Stage I + Stage II$^{\star}$) &
\textbf{85.53} & \textbf{89.50} &
{\textbf{49.25}} & {\textbf{53.50}} & {\textbf{74.07}} & {\textbf{80.70}} & {\textbf{10.73}} & {\textbf{36.26}} & {\textbf{48.68}} &
\textbf{45.45} & \textbf{48.50} & \textbf{72.49} & \textbf{79.89} & \textbf{9.27} & \textbf{32.97} & \textbf{42.79}\\

\bottomrule
\end{tabular}
\end{table*}

To complement the test-set evaluation in Table~2 of the main paper, we report the validation-set results; the trends are consistent with those on the test set.

\subsection{Two-stage Overall VGD}
\begin{table*}[t]
\centering
\caption{
Two-stage Overall VGD on VGD-Bench (validation and test set).
Each method first localizes the chart (Stage I), then crops the predicted region and localizes the target element within it (Stage II).
{All methods use the same released DeepSeek-API sub-queries (\textbf{s1\_text}/\textbf{s2\_text}) after manual verification; no method uses a human-rewritten decomposition.}
The full VGD-Net comprises ``RT-DETR-x + Ranking + MLLM check'' for Stage I and ``SimVG + MLAD + native patch 16'' for Stage II.
Methods marked with~$^{\star}$ use resized inputs rather than the original image resolution. All rows use the same 200-target validation list.
}
\label{tab:two_stage}
\fontsize{9}{10.8}\selectfont
\setlength{\tabcolsep}{2.2pt}
\begin{tabular}{l|ccccccc|ccccccc}
\toprule
&
\multicolumn{7}{c|}{\textbf{Validation (two-stage)}}
&
\multicolumn{7}{c}{\textbf{Test (two-stage)}}
\\
\cmidrule(lr){2-8}
\cmidrule(lr){9-15}
\textbf{Method}
&
\textbf{IoU} & \textbf{Acc} & \textbf{DotD} & \textbf{NWD} & \textbf{AP$_s$} & \textbf{AP$_m$} & \textbf{AP$_l$}
&
\textbf{IoU} & \textbf{Acc} & \textbf{DotD} & \textbf{NWD} & \textbf{AP$_s$} & \textbf{AP$_m$} & \textbf{AP$_l$}
\\
\midrule

\multicolumn{15}{l}{\emph{General MLLMs}}\\

Mistral Medium 3 &
3.37 & 0.50 & 23.11 & 39.52 & 0.00 & 0.00 & 1.01 &
5.12 & 1.40 & 32.77 & 49.14 & 0.00 & 0.00 & 0.51\\

InternVL2.5-8B &
1.21 & 0.50 & 9.13 & 18.34 & 0.00 & 0.00 & 0.87 &
1.60 & 0.93 & 12.56 & 23.18 & 0.00 & 0.19 & 0.38\\

GPT-4o &
5.64 & 1.49 & 27.06 & 42.72 & 0.00 & 0.11 & 1.74 &
6.70 & 3.27 & 36.19 & 50.74 & 0.00 & 0.00 & 2.31\\

GPT-5 &
10.40 & 9.41 & 43.12 & 59.20 & 0.49 & 1.20 & 5.51 &
14.58 & 8.41 & 54.44 & 67.93 & 0.00 & 0.76 & 7.18\\

Gemini-2.5 Pro &
29.14 & 26.73 & 63.91 & 72.77 & 1.22 & 7.72 & 18.99 &
29.35 & 30.84 & 61.53 & 68.76 & 0.00 & 13.81 & 13.33\\

Claude 4 Sonnet &
4.00 & 1.98 & 23.37 & 41.12 & 0.00 & 0.54 & 1.74 &
4.21 & 1.87 & 30.61 & 48.55 & 0.00 & 0.00 & 2.18\\

Llama-4 Maverick &
8.19 & 2.97 & 30.93 & 47.34 & 0.00 & 0.00 & 3.04 &
8.33 & 5.61 & 38.01 & 51.48 & 0.00 & 0.76 & 3.08\\

GLM-4.5V &
26.94 & 26.73 & 54.22 & 61.49 & 0.49 & 9.57 & 24.06 &
33.29 & 37.38 & 60.62 & 66.74 & 0.32 & 18.00 & 19.36\\

Qwen2.5-VL-7B &
8.00 & 6.93 & 29.99 & 39.21 & 0.00 & 1.74 & 3.77 &
12.07 & 8.88 & 37.63 & 46.86 & 0.00 & 2.48 & 6.79\\

Qwen3-VL-Plus &
4.64 & 2.97 & 17.90 & 28.80 & 0.73 & 0.00 & 2.46 &
6.25 & 4.21 & 21.62 & 31.17 & 0.00 & 0.19 & 4.49\\

\midrule

\multicolumn{15}{l}{\emph{Chart-specific / Literature Models (no fine-tuning)}}\\

CLIP (ViT-B/32)$^{\star}$ &
4.93 & 1.98 & 19.23 & 25.91 & 0.00 & 1.09 & 1.59 &
7.72 & 5.61 & 27.12 & 33.98 & 0.00 & 1.14 & 4.74\\

OWLv2 (ViT-B/16)$^{\star}$ &
7.89 & 5.94 & 18.24 & 24.20 & 0.00 & 3.91 & 6.81 &
8.87 & 7.01 & 22.18 & 29.40 & 0.00 & 2.19 & 8.46\\

KOSMOS-2 (1.6B)$^\star$ &
1.51 & 0.50 & 8.18 & 7.58 & 0.00 & 0.00 & 1.30 &
2.23 & 0.93 & 9.63 & 10.27 & 0.00 & 0.00 & 1.67\\

Florence-2 (0.77B)$^{\star}$ &
3.72 & 2.48 & 12.22 & 14.03 & 0.00 & 0.43 & 3.62 &
5.64 & 4.67 & 16.86 & 20.03 & 0.00 & 1.90 & 5.00\\

\midrule

\multicolumn{15}{l}{\emph{Fine-tuned on VGD-Bench}}\\

TransVG (ResNet50)$^\star$ &
11.65 & 9.90 & 32.97 & 45.33 & 0.00 & 2.71 & 8.36 &
12.50 & 10.75 & 34.69 & 45.75 & 0.00 & 2.33 & 10.40\\

MDETR (ResNet101)$^\star$ &
3.42 & 2.97 & 11.92 & 18.91 & 0.00 & 0.87 & 3.48 &
3.03 & 2.80 & 14.29 & 23.11 & 0.00 & 0.95 & 2.56\\

SeqTR (DarkNet53)$^\star$ &
15.53 & 16.75 & 35.67 & 45.54 & 0.00 & 6.34 & 15.80 &
17.11 & 17.13 & 38.90 & 47.69 & 0.00 & 5.23 & 17.59\\

Grounding DINO (Swin-T)$^{\star}$ &
25.43 & 27.23 & 41.13 & 48.82 & 8.05 & 23.91 & 20.72 &
27.05 & 28.50 & 45.77 & 54.31 & 7.10 & 17.90 & 29.23\\

LLaVA (ViT-L/14)$^{\star}$ &
15.42 & 12.87 & 41.27 & 53.21 & 0.73 & 2.72 & 11.01 &
18.96 & 15.42 & 48.19 & 59.61 & 0.00 & 3.33 & 10.13\\

LLaVA-NeXT (ViT-L/14)$^{\star}$ &
25.14 & 25.25 & 56.20 & 66.35 & 0.73 & 6.52 & 18.84 &
31.28 & 35.05 & 64.04 & 72.71 & 1.61 & 11.05 & 25.51\\

PaliGemma (SigLIP)$^{\star}$ &
9.18 & 6.93 & 30.09 & 46.76 & 0.00 & 1.63 & 5.51 &
15.05 & 12.62 & 45.01 & 59.35 & 0.32 & 1.81 & 8.85\\

OneRef (BEiT-3)$^\star$ &
25.66 & 27.23 & 48.79 & 57.78 & 1.71 & 14.46 & 23.33 &
31.47 & 35.98 & 55.60 & 63.44 & 2.26 & 14.76 & 29.87\\

InstanceVG (BEiT-3)$^\star$ &
21.24 & 23.76 & 39.67 & 49.24 & 1.46 & 12.61 & 19.86 &
24.25 & 28.04 & 45.09 & 53.95 & 3.23 & 14.00 & 24.49\\

InternVL2.5-8B$^{\star}$ &
28.74 & 26.24 & 64.20 & 71.30 & 0.98 & 9.46 & 22.46 &
38.86 & 40.65 & 71.75 & 77.79 & 0.97 & 15.33 & 30.00\\

Qwen2.5-VL-7B$^{\star}$ &
32.67 & 30.20 & 63.92 & 72.72 & 2.20 & 8.91 & 19.86 &
37.72 & 39.72 & 69.45 & 76.92 & 0.65 & 12.67 & 21.79\\

BARE (BEiT-3)$^\star$ &
14.80 & 13.37 & 39.21 & 50.05 & 0.00 & 2.93 & 10.43 &
19.87 & 19.63 & 46.49 & 56.82 & 0.32 & 4.76 & 13.08\\

SimVG (BEiT-3)$^\star$ &
29.78 & 34.65 & 53.45 & 61.58 & 1.46 & 18.04 & 28.99 &
36.43 & 42.52 & 57.08 & 64.90 & 4.52 & 19.33 & 42.44\\

VGD-Net (Stage I + Stage II$^{\star}$) &
45.45 & 48.50 & 72.49 & 79.89 & 9.27 & 32.97 & 42.79 &
57.06 & 66.82 & 82.40 & 86.79 & 6.13 & 43.14 & 53.33\\

\bottomrule
\end{tabular}
\end{table*}

\subsection{Stage I: YOLOv8x vs.\ RT-DETR-x}

In Stage I (query-aware chart retrieval), our VGD-Net employs RT-DETR-x as the object detector combined with a ranking model and an MLLM check (RM). To validate this design choice, we also experimented with YOLOv8x as an alternative detector under the same RM pipeline. As reported in Table~2 of the main paper, RT-DETR-x achieves 86.78\% IoU and 92.06\% Acc on the test set. YOLOv8x with the same RM pipeline achieves 78.73\% IoU and 84.04\% Acc on the test set, and 81.06\% IoU and 85.53\% Acc on the validation set. Both detectors are effective for chart-level grounding, but RT-DETR-x is consistently better ($+8.05$ IoU and $+8.02$ Acc on the test set), and its deformable-attention architecture generalizes better to unseen dashboards, supporting our choice for Stage I.

\subsection{Per-Class Analysis of VGD-Net}

\begin{table*}[htbp]
\centering
\setlength{\tabcolsep}{3pt}
\caption{Detailed performance of VGD-Net on validation and test sets. Accuracy is reported as Acc (H/T), where H is the number of Hits (correct predictions) and T is the total number of instances. Validation totals use the same 200-target list as every other method.}
\label{tab:val_test_performance_vgd_net}
\begin{tabular}{l cccc cccc}
\toprule
& \multicolumn{4}{c}{Validation} & \multicolumn{4}{c}{Test} \\
\cmidrule(lr){2-5} \cmidrule(lr){6-9}
Chart Type & IoU & Acc (H/T) & DotD & NWD & IoU & Acc (H/T) & DotD & NWD \\
\midrule
Card & 55.42 & 65.62 (21/32) & 75.36 & 80.10 & 70.10 & 91.11 (41/45) & 88.99 & 89.21 \\
Share & 80.66 & 100.00 (4/4) & 96.86 & 97.00 & 81.47 & 100.00 (4/4) & 95.72 & 95.03 \\
Bar Chart & 41.28 & 43.24 (32/74) & 66.52 & 76.46 & 63.30 & 75.41 (46/61) & 82.48 & 87.50 \\
Line Chart & 24.76 & 20.69 (6/29) & 70.04 & 73.78 & 29.39 & 25.93 (7/27) & 80.90 & 87.38 \\
Pie Chart & 86.19 & 100.00 (6/6) & 96.73 & 96.75 & 62.51 & 63.64 (7/11) & 82.97 & 85.87 \\
Donut Chart & 73.74 & 77.78 (14/18) & 87.19 & 89.62 & 81.22 & 100.00 (13/13) & 94.26 & 94.03 \\
Area Chart & 37.62 & 50.00 (3/6) & 64.57 & 67.98 & 33.34 & 12.50 (1/8) & 85.60 & 89.05 \\
Scatter Plot & 25.51 & 0.00 (0/2) & 60.76 & 66.53 & 23.33 & 0.00 (0/3) & 53.09 & 49.88 \\
Bubble Chart & 63.24 & 60.00 (3/5) & 88.83 & 87.78 & 50.98 & 57.14 (4/7) & 85.51 & 87.98 \\
Heatmap & 15.74 & 0.00 (0/3) & 59.05 & 88.48 & 70.22 & 75.00 (3/4) & 88.75 & 87.98 \\
Data Table & 37.82 & 50.00 (4/8) & 73.57 & 90.06 & 51.47 & 66.67 (8/12) & 79.24 & 88.33 \\
Map & 48.32 & 44.44 (4/9) & 63.48 & 66.46 & 51.86 & 57.14 (4/7) & 75.06 & 80.64 \\
Radar Chart & 0.00 & 0.00 (0/1) & 20.72 & 32.42 & 27.51 & 33.33 (1/3) & 34.31 & 36.15 \\
Dial Chart & n/a & n/a & n/a & n/a & 77.02 & 100.00 (2/2) & 94.24 & 92.48 \\
Progress Bar & 0.00 & 0.00 (0/1) & 0.11 & 2.10 & 8.07 & 0.00 (0/1) & 11.78 & 26.29 \\
Calendar & n/a & n/a & n/a & n/a & 57.53 & 100.00 (1/1) & 88.52 & 84.30 \\
UI & 0.00 & 0.00 (0/2) & 36.64 & 49.64 & 0.00 & 0.00 (0/4) & 12.45 & 33.86 \\
Image & n/a & n/a & n/a & n/a & 75.34 & 100.00 (1/1) & 93.16 & 91.28 \\
\midrule
\textbf{VGD-Net} & \textbf{45.45} & \textbf{48.50 (97/200)} & \textbf{72.49} & \textbf{79.89} & \textbf{57.06} & \textbf{66.82 (143/214)} & \textbf{82.40} & \textbf{86.79} \\
\bottomrule
\end{tabular}
\end{table*}

Table~\ref{tab:val_test_performance_vgd_net} presents the detailed per-class performance of VGD-Net on the validation and test sets.
Compared with the overall results reported in the main paper, the per-class analysis reveals substantial performance variation across chart types. Categories with regular structures (e.g., Card and Pie Chart) generally achieve higher localization accuracy, whereas chart types involving small or sparse targets (e.g., Line Chart, Scatter Plot, and Radar Chart) remain considerably more challenging.

Performance on several chart types is less stable because only a limited number of instances are available for evaluation. Categories such as Progress Bar, Calendar, Scatter Plot, and Radar Chart contain relatively few samples, resulting in larger fluctuations between the two splits.

We also observe that some chart types (e.g., Area Chart and Scatter Plot) achieve relatively low IoU but comparatively high DotD and NWD, indicating that the predicted regions are spatially close to the ground truth despite limited overlap. This further suggests that distance-based metrics provide a useful complement to IoU when evaluating small or thin targets.

Finally, we do not report AP$_s$, AP$_m$, and AP$_l$ in this per-class analysis. Unlike standard object detection benchmarks, the target regions in our task vary substantially in scale across different chart types (e.g., small points in Line Charts versus large sectors in Pie Charts). Consequently, size-based AP metrics are less informative for per-class evaluation and may lead to misleading comparisons.

\subsection{Error Propagation Analysis}
\label{sec:error_propagation}

\begin{table}[htbp]
\centering
\caption{Error propagation analysis of the two-stage pipeline. \mbox{``Total''} is the per-split sum of the four preceding cells.}
\label{tab:error_propagation}
\small
\setlength{\tabcolsep}{7pt}
\begin{tabular}{lcccc}
\toprule
\textbf{Split} & \textbf{S1} & \textbf{S2 wrong} & \textbf{S2 right} & \textbf{Total} \\
\midrule
\multicolumn{5}{l}{\itshape Pipeline incorrect} \\
test & wrong & 3 & 5 & \\
     & right & 59 & 4 & 71 \\
val  & wrong & 6 & 10 & \\
     & right & 76 & 11 & 103 \\
comb.& wrong & 9 & 15 & \\
     & right & 135 & 15 & 174 \\
\midrule
\multicolumn{5}{l}{\itshape Pipeline correct} \\
test & wrong & 1 & 8 & \\
     & right & 6 & 128 & 143 \\
val  & wrong & 1 & 4 & \\
     & right & 10 & 82 & 97 \\
comb.& wrong & 2 & 12 & \\
     & right & 16 & 210 & 240 \\
\bottomrule
\end{tabular}
\end{table}

To understand how errors propagate through the two-stage VGD-Net pipeline, we conduct a fine-grained error attribution on the validation and test sets. For each sample, we evaluate Stage I and Stage II independently against the ground truth and compare their individual correctness with the final pipeline output; a stage is correct when its predicted box achieves IoU $>0.5$ ($>50$ on the 0--100 scale used throughout). In Table~\ref{tab:error_propagation}, S1 and S2 denote these two independent evaluations, S2 refers to Stage II on the ground-truth crop, and the pipeline output is the final end-to-end prediction. All counts come from the released per-sample predictions used in Table~\ref{tab:two_stage} and Table~\ref{tab:val_test_performance_vgd_net}: the pipeline-correct totals (143/214 test, 97/200 validation) match those tables exactly, and the S2-correct totals (145/214, 107/200) match the Stage II isolation results. The decomposition covers 414 samples (174 pipeline-incorrect and 240 pipeline-correct); among the 376 samples where Stage I is correct the pipeline succeeds in 226 (60.11\%), and among the 38 where Stage I fails it succeeds in 14 (36.8\%).

Several observations emerge from this decomposition.

\noindent\textbf{Stage I correct but pipeline incorrect: the dominant error source.} The largest category of pipeline errors occurs when Stage I is individually correct yet the pipeline output is wrong (150 out of 174 errors, 86.2\%). Within this category, 135 cases (77.6\%) correspond to Stage II being individually incorrect on the ground-truth crop, and 15 cases (8.6\%) correspond to Stage II being individually correct. The two sub-cases reflect different underlying mechanisms.

For the 15 cases where Stage II is individually correct yet the pipeline fails, the root cause is clear: the IoU-based correctness criterion for Stage I (IoU $> 50$) does not guarantee that the cropped region passed to Stage II actually contains the target object. Although Stage I's predicted box meets the IoU threshold, the crop may have already excluded critical visual information. When evaluated independently with the ground-truth crop, Stage II correctly localizes the target; however, in the actual pipeline, Stage II receives the crop produced by Stage I, which may not cover the target, causing Stage II to produce an output that fails to combine correctly with Stage I's prediction.

For the 135 cases where Stage II is individually incorrect, Stage II fails to localize the target even when provided with the oracle (ground-truth) crop, indicating inherent capacity limitations of Stage II on these challenging samples. The gap between Stage I's actual crop and the oracle crop further degrades an already insufficient Stage II performance, but the primary limitation lies in Stage II itself rather than in the crop quality.

\noindent\textbf{Stage I incorrect but pipeline correct: Stage II can compensate.} Conversely, there are 14 cases where Stage I is individually incorrect (IoU $< 50$) yet the pipeline ultimately succeeds. This is expected because the Acc@0.5 criterion measures the box overlap with the chart region, not whether the crop actually covers the target: a Stage I box can fail the IoU threshold while still including the target, so Stage II can succeed on that crop even though Stage I is counted as an error. Of these 14 cases, 12 are rescued by Stage II when it is given the ground-truth crop: Stage I's predicted box, despite failing the IoU threshold, still crops a region that covers the target object. The remaining 2 cases are instances where both Stage I and Stage II are individually incorrect, yet the pipeline fusion produces a correct output. Stage II is therefore robust to imperfect crops as long as the target is not entirely excluded.

\noindent\textbf{Error propagation rate.} Summing across the combined row of Table~\ref{tab:error_propagation}, Stage I is incorrect in $9 + 15 + 2 + 12 = 38$ cases. Of these, $9 + 15 = 24$ result in a pipeline error, yielding an error propagation rate of $24/38 \approx 63.2\%$; the remaining $12 + 2 = 14$ (36.8\%) are rescued by Stage II or the end-to-end fusion. Conversely, Stage I is correct in $135 + 15 + 16 + 210 = 376$ cases, yet $135 + 15 = 150$ of them ($150/376 \approx 39.9\%$) still result in a pipeline error. 

\noindent\textbf{Implications.} These findings highlight two distinct bottlenecks. First, the crop interface between Stage I and Stage II: when Stage I's predicted box meets the IoU threshold but its crop fails to cover the target region, Stage II cannot recover even if it is individually capable. Expanding the crop margin, incorporating an explicit coverage check, or adopting a stricter correctness criterion for Stage I that accounts for crop completeness would directly address the 15 cases where both stages are individually correct yet the pipeline fails. Second, the 135 cases where Stage II underperforms even with oracle crops suggest that strengthening Stage II's localization capacity (particularly on challenging chart types with small or sparse targets) is essential to reducing the pipeline errors in the Stage I-correct regime.

\section{Evaluation via Bounding Box Outputs}

\begin{figure*}[htbp]
  \centering
  \includegraphics[width=\textwidth]{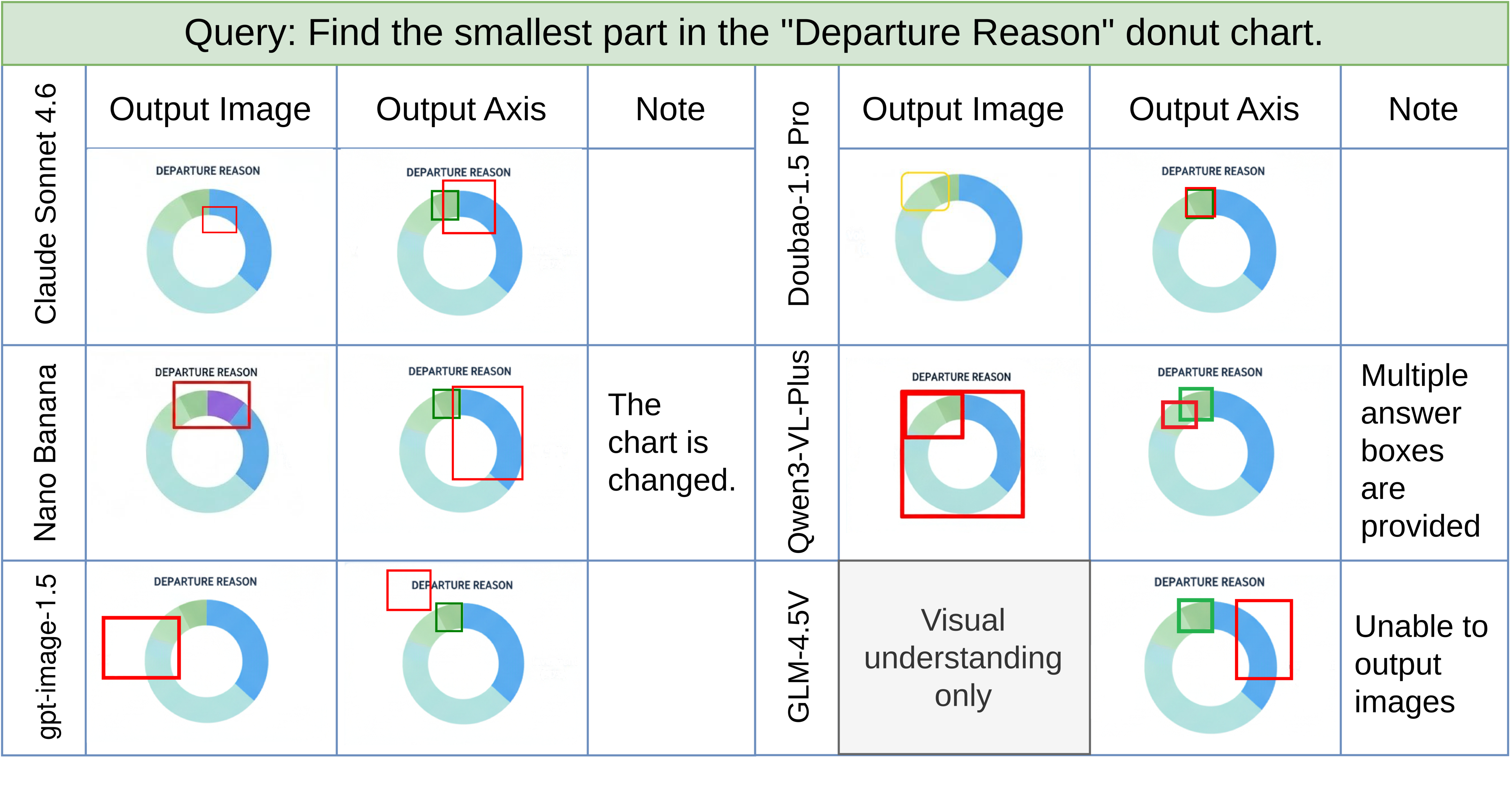}
  \caption{BBox outputs versus directly generated annotated images. Output Image: the image returned by each provider's official web interface. Output Axis: green = ground truth, red = prediction. Requiring box outputs does not reduce performance.}
  \label{fig:prove_axis}
\end{figure*}

We adopt bounding box (BBox) outputs as the evaluation interface for MLLM-based baselines and render the predicted boxes only for visualization, using the prompts in Section~\ref{sec:prompts} for consistency. Bounding boxes are the conventional detection representation and permit precise, standardized scoring through Intersection-over-Union (IoU) and mean Average Precision (mAP).

\textbf{Source of the generated images.} The Output Image column of Figure~\ref{fig:prove_axis} was collected by submitting the same query to each provider's official website (its chat interface), not to a fixed API. Such products may route a request to different internal tools: the images returned by GPT-5 and Gemini, for instance, are produced by their image-generation models (gpt-image-1.5 and Nano Banana, respectively), while other platforms may return edited or synthesized images or decline the request. This comparison is therefore not strictly controlled and is only illustrative.

\textbf{Why image-generation models are not used as baselines.} First, a generative model may alter the input image: it can redraw elements, change labels or colours, or regenerate the whole dashboard, so the returned picture is no longer the visual evidence to be grounded. Second, our task is visual grounding (predicting where the queried element is), which requires a location rather than a newly synthesized image; the generative paradigm measures a different capability and cannot be scored with the same IoU/Acc protocol. Requiring BBox outputs avoids both issues and, as Figure~\ref{fig:prove_axis} shows, does not reduce performance.

\section{Limitations}

Despite its strong performance, VGD-Net still has several limitations that point to directions for future work.

First, as quantified in the error propagation analysis (Section~\ref{sec:error_propagation}), errors in Stage I propagate to Stage II, with a propagation rate of 63.2\%. The 15 cases where both stages are individually correct yet the pipeline fails show that the crop interface is fragile, and the 135 cases where Stage II fails even with oracle crops reveal capacity limits on small or sparse targets. These findings highlight the need for more robust cross-stage reasoning, such as iterative refinement between stages or auxiliary training objectives that explicitly encourage Stage I to produce Stage II-friendly crops.

Second, VGD-Net requires densely annotated bounding boxes for both chart-level and in-chart grounding, which limits its applicability in data-sparse settings. Constructing high-quality annotations for chart grounding is particularly labor-intensive, as it demands annotators who understand diverse chart semantics and dashboard layouts. This reliance on annotations suggests the need for few-shot or weakly supervised approaches that can generalize from limited labeled examples. Self-training with unlabeled dashboard images, leveraging synthetic chart data for pre-training, or using MLLMs to generate pseudo-labels are promising directions to reduce annotation cost without sacrificing grounding accuracy.

Third, the evaluation uses a single random split and small-object grounding remains a weakness; source- or template-held-out splits would better characterize the applicability of the method.

Finally, VGD-Net processes each query independently; sharing intermediate features across queries on the same dashboard could improve deployment efficiency.

\section{Annotated Examples from VGD-Bench}
\label{sec:dataset_examples}

\begin{figure*}[t]
  \centering
  \begin{subfigure}{\textwidth}
    \centering
    \includegraphics[width=\linewidth,height=0.452\textheight,keepaspectratio]{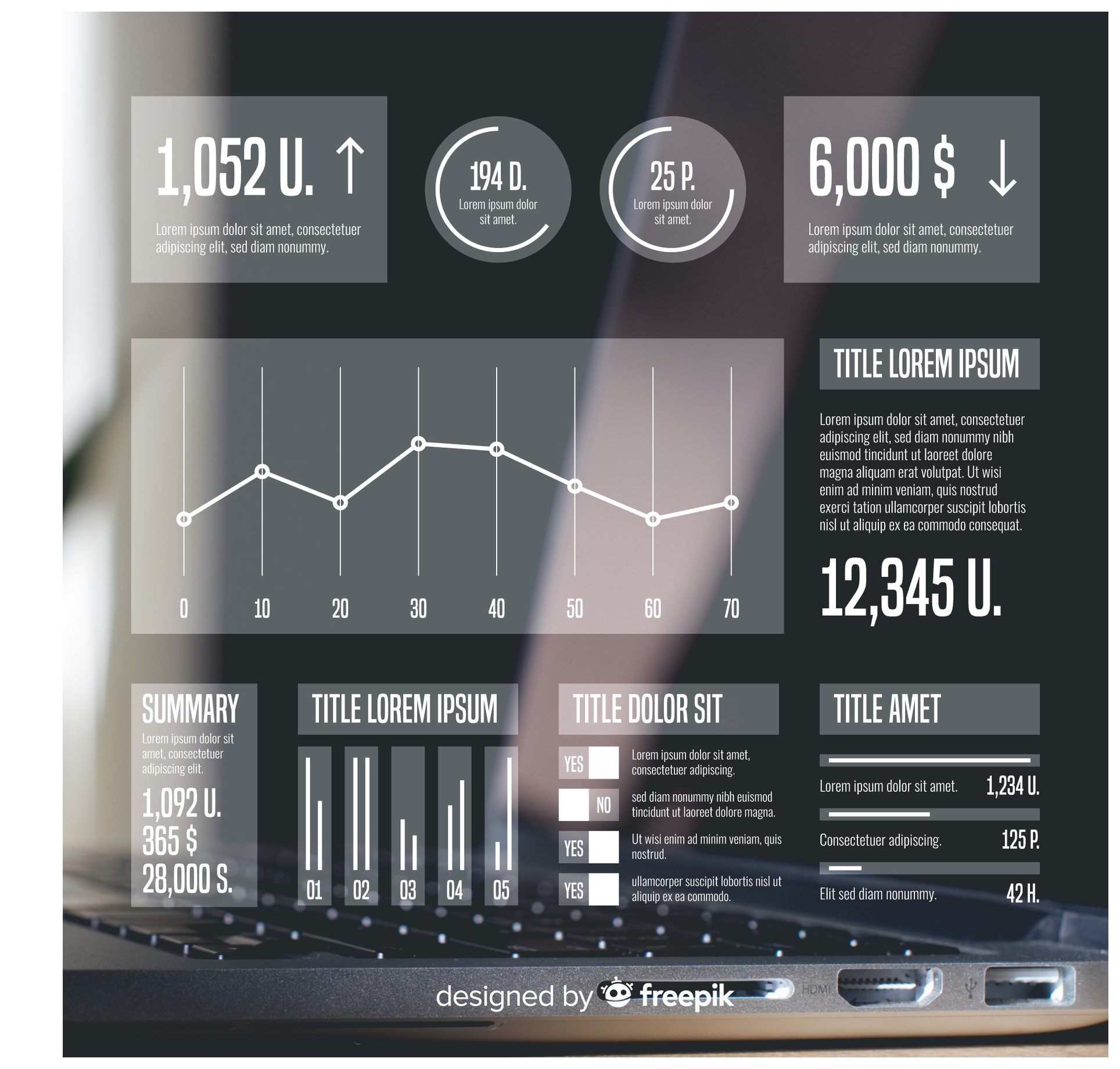}
    \caption{Input.}
  \end{subfigure}

  \vspace{0.2em}
  \begin{subfigure}{\textwidth}
    \centering
    \includegraphics[width=\linewidth,height=0.452\textheight,keepaspectratio]{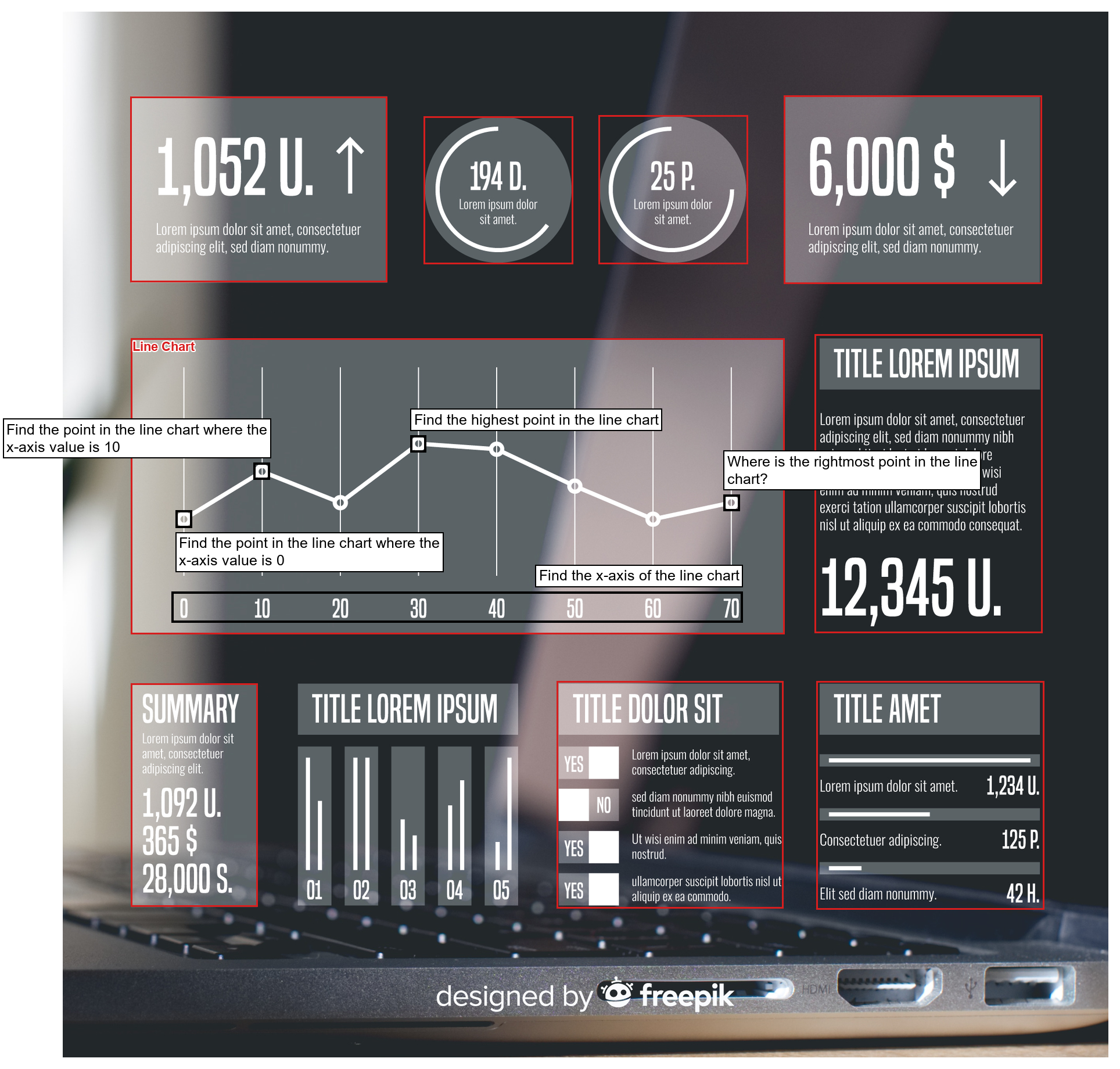}
    \caption{With region (red) and target (black) annotations.}
  \end{subfigure}
  \caption{Business infographic dashboard. Source: \url{https://img.magnific.com/free-vector/business-infographic-template-with-photo_23-2148234013.jpg} (Designed by Freepik, free license).}
  \label{fig:ex_a}
\end{figure*}

\begin{figure*}[t]
  \centering
  \begin{subfigure}{\textwidth}
    \centering
    \includegraphics[width=\linewidth,height=0.452\textheight,keepaspectratio]{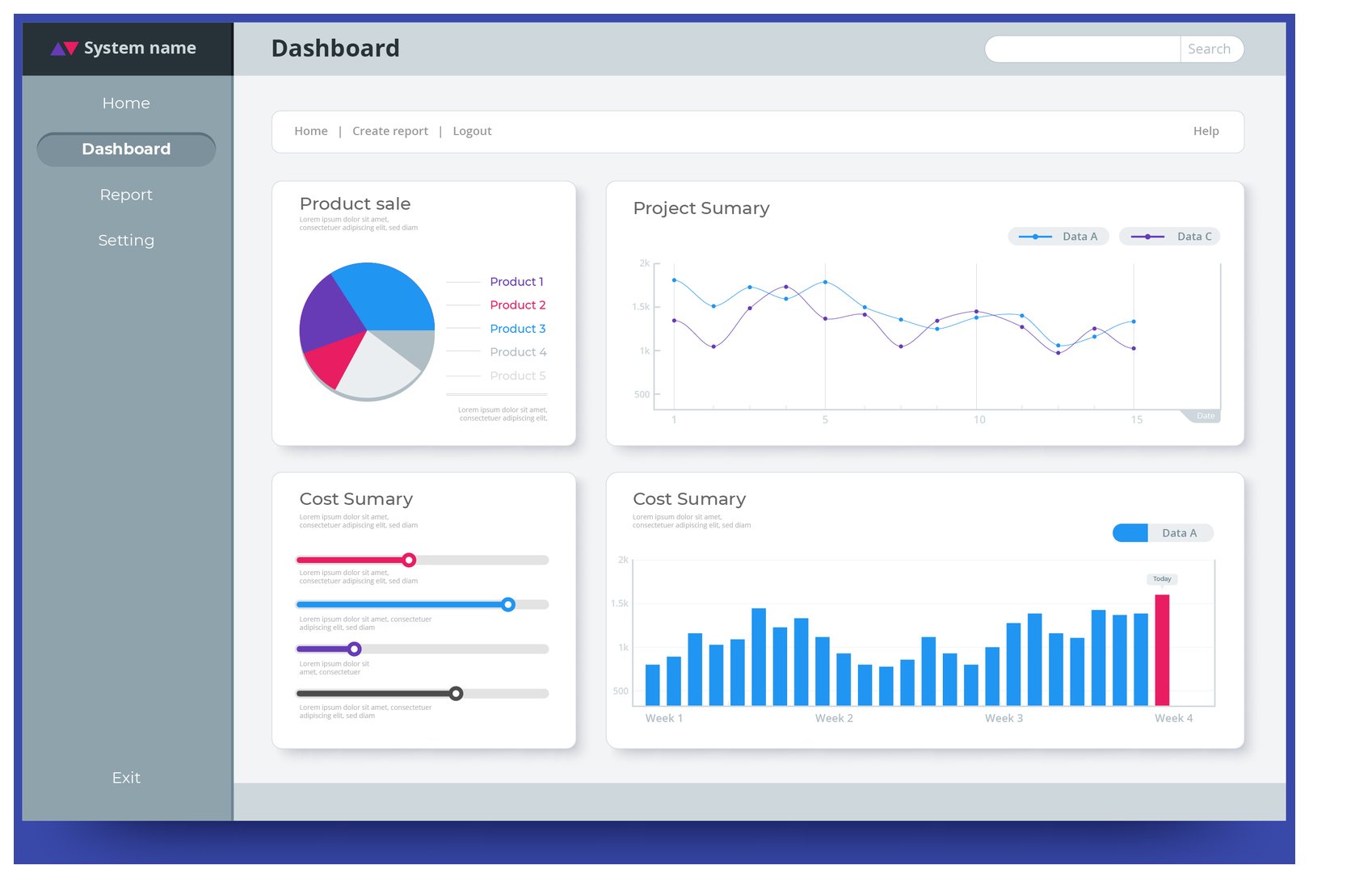}
    \caption{Input.}
  \end{subfigure}

  \vspace{0.2em}
  \begin{subfigure}{\textwidth}
    \centering
    \includegraphics[width=\linewidth,height=0.452\textheight,keepaspectratio]{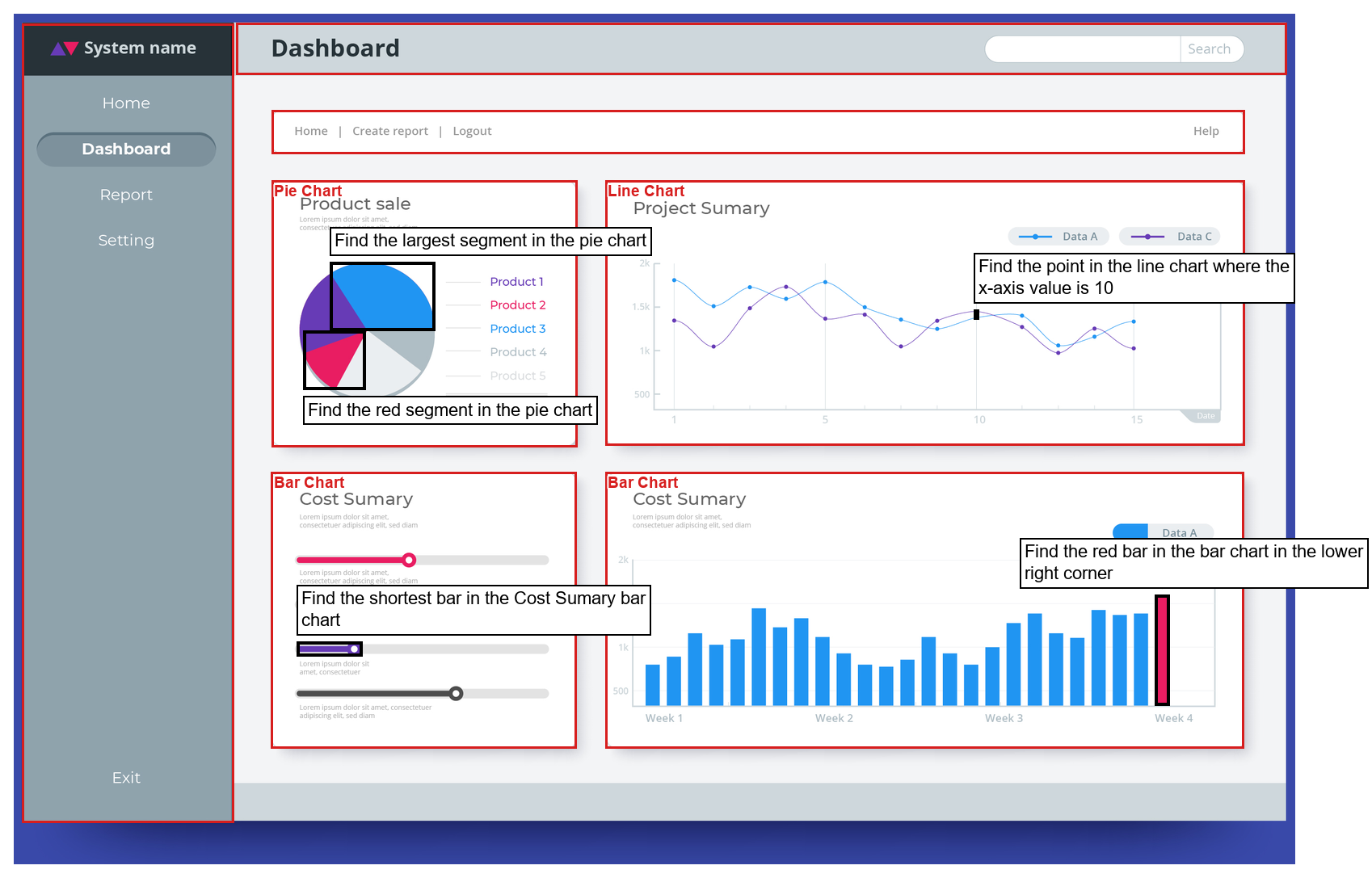}
    \caption{With annotations.}
  \end{subfigure}
  \caption{System dashboard. Source: \url{https://img.magnific.com/free-vector/dashboard-user-panel-template_23-2148375535.jpg} (Designed by Freepik, free license).}
  \label{fig:ex_b}
\end{figure*}

\begin{figure*}[t]
  \centering
  \begin{subfigure}{\textwidth}
    \centering
    \includegraphics[width=\linewidth,height=0.452\textheight,keepaspectratio]{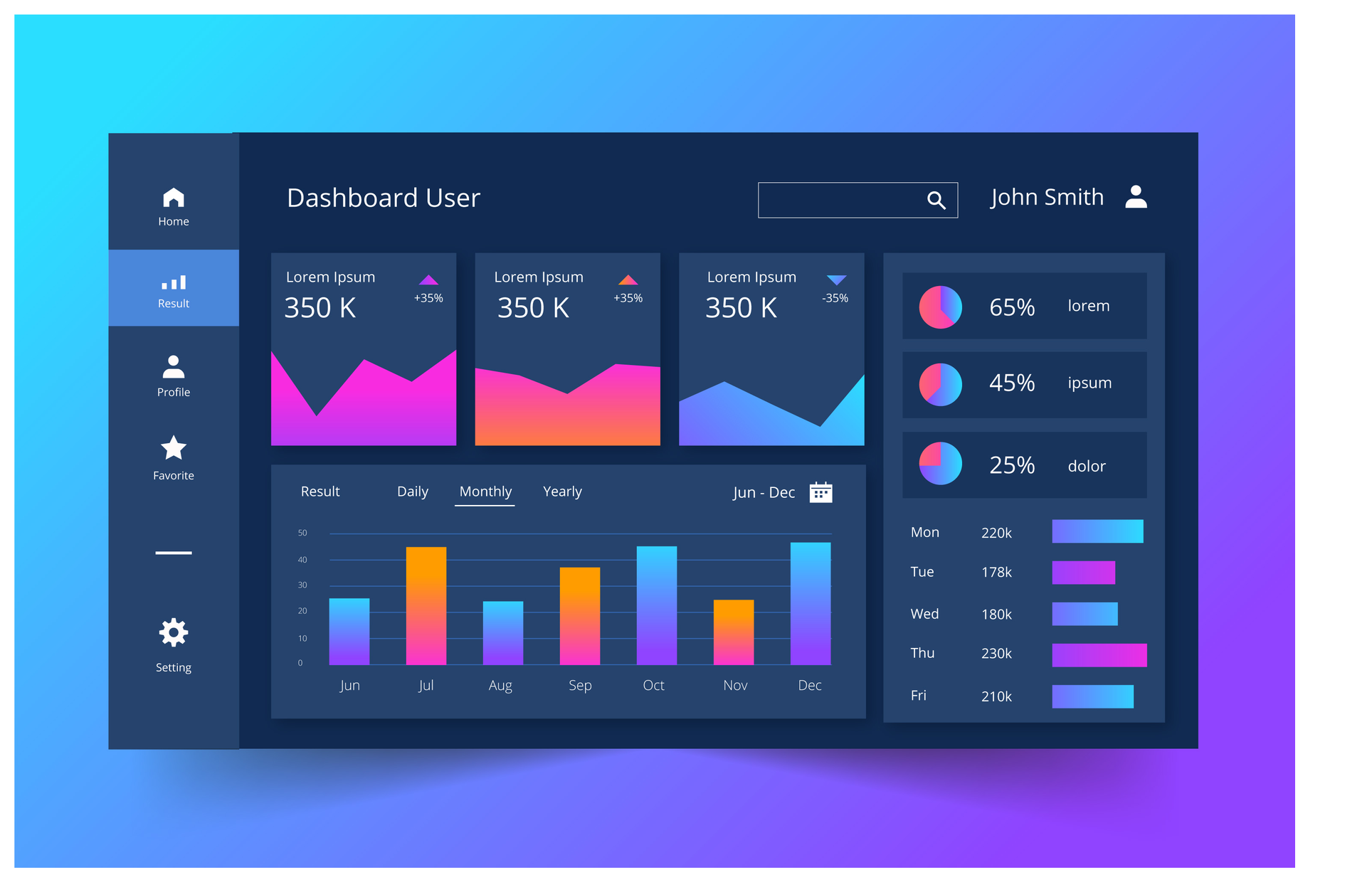}
    \caption{Input.}
  \end{subfigure}

  \vspace{0.2em}
  \begin{subfigure}{\textwidth}
    \centering
    \includegraphics[width=\linewidth,height=0.452\textheight,keepaspectratio]{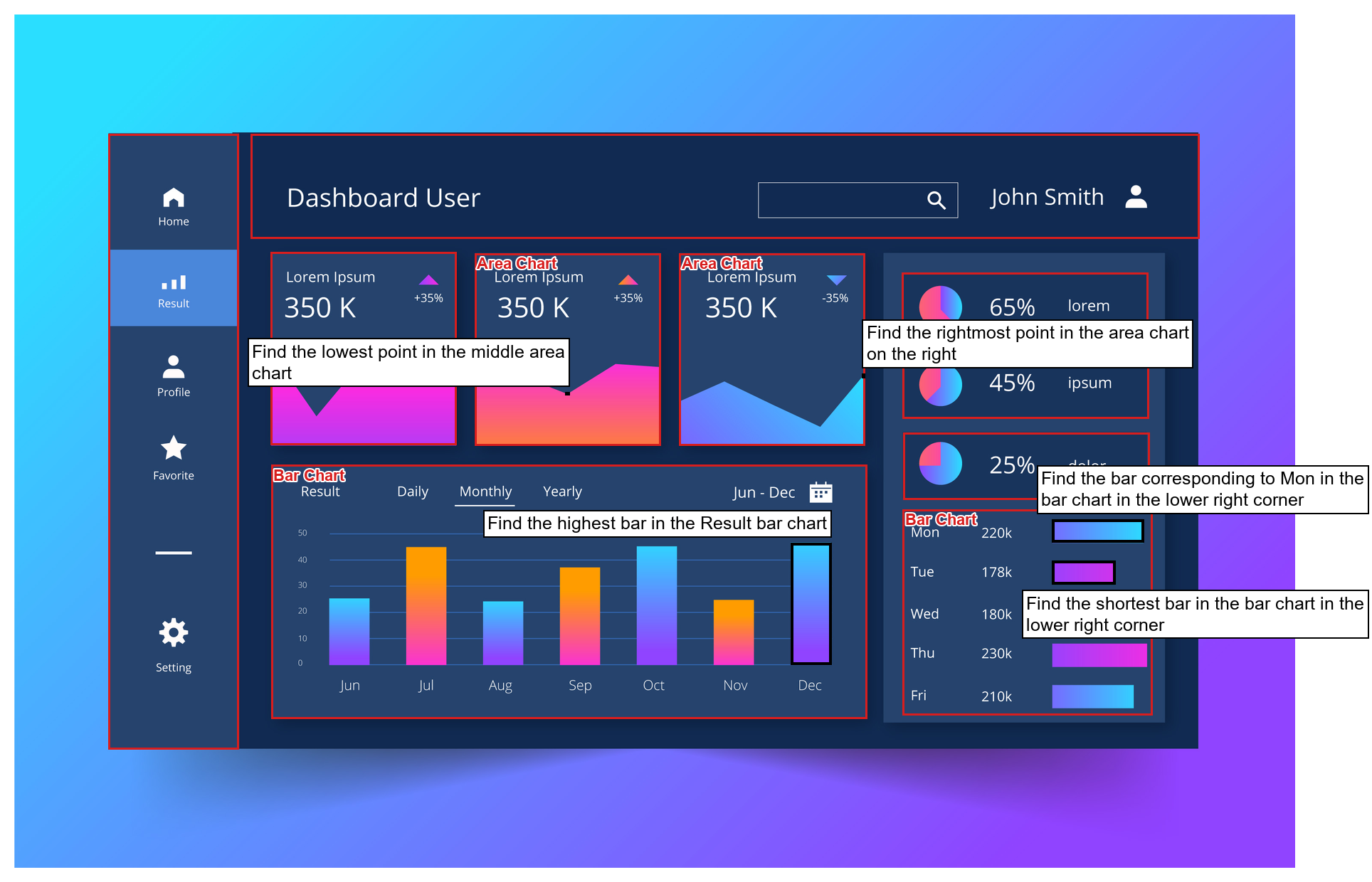}
    \caption{With annotations.}
  \end{subfigure}
  \caption{User panel dashboard. Source: \url{https://img.magnific.com/free-vector/user-panel-infographic-template-dashboard_23-2148378205.jpg} (Designed by Freepik, free license).}
  \label{fig:ex_c}
\end{figure*}

\begin{figure*}[t]
  \centering
  \begin{subfigure}{\textwidth}
    \centering
    \includegraphics[width=\linewidth,height=0.452\textheight,keepaspectratio]{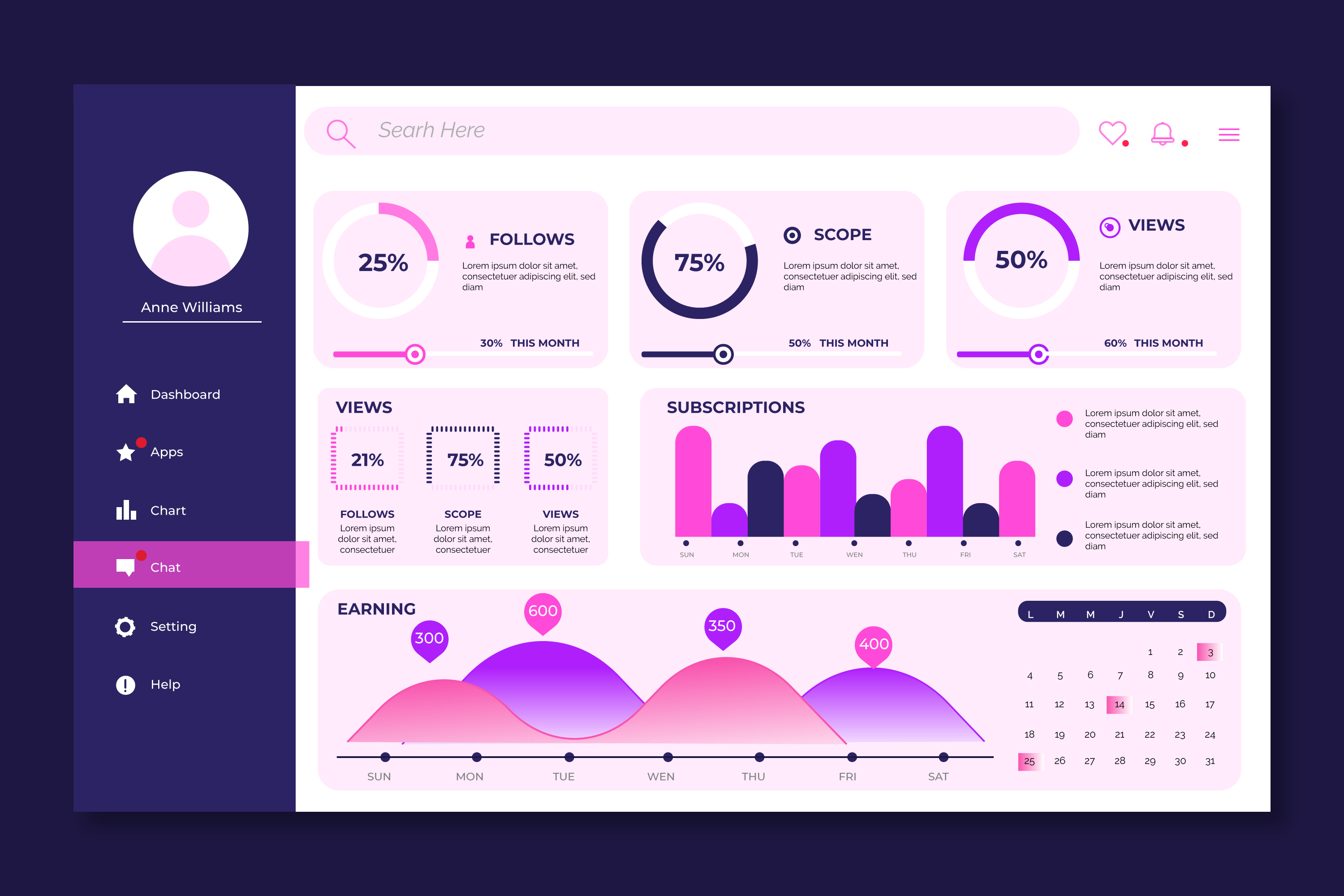}
    \caption{Input.}
  \end{subfigure}

  \vspace{0.2em}
  \begin{subfigure}{\textwidth}
    \centering
    \includegraphics[width=\linewidth,height=0.452\textheight,keepaspectratio]{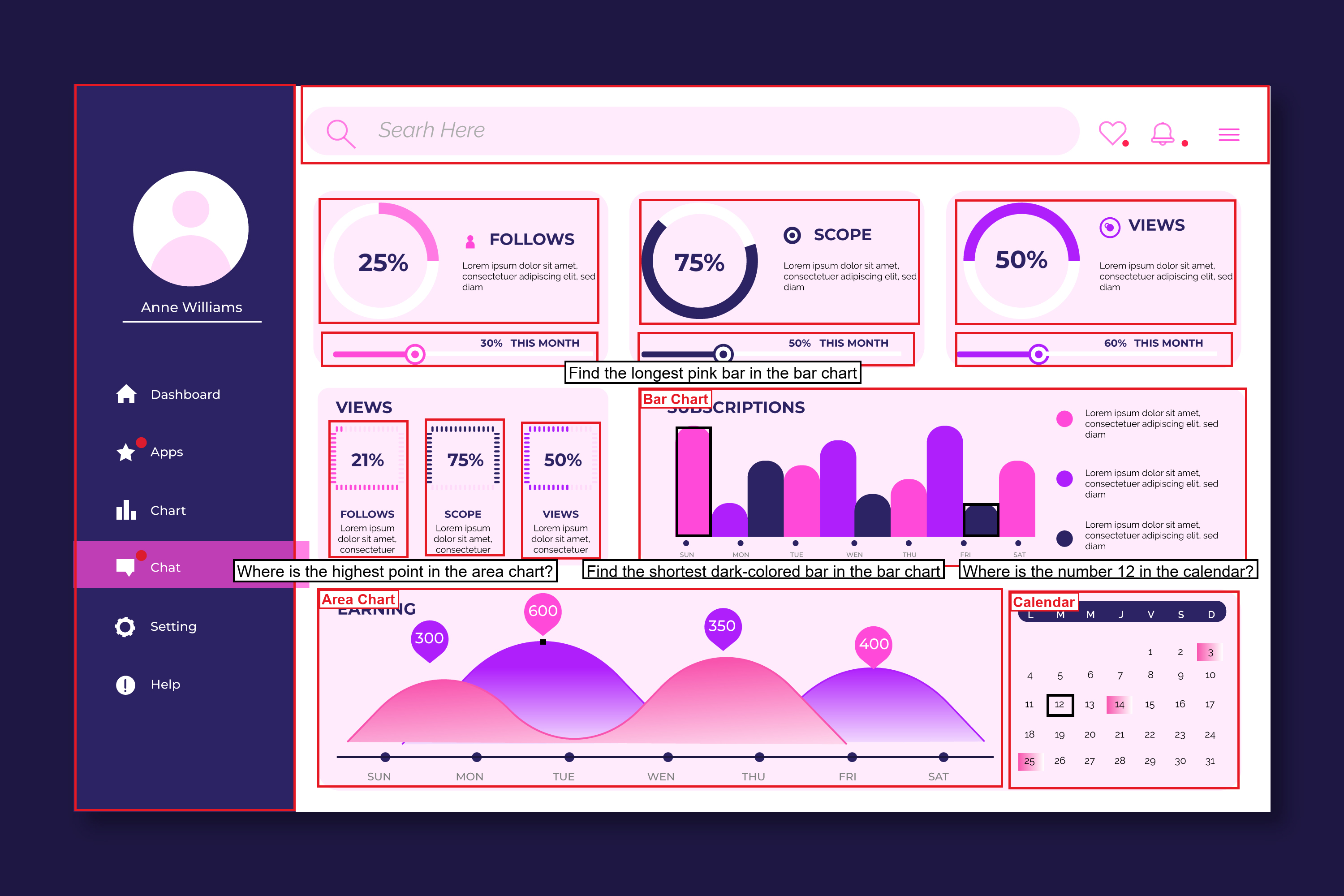}
    \caption{With annotations.}
  \end{subfigure}
  \caption{User analytics dashboard. Source: \url{https://img.magnific.com/free-vector/dashboard-user-panel-template_23-2148382158.jpg} (Designed by Freepik, free license).}
  \label{fig:ex_d}
\end{figure*}

Figures~\ref{fig:ex_a}--\ref{fig:ex_d} show four annotated dashboards from VGD-Bench, each with the input on top and the annotated version below. All four are built from Freepik templates under the Freepik free license, which requires attribution: \emph{Designed by Freepik}. Following the convention of Figure~2 in the main paper, red boxes denote region annotations (the chart, card, or UI component associated with a query) and black boxes denote target annotations (the element that must be grounded); each target is labeled with its query and each region with its chart type. All boxes are drawn directly from the JSON annotations in COCO $[x,y,w,h]$ format.

The four examples contain 9, 7, 10, and 14 region annotations and 5, 5, 5, and 4 target annotations, respectively.
Regions without a target (e.g., navigation bars and empty UI areas) are still annotated, which allows the same dashboards to support chart-retrieval, in-chart and overall VGD.

\section{Public Resource Used}

In this section, we acknowledge the use of the following public resources during the course of this work.

\textbf{Software and libraries}
\begin{itemize}
  \item PyTorch \dotfill\ PyTorch License
  \item MMDetection / MMCV \dotfill\ Apache License 2.0
  \item PaddleOCR (PP-OCRv5) \dotfill\ Apache License 2.0
  \item OpenCV \dotfill\ Apache License 2.0
  \item Hugging Face Transformers \dotfill\ Apache License 2.0
  \item timm \dotfill\ Apache License 2.0
  \item PEFT \dotfill\ Apache License 2.0
  \item FairScale \dotfill\ BSD-3-Clause License
  \item TorchMetrics \dotfill\ Apache License 2.0
  \item einops \dotfill\ MIT License
  \item NumPy \dotfill\ BSD-3-Clause License
\end{itemize}

\textbf{Pretrained models and checkpoints}
\begin{itemize}
  \item BEiT-3 \dotfill\ MIT License
  \item RT-DETR \dotfill\ Apache License 2.0
  \item YOLOv8 (Ultralytics) \dotfill\ AGPL-3.0 License
  \item Grounding DINO \dotfill\ Apache License 2.0
  \item MDETR \dotfill\ Apache License 2.0
  \item OneRef \dotfill\ Apache License 2.0
  \item SimVG \dotfill\ MIT License
  \item CLIP \dotfill\ MIT License
  \item OWLv2 \dotfill\ Apache License 2.0
  \item KOSMOS-2 \dotfill\ MIT License
  \item Florence-2 \dotfill\ MIT License
  \item LLaVA-1.5 / LLaVA-NeXT \dotfill\ Apache License 2.0
  \item PaliGemma \dotfill\ Gemma Terms of Use
  \item Qwen2.5-VL \dotfill\ Apache License 2.0
  \item Qwen3-VL \dotfill\ Apache License 2.0
  \item InternVL2.5 \dotfill\ MIT License
\end{itemize}

\textbf{Data}
\begin{itemize}
  \item Objects365 \dotfill\ Academic research use
  \item RefCOCO / RefCOCOg \dotfill\ Research use
  \item Freepik \dotfill\ Freepik License
  \item Dribbble / Tableau Public \dotfill\ Platform Terms of Use
\end{itemize}

\textbf{Commercial APIs} GPT-4o, GPT-5, Gemini-2.5 Pro, Claude, Mistral Medium 3 and DeepSeek-API were queried through their providers' public APIs under the corresponding proprietary API terms; no commercial data or weights are redistributed with this work.

\end{document}